\pdfoutput=1
\documentclass[a4paper,11pt]{article}

\usepackage{heppub}
\usepackage{xcolor}
\usepackage{subfigure}
\usepackage{booktabs}

\graphicspath{{figures/}}

\newcommand{\HeightTwoSubfigs}{0.36\textwidth}
\newcommand{\WidthTwoSubfigs}{0.49\textwidth}

\newcommand{\GeV}{\,\mathrm{GeV}}
\newcommand{\TeV}{\,\mathrm{TeV}}

\newcommand{\nn}{\nonumber}

\newcommand{\ymin}{y_{\min}}
\newcommand{\Fspl}[1]{\widehat{F}_{#1}^{\ms\text{spl}}}
\newcommand{\Fint}[2]{\widehat{F}_{#1}^{\ms\text{int}\ms #2}}
\newcommand{\muLow}{\mu_{\text{low}}}
\newcommand{\muHi}{\mu_{\text{high}}}

\newcommand{\qbar}{\bar{q}}
\newcommand{\ubar}{\bar{u}}
\newcommand{\dbar}{\bar{d}}
\newcommand{\sbar}{\bar{s}}
\newcommand{\cbar}{\bar{c}}

\newcommand{\tbar}{\bar{t}}

\newcommand{\nf}[1]{n_{#1}}

\newcommand{\conv}[1]{\underset{#1}{\otimes}}

\newcommand{\dd}{\mathrm{d}}

\newcommand{\ms}{\mskip 1.5mu}
\newcommand{\bs}{\mskip -1.5mu}

\newcommand{\chili}{\textsc{ChiliPDF}}
\newcommand{\dove}{\texttt{DOVE}}

\newcommand{\msbar}{$\overline{\text{MS}}$ }
\newcommand{\msbars}{\scalebox{0.6}{$\overline{\text{MS}}$}}

\allowdisplaybreaks[1]

\newcommand{\rev}[1]{#1}

\title{Evolution and interpolation of double parton distributions using
Chebyshev grids}

\author[a]{Markus Diehl,}
\author[b]{Riccardo Nagar,}
\author[a]{Peter Pl{\"o}{\ss}l,}
\author[a]{and Frank J.~Tackmann}

\affiliation[a]{Deutsches Elektronen-Synchrotron DESY,
Notkestr.~85, 22607 Hamburg, Germany}
\affiliation[b]{Universit\`a degli Studi di Milano-Bicocca \& INFN
Sezione di Milano-Bicocca,\\
Piazza della Scienza~3, Milano 20126, Italy
}

\emailAdd{markus.diehl@desy.de, riccardo.nagar@unimib.it,
peter.ploessl@desy.de, frank.tackmann@desy.de}

\abstract{%
Double parton distributions are the nonperturbative ingredients needed for
computing double parton scattering processes in hadron-hadron collisions.  They
describe a variety of correlations between two partons in a hadron and depend on
a large number of variables, including two independent renormalization scales.
This makes it challenging to compute their scale evolution with satisfactory
numerical accuracy while keeping computational costs at a manageable level. We
show that this problem can be solved using interpolation on Chebyshev grids,
extending the methods we previously developed for ordinary single-parton
distributions. Using an implementation of these methods in the C++ library
\textsc{ChiliPDF}, we study for the first time the evolution of double parton
distributions beyond leading order in perturbation theory.
}
\date{v1: May 8, 2023 \\ v2: June 26, 2023}

\preprint{\vbox{%
\hbox{DESY 23-062}}
}

\arxivnumber{2305.04845}

\begin{document}

\maketitle

%%%%%%%%%%%%%%%%%%%%%%%%%%%%%%%%%%%%%%%%%%%%%%%%%%%%%%%%%%%%%%%%%%%%%%%%%%%%%%%%
\section{Introduction}
\label{sec:intro}
%%%%%%%%%%%%%%%%%%%%%%%%%%%%%%%%%%%%%%%%%%%%%%%%%%%%%%%%%%%%%%%%%%%%%%%%%%%%%%%%

Double parton scattering (DPS) is a mechanism in hadron-hadron collisions in
which two partons from each hadron initiate two separate hard-scattering
processes.  The Tevatron and the LHC have provided evidence for this mechanism
for a wide range of final states with heavy flavors, jets, photons, or
electroweak gauge bosons \cite{Abe:1997xk, Abazov:2015nnn, Aaij:2016bqq,
ATLAS:2019jzd, CMS:2019jcb, CMS:2021lxi}.  While often suppressed compared with
single-parton scattering, DPS can be significant and even dominate in specific
final states or  kinematic regions.  A prominent example is like-sign $W$
production \cite{CMS:2019jcb, CMS:2022pio, Kulesza:1999zh, Gaunt:2010pi,
Golec-Biernat:2014nsa, Ceccopieri:2017oqe, Cotogno:2018mfv, Cotogno:2020iio},
which (via its leptonic decay) also provides a background to searches for
physics beyond the Standard Model \cite{Khachatryan:2016kod, Sirunyan:2018yun}.
Significant progress in the theory description of DPS has been made in the last
decade \cite{Blok:2010ge, Gaunt:2011xd, Ryskin:2011kk, Diehl:2011yj,
Manohar:2012jr, Ryskin:2012qx, Gaunt:2012dd, Blok:2013bpa, Diehl:2017kgu}, and
there is a sustained interest in phenomenological aspects (see for instance the
recent study in \cite{Fedkevych:2020cmd} and the references therein).  A detailed
account of theoretical and experimental aspects of DPS is given
in \refcite{Bartalini:2017jkk}.

Double parton distributions (DPDs) are the nonperturbative quantities needed to
compute DPS cross sections.  They quantify a variety of correlations between two
partons in the proton \cite{Diehl:2011yj, Manohar:2012jr, Kasemets:2017vyh} and
thus reveal aspects of hadron structure that are not accessible in ordinary
parton distributions (PDFs).  Our knowledge of DPDs is still quite limited,
although there is considerable activity in devising theory guided  ans\"atze
\cite{Gaunt:2009re, Golec-Biernat:2015aza, Diehl:2020xyg, Golec-Biernat:2022wkx}
and in computing DPDs using lattice QCD \cite{Bali:2021gel} or quark models
\cite{Chang:2012nw, Rinaldi:2013vpa, Broniowski:2013xba,  Rinaldi:2014ddl,
Broniowski:2016trx, Kasemets:2016nio, Rinaldi:2016jvu, Rinaldi:2016mlk}.

Reflecting the multitude of information they contain, DPDs depend on a large
number of variables, namely the momentum fractions $x_1$ and $x_2$ of the two
partons, the transverse distance $y$ between them, and the renormalization
scales $\mu_1$ and $\mu_2$ associated with each parton (corresponding to the
factorization scales in the two hard-scattering processes that the partons
initiate).  Their scale dependence is well understood (see \eq{double-dglap}
below), but the numerical delivery of evolved DPDs is a computational challenge.
 We believe that this represents one of the bottlenecks for using realistic
forms of DPDs in predictions, either at the analytical level or in the form of
Monte Carlo event generators \cite{Corke:2011yy, Blok:2015afa, Blok:2015rka,
Cabouat:2019gtm, Cabouat:2020ssr}.  In the present work, we present a method
that addresses this challenge, allowing for high numerical accuracy at a
moderate computational cost.

We recall that the evolution equations for PDFs can be solved numerically either
by discretization in the momentum fraction $x$ or in Mellin space, and that
public codes are available for both options, see for instance
\refscite{Cafarella:2008du, Salam:2008qg, Botje:2010ay, Bertone:2013vaa,
Bertone:2017gds} and \cite{Weinzierl:2002mv, Vogt:2004ns, Candido:2022tld}.  For
DPDs, even the simplest realistic initial conditions for evolution have a
correlated dependence on $x_1$ and $x_2$ (see \eqs{F-spl}{F-int} below), which
precludes the analytic computation of complex Mellin moments in these two
variables.  As far as we can see, this makes a discretization in $x_1$ and $x_2$
inevitable.

The main reason why the handling of evolved distributions is much more demanding
for DPDs than for PDFs is the sheer amount of computer memory needed to store the
former. As an example, consider interpolation grids for $x_1$ and $x_2$ with $64$
points (for PDFs, this is at the lower end of typical grid sizes in the LHAPDF
interface \cite{Buckley:2014ana}), and grids with $48$ points for $y$, $\mu_1$,
and $\mu_2$.  This corresponds to $64^{2} \times 48^3 \approx 4.5 \times 10^8$
real numbers and thus (with 8 bytes for a double precision floating-point number)
to $3.4 \operatorname{GiB}$ per parton flavor combination.  A full unpolarized
DPD with $5$ active quark flavors has $11^2 = 121$ parton flavor combinations.
Getting evolved DPDs from pre-computed grids would thus require an enormous
amount of memory (with an added penalty of accessing this memory for
interpolation in the five variables).  The path we have chosen instead is to
store a DPD only for a pair of initial scales, and to evolve ``on the fly'' to
the desired values of $\mu_1$ and $\mu_2$.  In the above example, this reduces
the memory imprint to $64^2 \times 48$ points and thus to $1.5
\operatorname{MiB}$ per flavor combination.

We will see in \subsec{dglap_evolution} that the computational effort for
evolving a DPD with $p_x$ grid points in $x_1$ and $x_2$ scales like $p_x^3$,
whilst the scaling with the number $p_y$ of points in $y$ is at most linear.  It
is therefore of great importance to limit the size of interpolation grids for
the momentum fractions, without sacrificing numerical accuracy.  In our previous
work \cite{Diehl:2021gvs} we found that a highly accurate interpolation and
evolution of PDFs in $x$ space is possible using Chebyshev grids.  In the
present paper, we show how to adapt this approach to the case of DPDs.  We find
that with $p_x \sim 54$ and $p_y \sim 40$ to $56$ points, one can achieve high
accuracy for the interpolation and integration of DPDs with $x_1, x_2 \ge
10^{-5}$ and $y \ge (7 \TeV)^{-1}$.

We have implemented our approach in the C++ library \chili,%
\footnote{\underline{Ch}ebyshev \underline{I}nterpolation
\underline{Li}brary for \underline{PDF}s}
which is under development and which will be made public in the future.  Our
implementation demonstrates that the methods we will describe do work in
practice.
It also allows us to explore a number of physics aspects.  So far, DPD evolution
has been studied only with DGLAP kernels at leading order (LO). We extend this
to next-to-leading and next-to-next-to-leading order (NLO and NNLO).  Moreover,
our implementation includes the change in the number of active quark flavors at
specified matching scales, and access to distributions with different scales
($\mu_1$ and $\mu_2$) and flavor numbers ($\nf{1}$ and $\nf{2}$) for the two
partons.  Last but not least, we can evolve DPDs for polarized partons, which
may for instance have measurable impact on observables in like-sign $W$
production according to the studies in \refscite{Cotogno:2018mfv,
Cotogno:2020iio}.

This paper is organized as follows.  In \sec{theory-basics} we recall some
basics about DPDs and set up the corresponding notation.  Using our
implementation for a physics study, we compare in \sec{result-plots} the
evolution of DPDs at LO, NLO, and NNLO.  In \sec{chilipdf-basics}, we summarize
the method we developed in \refcite{Diehl:2021gvs} for interpolating and evolving
PDFs and describe its extension to DPDs.  We study the joint interpolation in
the two momentum fractions $x_1$ and $x_2$ in \sec{x1-x2-interpolation}, and the
evolution in the two scales $\mu_1$ and $\mu_2$ in \sec{dpd-evolution}.
\Sec{y-interpolation} is devoted to the interpolation in the distance $y$
between the partons, from some minimum value to infinity.  In \sec{dove} we
cross check our implementation against the DPD evolution code \dove, which was
originally introduced in \refcite{Gaunt:2009re} and further developed in
subsequent work.  We conclude in \sec{conclusions}.  In \app{path-indep} we
prove an important statement about the independence of DPD evolution and flavor
matching on the path taken in the $(\mu_1, \mu_2)$ plane.

%%%%%%%%%%%%%%%%%%%%%%%%%%%%%%%%%%%%%%%%%%%%%%%%%%%%%%%%%%%%%%%%%%%%%%%%%%%%%%%%
\section{Basics and main results}
\label{sec:basics-results}
%%%%%%%%%%%%%%%%%%%%%%%%%%%%%%%%%%%%%%%%%%%%%%%%%%%%%%%%%%%%%%%%%%%%%%%%%%%%%%%%

%===============================================================================
\subsection{Theory framework}
\label{sec:theory-basics}
%===============================================================================

To begin with, let us recall some properties of DPDs that are relevant to this
work.  A DPD $F^{\nf{1}, \nf{2}}_{a_1 a_2}(x_1, x_2, y; \mu_1, \mu_2)$ depends on
the longitudinal momentum fractions $x_1$ and $x_2$ of the two partons, on their
distance $y$ in the transverse plane, and on the factorization scales $\mu_1$ and
$\mu_2$ associated with each parton.  The labels $a_1$ and $a_2$ specify the
flavor and polarization of each parton.  DPDs involving transverse quark or
linear gluon polarization have open Lorenz indices and depend on the  difference
$\vec{y}$ of the transverse parton positions rather than on the length $y$ of
this vector.  As specified in section~2 of \refcite{Diehl:2013mla}, they can be
decomposed into scalar distributions that depend on $y$.  The evolution and
matching equations discussed below hold separately for these scalar
distributions.  Throughout this work, we consider only DPDs in the color-singlet
channel, i.e.\ where the color is summed over separately for each of the two
extracted partons.

Notice that we allow different numbers $\nf{1}$ and $\nf{2}$ of active quark
flavors for the two partons, which is useful when the hard-scattering processes
initiated by $a_1$ and $a_2$ take place at very different scales.  Technically,
$\nf{1}$ and $\nf{2}$ are specified by the renormalisation prescription for the
twist-two operator pertaining to parton $a_1$ and $a_2$, respectively.

The scale dependence of a DPD is given by the evolution equations
\begin{align}
   \label{eq:double-dglap}
   \frac{\dd}{\dd \ln \mu_1^2} \,
      F^{\nf{1}, \nf{2}}_{a_1 a_2}(x_1,x_2,y; \mu_1, \mu_2)
   &= \sum_{b_1}
      \Bigl[ P_{a_1 b_1}^{\nf{1}}(\mu_1)
      \conv{1} F^{\nf{1}, \nf{2}}_{b_1 a_2}(x_2, y; \mu_1, \mu_2) \Bigr](x_1)
   \,,
   \nonumber \\
   \frac{\dd}{\dd \ln \mu_2^2} \,
      F^{\nf{1}, \nf{2}}_{a_1 a_2}(x_1,x_2,y; \mu_1, \mu_2)
   &= \sum_{b_2}
      \Bigr[ P_{a_2 b_2}^{\nf{2}}(\mu_2)
      \conv{2} F^{\nf{1}, \nf{2}}_{a_1 b_2}(x_1, y; \mu_1, \mu_2) \Bigr](x_2)
\end{align}
with separate Mellin convolutions
\begin{align}
   \label{eq:convol-def}
   \Bigr[ P(\mu_1) \conv{1} F(x_2, y; \mu_1, \mu_2) \Bigr](x_1)
   &= \int_{x_1}^{1} \frac{\dd z}{z}\; P(z; \mu_1)\,
      F\Bigl( \frac{x_1}{z}, x_2, y; \mu_1, \mu_2 \Bigr)
   \,,
   \nn \\[0.2em]
   \Bigr[ P(\mu_2) \conv{2} F(x_1, y; \mu_1, \mu_2) \Bigr](x_2)
   &= \int_{x_2}^{1} \frac{\dd z}{z}\; P(z; \mu_2)\,
      F\Bigl( x_1, \frac{x_2}{z}, y; \mu_1, \mu_2 \Bigr)
\end{align}
for the two momentum fractions.
Although we define each convolution integral in \eq{convol-def}
with the upper integration limit equal to $1$, the effective integration range
is reduced by the support property
\begin{align}
   \label{eq:F-support-region}
   F^{\nf{1}, \nf{2}}_{a_1 a_2}(x_1,x_2,y; \mu_1, \mu_2) &= 0
   &
   \text{for } x_1 + x_2 > 1
\end{align}
of DPDs, which is conserved by the evolution equations \eqref{eq:double-dglap}.
The evolution kernels $P^{\nf{}}_{a b}(z; \mu)$ in \eq{double-dglap} are
identical to the familiar DGLAP kernels for the evolution of PDFs with $\nf{}$
active quark flavors.  Note also that the distance $y$ plays no active role in
the scale evolution, so that each ``slice'' of a DPD at fixed $y$ evolves by
itself.

We remark in passing that the homogeneous evolution equations
\eqref{eq:double-dglap} hold for DPDs depending on the interparton distance $y$.
 An inhomogeneous term appears if one integrates the distributions over
$\vec{y}$ or performs a Fourier transform from $\vec{y}$ to its conjugate
transverse momentum \cite{Diehl:2011yj, Diehl:2017kgu}.  This term is closely
related with the splitting contribution \eqref{eq:F-spl} discussed below and has
been extensively studied in the earlier literature \cite{Kirschner:1979im,
Shelest:1982dg, Snigirev:2003cq, Gaunt:2009re, Ceccopieri:2010kg}.  Throughout
this work, we will exclusively work with DPDs that depend on $y$ and evolve as
given in \eq{double-dglap}.

The transition from $\nf{} - 1$ to $\nf{}$ active flavors in a DPD is described
by matching equations
\begin{align}
   \label{eq:dpd-flavor-matching}
   F_{a_1 a_2}^{\nf{1}, \nf{2}}(x_1,x_2,y; \mu_1, \mu_2)
   &= \sum_{b_1}
      \Bigl[ A_{a_1 b_1}^{\nf{1}}(m_{\nf{1}}; \mu_1)
      \conv{1} F_{b_1 a_2}^{\nf{1} - 1, \nf{2}}(x_2, y;\mu_1, \mu_2) \Bigr](x_1)
   \,,
   \nonumber \\
   F_{a_1 a_2}^{\nf{1}, \nf{2}}(x_1,x_2,y; \mu_1, \mu_2)
   &= \sum_{b_2}
      \bigl[ A_{a_2 b_2}^{\nf{2}}(m_{\nf{2}}; \mu_2)
      \conv{2} F_{a_1 b_2}^{\nf{1}, \nf{2} - 1}(x_1, y;\mu_1, \mu_2) \Bigr](x_2)
   \,,
\end{align}
where $m_{\nf{}}$ is the mass of the quark with flavor number $\nf{}$ and the
matching kernels $A^{\nf{}}_{a b}(z, m_{\nf{}}; \mu)$ are the same as the ones
for PDFs \cite{Buza:1996wv}:
\begin{align}
   \label{eq:pdf-flavor-matching}
   f_{a}^{\nf{}}(x; \mu)
   &= \sum_{b}
      \Bigl[ A_{a b}^{\nf{}}(m_{\nf{}}; \mu)
      \conv{} f_{b}^{\nf{} - 1}(\mu) \Bigr](x)
   \,.
\end{align}

\Eq{double-dglap} is a coupled set of differential equations in $\mu_1$ and
$\mu_2$.  In \app{path-indep} we show that with an initial condition at some
point $(\mu_{0 1}, \mu_{0 2})$ one obtains a unique DPD
regardless of the evolution path taken in the $(\mu_1, \mu_2)$ plane.
We will also show that the result of successive flavor matching does not depend
on the order in which the flavor thresholds for the two partons are crossed.
Note that the path independence of evolution and flavor matching in the $(\mu_1,
\mu_2)$ plane is exact even if the perturbative expansions of the DGLAP and
matching kernels are truncated.  This is in contrast to the independence on the
scale at which the flavor matching is performed, which only holds to the
perturbative order of the truncation.

The following evolution and matching kernels are currently implemented in
\chili\ and used in the present work:
\begin{itemize}
\item DGLAP evolution for unpolarized and longitudinally polarized partons up to
NNLO (order $\alpha_s^3$).  For the NNLO kernels we use the approximate
parameterized forms given in \refscite{Moch:2004pa,Vogt:2004mw} and
\refscite{Moch:2014sna, Moch:2015usa}.  The kernels in these references were
confirmed by the calculations in \refscite{Blumlein:2021enk, Blumlein:2021ryt}.
\item DGLAP evolution up to NLO for transversely polarized quarks and linearly
polarized gluons, with the kernels given in \refscite{Vogelsang:1997ak,
Vogelsang:1998yd}.
\item Flavor matching for unpolarized partons up to NNLO, i.e.\ order
$\alpha_s^2$.  The corresponding kernels were computed independently in
\refscite{Buza:1996wv} and \cite{Ablinger:2014lka, Ablinger:2014vwa,
Behring:2014eya}, and we verified that they agree with each other.
\item Flavor matching for polarized partons up to NLO.  At order $\alpha_s$, the
matching kernels are directly proportional to the DGLAP kernels for the same
polarization.
\end{itemize}

We now take a closer look at the dependence of DPDs on the kinematic variables
$x_1, x_2$, and $y$.  The $y$ dependence is not directly observable, because DPDs
enter cross sections (differential or integrated) via so-called double parton
luminosities
\begin{align}
   \label{eq:DPD-lumi-def}
   &
   \mathcal{L}_{a_1 a_2, b_1 b_2}^{\nf{1}, \nf{2}}(x_1, x_2,
      \bar{x}_1, \bar{x}_2; \mu_1, \mu_2, y_{\min})
   \nn \\
   &\quad
   = \int \dd^2 \vec{y} \;\,
      \theta(y - \ymin) \;
      F_{a_1 a_2}^{\nf{1}, \nf{2}}(x_1,x_2,y; \mu_1, \mu_2) \,
      F_{b_1 b_2}^{\nf{1}, \nf{2}}(\bar{x}_1,\bar{x}_2,y; \mu_1, \mu_2)
   \,,
\end{align}
where $y$ is integrated over with a lower cutoff $\ymin$ of order $1 /
\min(\mu_1, \mu_2)$.  The origin of this cutoff is explained in
\refcite{Diehl:2017kgu}, where it is also shown how the dependence on this cutoff
is removed when the cross sections for single and double parton scattering are
combined.

The form in \eq{DPD-lumi-def} holds for unpolarized
and longitudinally polarized partons; for transverse quark or linear gluon
distributions additional tensors depending on $\vec{y}$ appear under the
integral.

At sufficiently small distances $y$, the dominant contribution to a DPD arises
from the perturbative splitting of a single parton into the two observed partons
$a_1$ and $a_2$.  At leading order in the coupling, this mechanism gives
\cite{Diehl:2011yj, Diehl:2017kgu}
\begin{align}
   \label{eq:F-spl}
      F_{a_1 a_2}^{\text{spl}}(x_1,x_2, y; \mu_{y}, \mu_{y})
   &= \frac{1}{\pi y^2}\,
      \frac{\alpha_s(\mu_{y})}{2\pi}\,
      V^{(1)}_{a_1 a_2, a_0} \biggl( \frac{x_1}{x_1 + x_2} \biggr)\,
      \frac{f_{a_0}(x_1 + x_2;\mu_{y})}{x_1 + x_2}
      \,,
\end{align}%
where the scale $\mu_{y}$ should be taken of order $1/y$ so as to avoid large
logarithms $\ln (y \mu)$ in higher-order corrections.  In addition, there is an
``intrinsic'' two-parton contribution, which lacks the $1/y^2$ enhancement of
the splitting part at small $y$ but grows more strongly at small $x_1$ and $x_2$
for gluons and sea quarks.

At distances $y$ of nonperturbative size, our knowledge of DPDs is rather
limited, and in practice one needs to make a model ansatz.  Where needed, we
will take recourse to the model used in \refscite{Diehl:2017kgu, Diehl:2020xyg},
where a DPD is written as a sum $F^{\text{int}} + F^{\text{spl}}$ of intrinsic
and splitting parts at all $y$.  At a starting scale $\mu_0$, the intrinsic part
of the DPD is assumed to be proportional to the product of two PDFs:
\begin{align}
   \label{eq:F-int}
      F_{a_1 a_2}^{\text{int}}(x_1,x_2, y; \mu_0, \mu_0)
   &= \frac{1}{4\pi h_{a_1 a_2}}\,
      \exp\biggl[ -\frac{y^2}{4h_{a_1 a_2}} \biggr]\,
      \rho(x_1, x_2)\,
      f_{a_1}(x_1;\mu_0)\, f_{a_2}(x_2;\mu_0)\;
      \,,
\end{align}%
with a phase space factor
\begin{align}
   \label{eq:rho-def}
   \rho(x_1, x_2)
   &=
   \frac{(1 - x_1 - x_2)^r}{(1 - x_1)^r (1 - x_2)^r}
\end{align}
to ensure that the DPD smoothly goes to zero at the kinematic boundary $x_1 +
x_2 \to 1$.  Other models in the literature take only a power $(1-x_1-x_2)^r$;
for the motivation of the form \eqref{eq:rho-def} we point to section~3.2 in
\refcite{Gaunt:2009re}.

The splitting term in the model of \refscite{Diehl:2017kgu, Diehl:2020xyg} is
given by the perturbative splitting expression \eqref{eq:F-spl} multiplied with
the same exponential factor $\exp\bigl[ - y^2 / (4 h_{a_1 a_2}) \bigr]$ as in the
intrinsic part.  This retains the correct small-$y$ limit whilst providing a more
realistic decrease at large~$y$.

%===============================================================================
\subsection{Quantitative impact of higher orders in DPD evolution}
\label{sec:result-plots}
%===============================================================================

With the methods presented in this work, we are able to evolve DPDs at different
perturbative orders, for all relevant polarization combinations and with
different scales and active flavor numbers of the two partons.  As a
demonstration, we now compare DPDs evolved at LO, NLO, and NNLO from a common
starting condition.

As starting condition we take the perturbative splitting form \eqref{eq:F-spl}
with $\nf{1} = \nf{2} = 3$ active flavors.  The splitting formula is evaluated
at $\mu_y = 2 \GeV$ and $y \approx 0.561 \GeV^{-1}$, which corresponds to the
scale choice $\mu_y = b_0 / y$ with $b_0 = 2 e^{-\gamma_E} \approx 1.12$, where
$\gamma_E$ is the Euler-Mascheroni constant.  This choice is suggested by the
form of the NLO corrections to the splitting formula \eqref{eq:F-spl} and was
found to keep these corrections at a moderate size \cite{Diehl:2019rdh,
Diehl:2021wpp}.  For the PDFs in the DPD splitting formula, we take the default
NNLO set of the MSHT20 parameterization \cite{Bailey:2020ooq}.\footnote{
We use the central PDF from the \texttt{MSHT20nnlo\_as118} set obtained via
LHAPDF \cite{Buckley:2014ana}.  The \chili\ library includes an interface for
importing data from LHAPDF.}
The parameterization assumes quark masses $m_c = 1.4 \GeV$, $m_b = 4.75 \GeV$
and a five-flavor coupling $\alpha_s(m_Z) = 0.118$.

We evolve these initial conditions to $\mu_1 = 9 \GeV$ with $\nf{1} = 3$ for the
first parton and to $\mu_2 = m_W$ with $\nf{2} = 5$ for the second one.  This is
a setting relevant for the production of a $J/\bs\Psi$ and a $W$, a channel for
which double parton scattering has in fact been observed experimentally
\cite{ATLAS:2014yjd, ATLAS:2019jzd}.
Flavor matching for the second parton is performed at the charm and bottom
quark masses, with kernels of the same order (LO, NLO, or NNLO) as the
evolution kernels.  An exception is longitudinally polarized evolution at NNLO,
where we use NLO flavor matching.

\begin{figure}
   \subfigure[
      $g g$
   ]{%
   \includegraphics[width=\WidthTwoSubfigs]{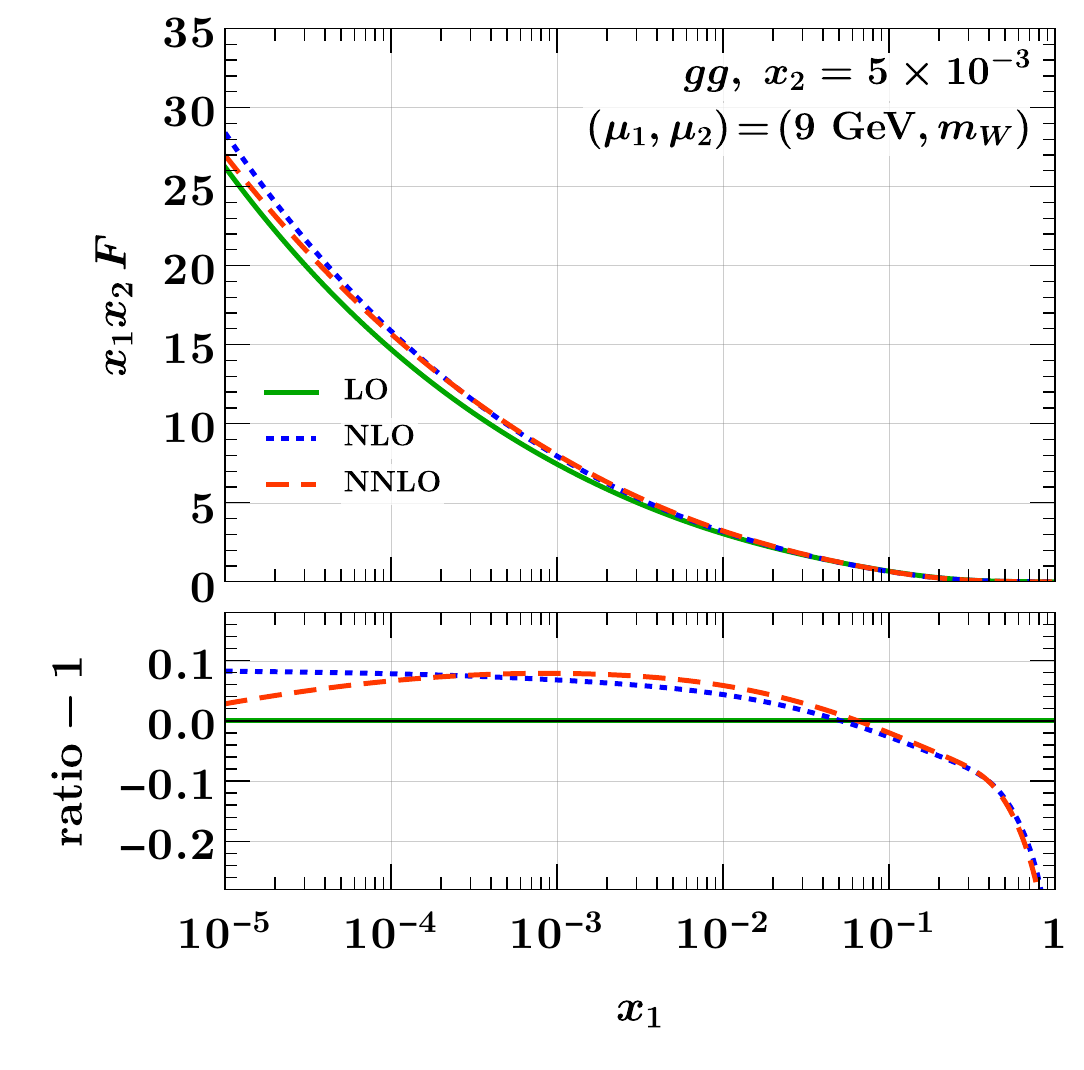}
   }
   \hfill
   \subfigure[
      $u \ubar$
   ]{%
   \includegraphics[width=\WidthTwoSubfigs]{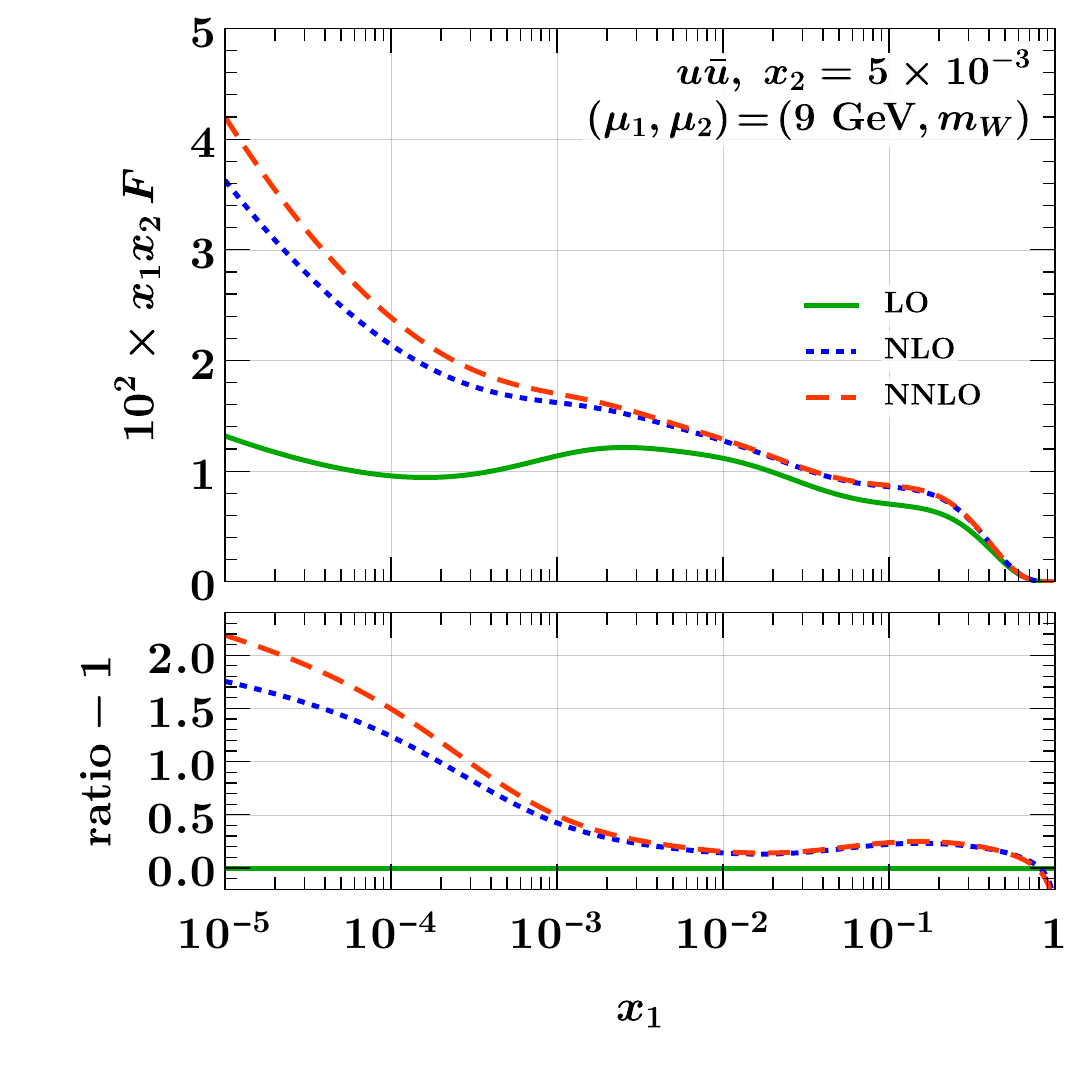}
   }%
   \\
   \subfigure[
      $u g$
   ]{%
   \includegraphics[width=\WidthTwoSubfigs]{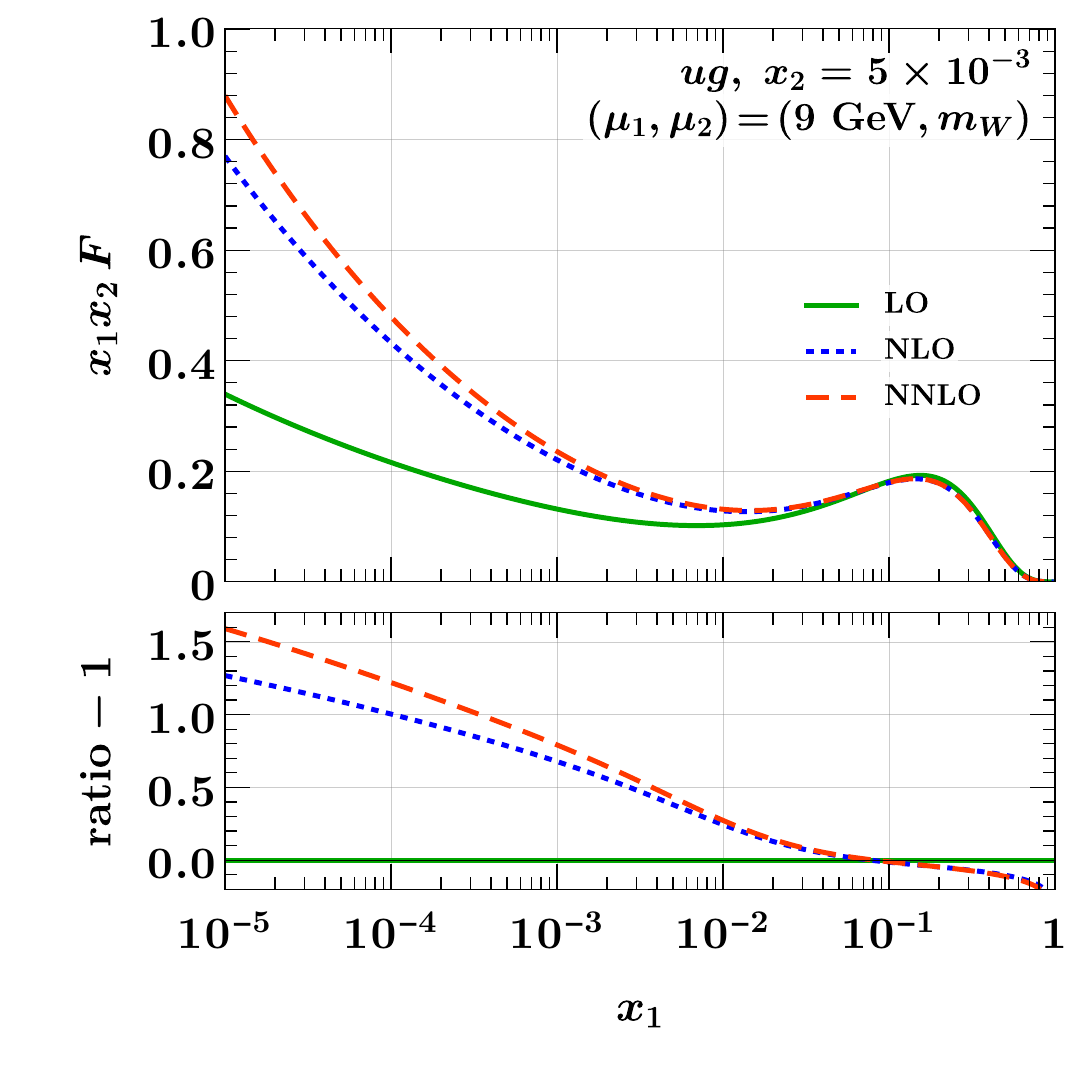}
   }
   \hfill
   \subfigure[
      $s c$
   ]{%
   \includegraphics[width=\WidthTwoSubfigs]{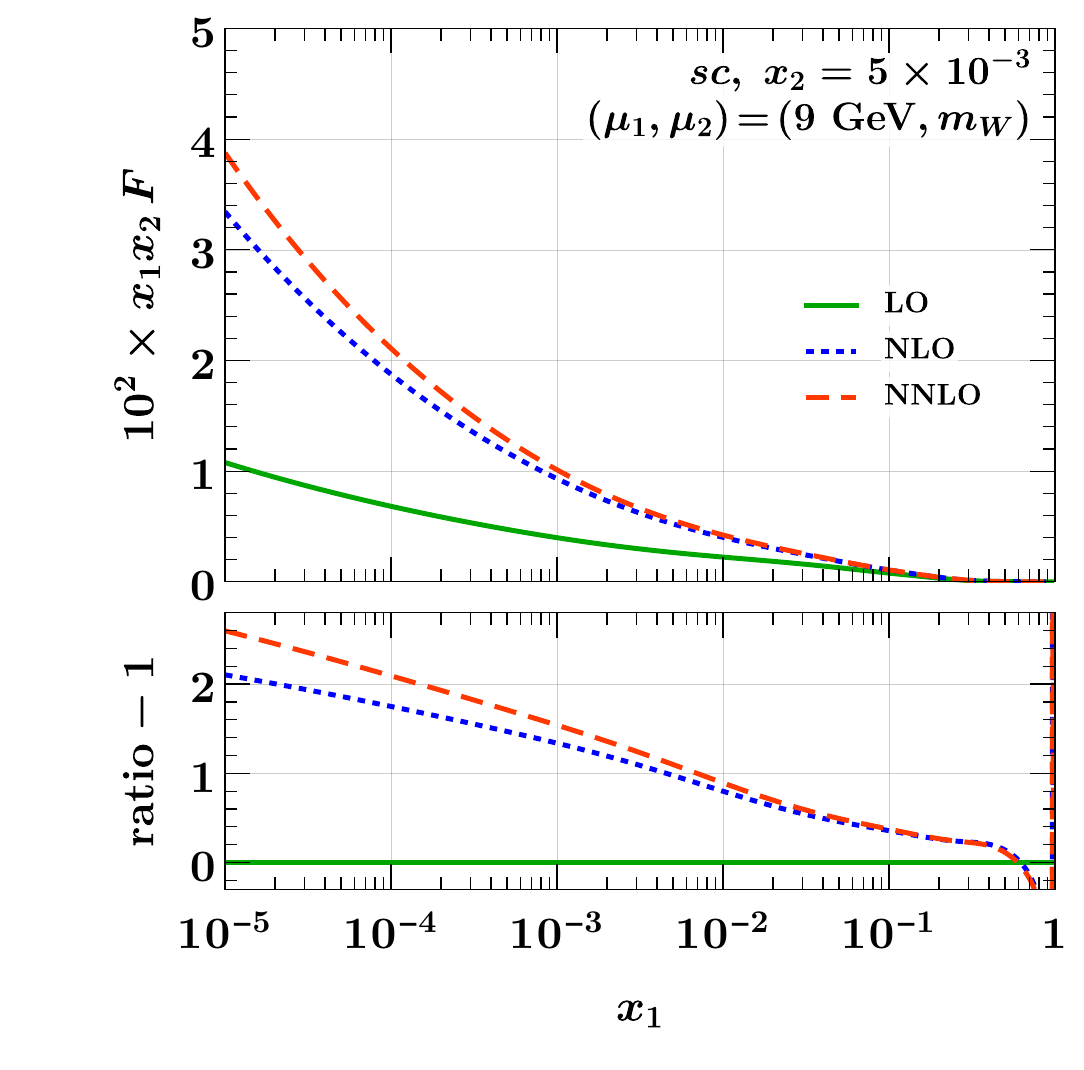}
   }
\caption{\label{fig:pert_orders_unpol}Unpolarized DPDs at $x_2 = 5 \times
10^{-3}$, evolved from $(\mu_1, \mu_2) = (2 \GeV, 2 \GeV)$ to $(9 \GeV, m_W)$ at
different perturbative orders.  The starting conditions and the procedure of
flavor matching are specified in the text.  The ratio shown in the small panels
is taken with respect to the LO result.}
\end{figure}

We show the effect of the higher-order DGLAP evolution for a selection of
unpolarized DPDs in \fig{pert_orders_unpol}. The distributions are given in the
top panel of each plot, and their ratios with respect to the leading-order
distribution in the lower panel.  At small momentum fractions, higher-order
evolution has an effect of order $10\%$ on $F_{g g}$, whereas for the other
flavor combinations shown in the figure, the effect is of order $1$. In all
cases, we find that the change from LO to NLO is much larger than the rather
moderate change from NLO to NNLO.

\begin{figure}
   \subfigure[
      $\Delta g \Delta g$
   ]{%
   \includegraphics[width=\WidthTwoSubfigs]{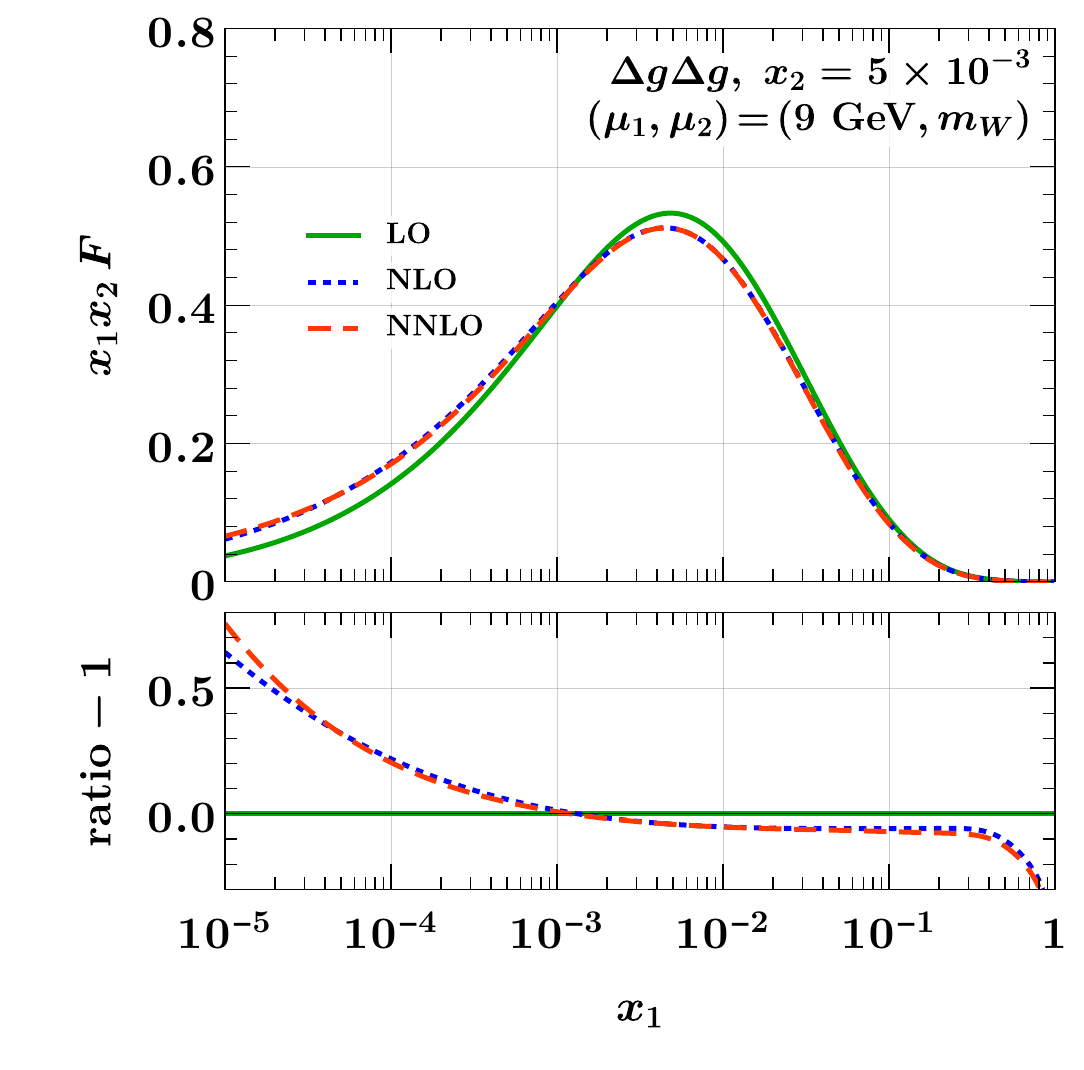}%
   }
   \hfill
   \subfigure[
      $\Delta u \Delta\ubar$
   ]{%
   \includegraphics[width=\WidthTwoSubfigs]{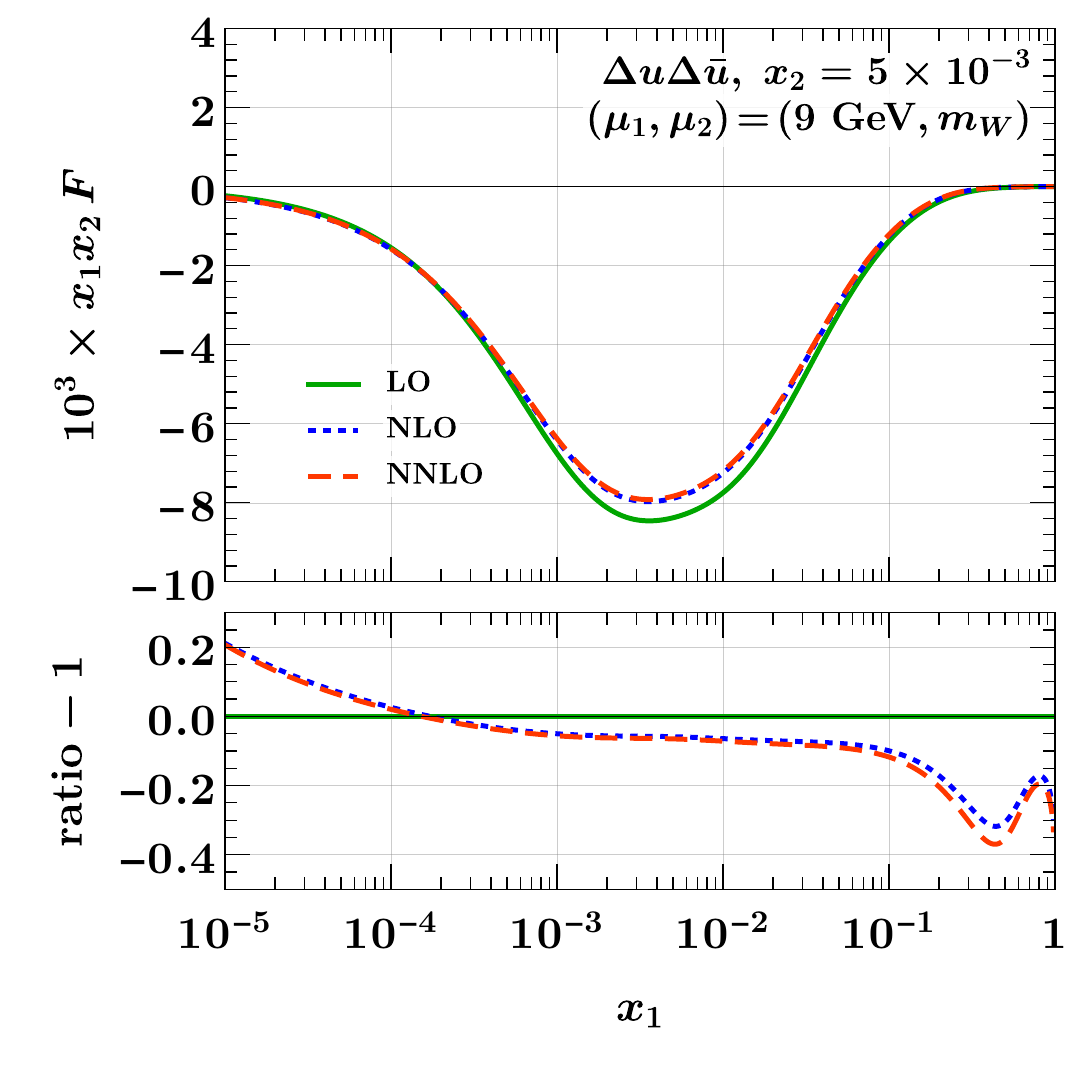}
   }
   \\
   \subfigure[
      $\delta g \delta g$
   ]{%
   \includegraphics[width=\WidthTwoSubfigs]{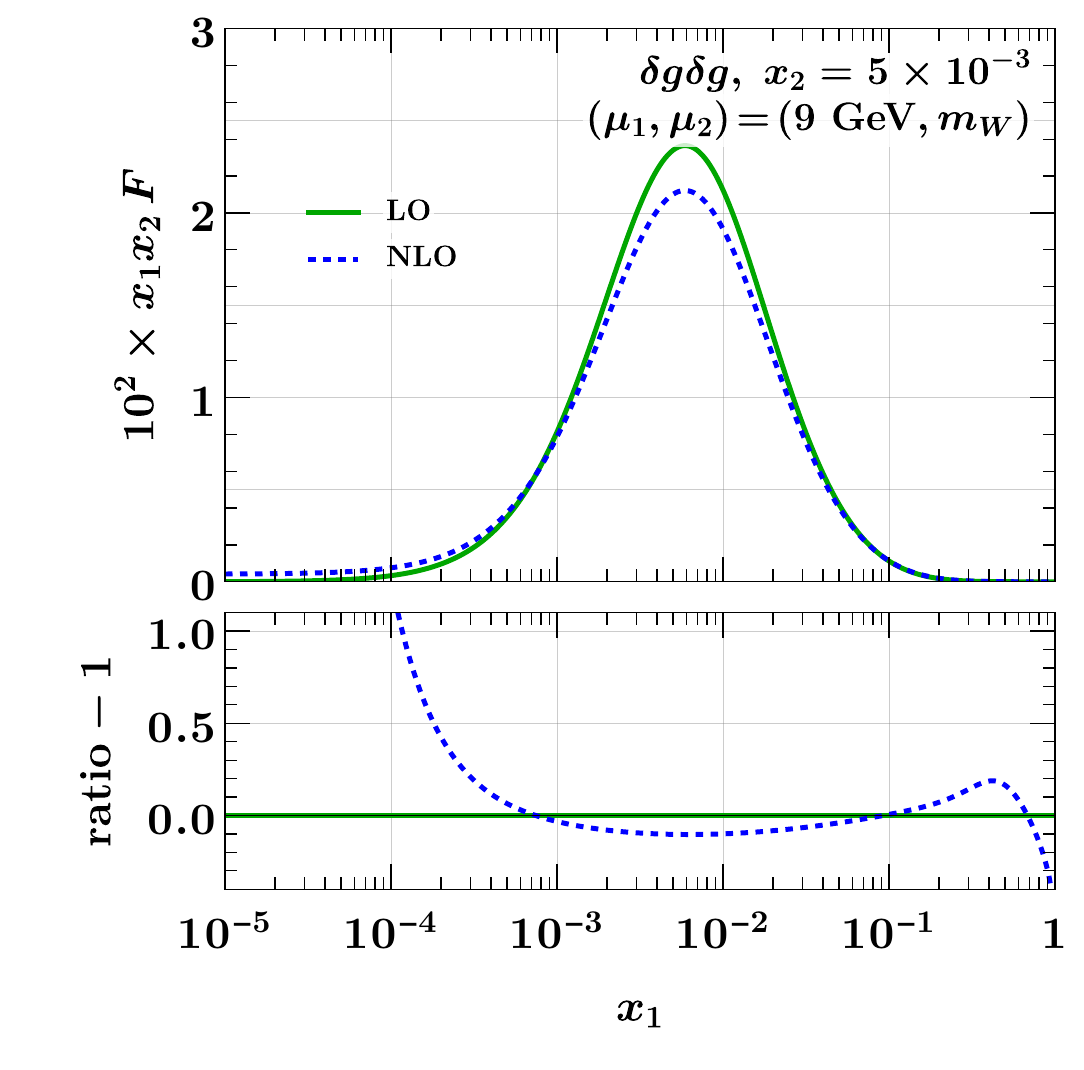}
   }
   \hfill
   \subfigure[
      $\delta u \delta\ubar$
   ]{%
   \includegraphics[width=\WidthTwoSubfigs]{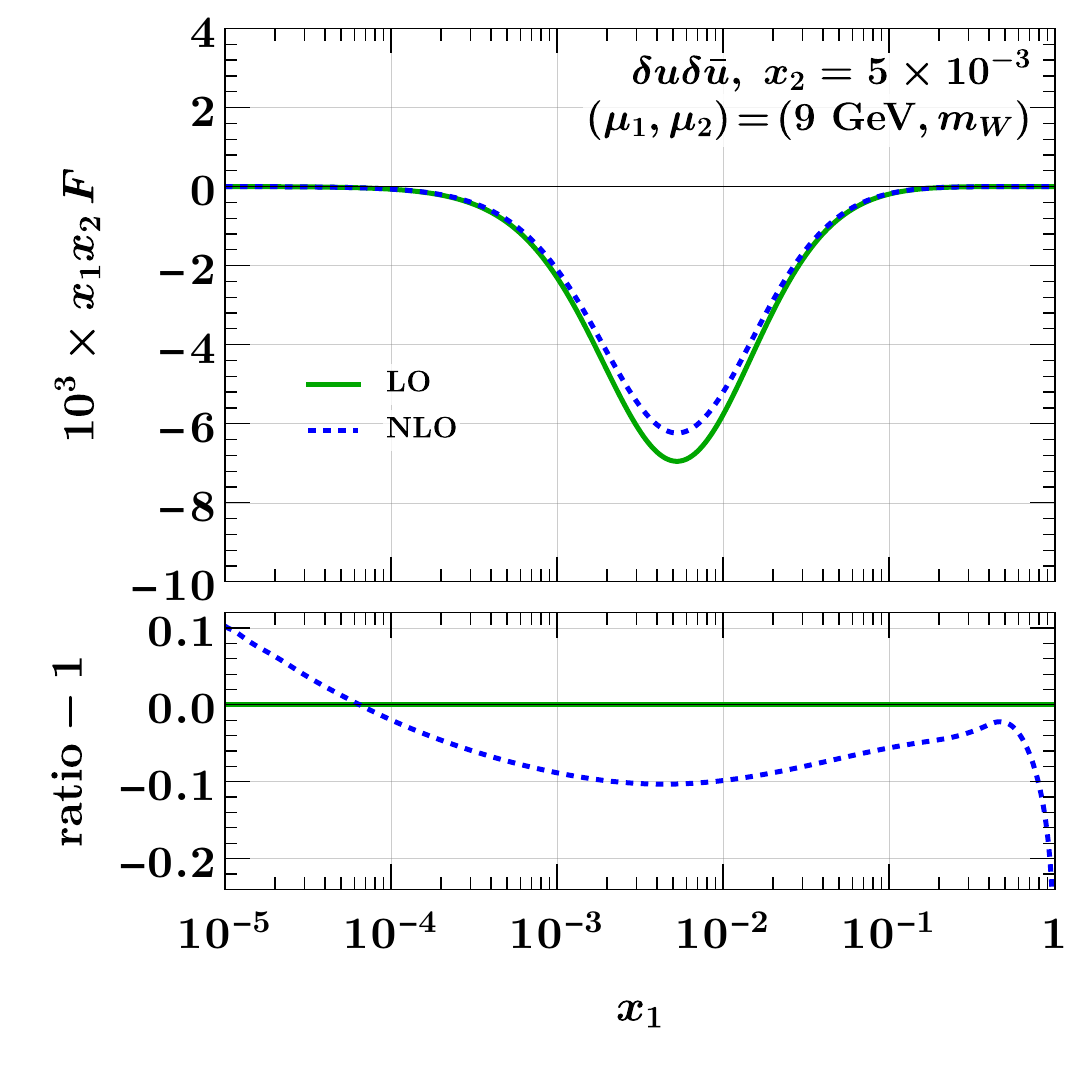}
   }
\caption{\label{fig:pert_orders_pol}As \fig{pert_orders_unpol} but for polarized
DPDs.  Panels (a) and (b) are for longitudinal, panel (c) for linear, and panel
(d) for transverse polarization.}
\end{figure}

A selection of polarized DPDs is shown in \fig{pert_orders_pol}.  For linear
gluon and transverse quark polarization, we have only implemented evolution at
LO and NLO.  We find that the NLO corrections to evolution introduce effects of
order $10\%$ for $\Delta u\Delta\ubar$ and $\delta u \delta\ubar$ in
different regions of $x_1$, where as for $\Delta g \Delta g$ they reach $70\%$
at small $x_1$.  For $\delta g \delta g$, one obtains a huge relative
difference between NLO and LO evolution in the limit $x_1 \ll x_2$.  This is
because for $z\to 0$ the splitting function $P_{\delta g \delta g}(z)$ vanishes
like $z$ at LO but grows like $1/z$ at NLO \cite{Vogelsang:1998yd}.  At both
orders, the evolved distribution at $x_1 \ll x_2$ remains however small compared
with its peak around $x_1 \sim x_2$.

As is the case for PDFs, DPDs are not observables, and distributions evolved at
different orders are combined with parton-level cross sections computed at
different orders.  Nevertheless, the strong evolution effects we observed in
several channels and kinematic regions suggest that NLO corrections in double
parton scattering can be of substantial size.

\begin{figure}
   \subfigure[
      $g g$
   ]{%
   \includegraphics[width=\WidthTwoSubfigs]{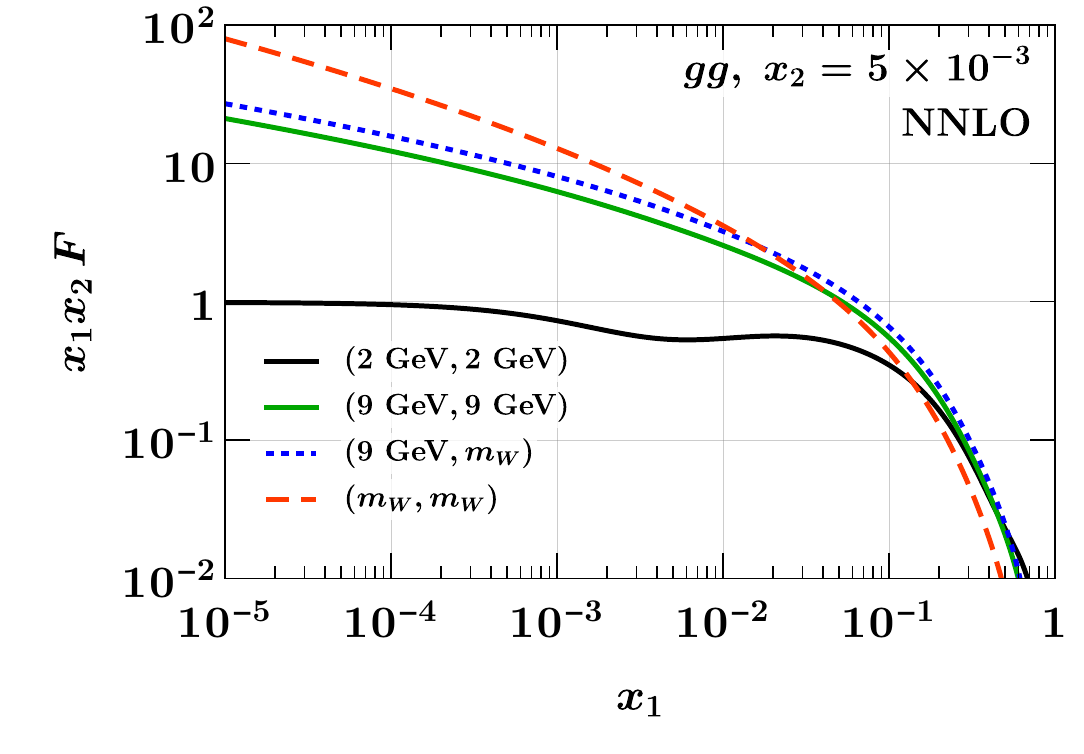}
   }
   \hfill
   \subfigure[
      $u \ubar$
   ]{%
   \includegraphics[width=\WidthTwoSubfigs]{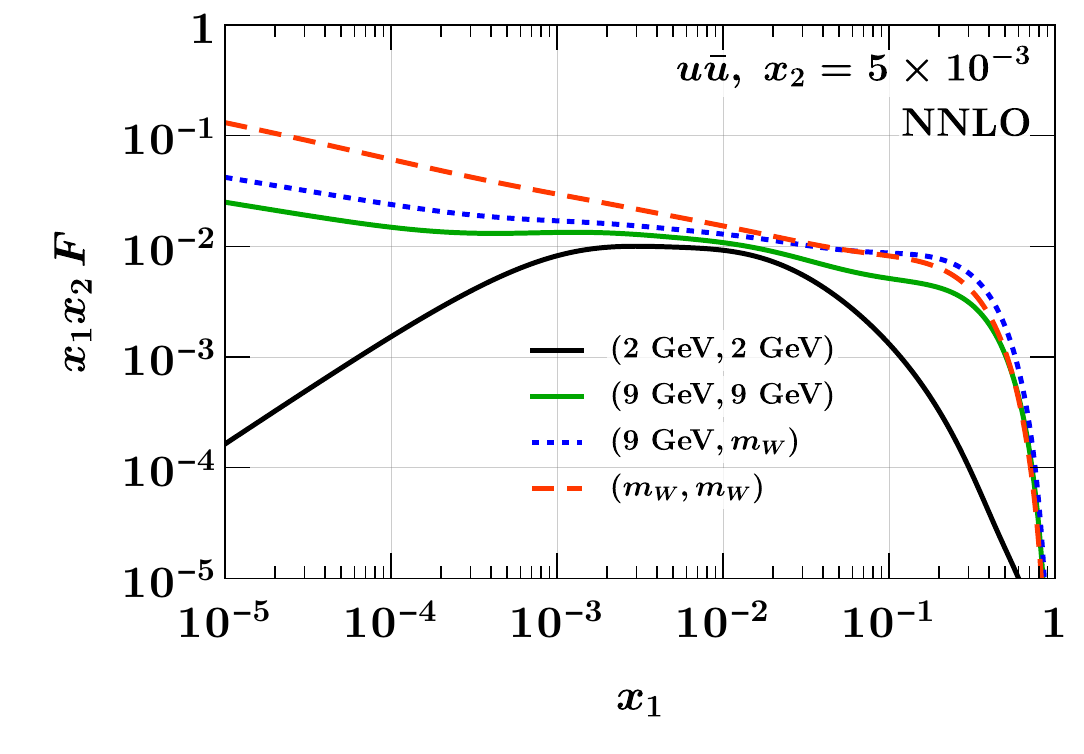}
   }
   \\
   \subfigure[
      $\Delta g \Delta g$
   ]{%
   \includegraphics[width=\WidthTwoSubfigs]{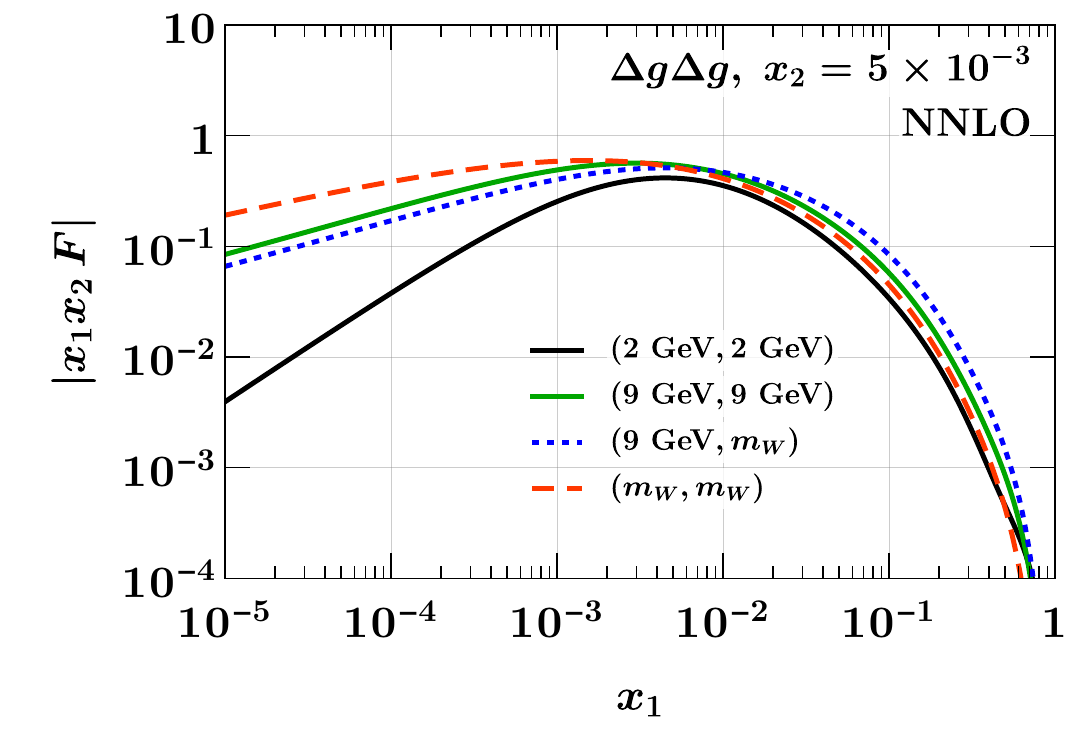}
   }
   \hfill
   \subfigure[
      $\Delta u \Delta\ubar$
   ]{%
   \includegraphics[width=\WidthTwoSubfigs]{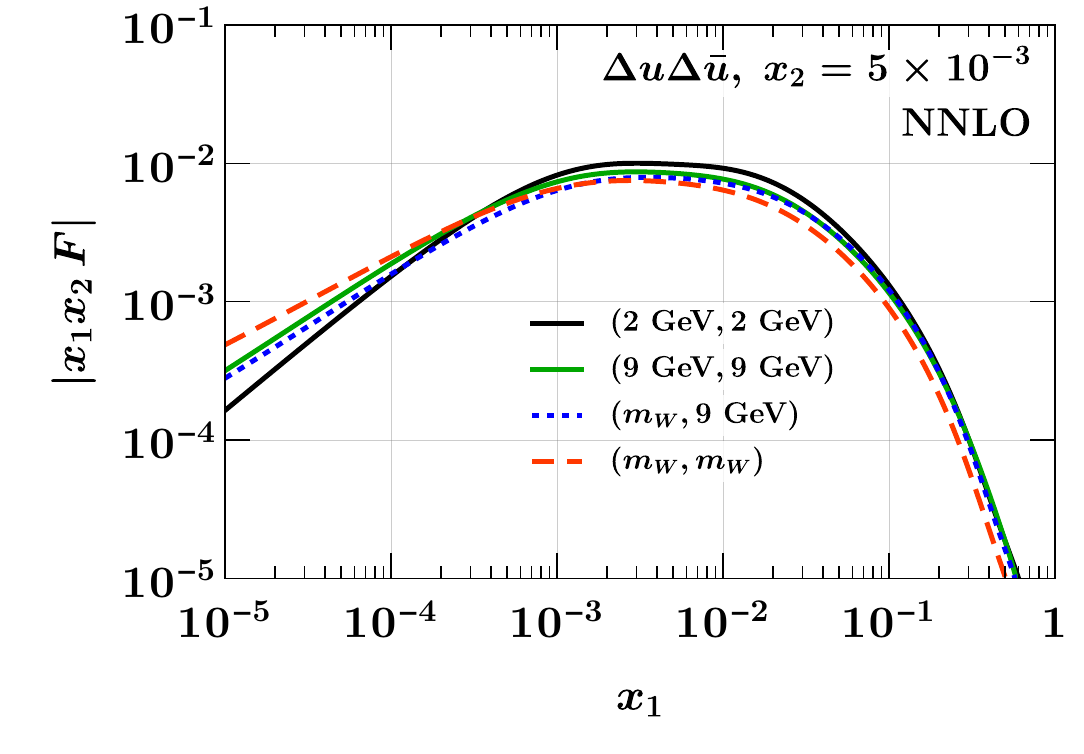}
   }
\caption{\label{fig:scale-dependence} Comparison of DPDs at different scales.  The starting conditions at $(\mu_1, \mu_2) = (2 \GeV, 2 \GeV)$ and $(\nf{1}, \nf{2}) = (3, 3)$ are the same as those used for the previous two figures.  At higher scales, the number of active flavors is $\nf{i} = 3$ for $\mu_i = 9 \GeV$ and $\nf{i} = 5$ for $\mu_i = m_W$.  Evolution and flavor matching are carried out at the highest available order.}
\end{figure}

\rev{In \fig{scale-dependence}, we illustrate the scale dependence of DPDs, using evolution at NNLO and flavor matching at the highest available order (NNLO for unpolarized partons and NLO for polarized ones).  We show the distributions at the starting point $(\mu_1, \mu_2) = (2 \GeV, 2 \GeV)$, and after evolution to different combinations of the scales $\mu_i = 9 \GeV$ or $\mu_i = m_W$ for the two partons, with associated flavor numbers $\nf{i} = 3$ in the first and $\nf{i} = 5$ in the second case.  For unpolarized partons, we find that the effect of evolution from $\mu_1 = \mu_2 = 2 \GeV$ to $9 \GeV$ is huge at small $x$, turning a shape that is flat or increasing with $x$ at the low scale into a shape decreasing with $x$ at the higher scale.  Subsequent evolution to $m_W$ for one or both partons has a significant, although less dramatic, effect.  For polarized partons, evolution is generally weaker, and we find DPDs increasing with $x$ at small $x$ even when both partons are taken at the high scale $m_W$.}

%

%%%%%%%%%%%%%%%%%%%%%%%%%%%%%%%%%%%%%%%%%%%%%%%%%%%%%%%%%%%%%%%%%%%%%%%%%%%%%%%%
\section{Interpolation and evolution of DPDs in ChiliPDF}
\label{sec:chilipdf-basics}
%%%%%%%%%%%%%%%%%%%%%%%%%%%%%%%%%%%%%%%%%%%%%%%%%%%%%%%%%%%%%%%%%%%%%%%%%%%%%%%%

In this section we sketch the mathematical framework underlying \chili,
summarizing the relevant methods presented in \refcite{Diehl:2021gvs} and
extending them from PDFs to DPDs.  In particular, we give explicit formulae for
interpolation and integration, and for the discretization of the DGLAP evolution
equations.  A detailed account of Chebyshev interpolation and its applications
can be found in \refcite{Trefethen}.\footnote{The first chapters of this book
are available on \url{https://people.maths.ox.ac.uk/trefethen/ATAP}.}
%
%===============================================================================
\subsection{Chebyshev interpolation}
\label{subsec:chebyshev}
%===============================================================================
%
We start by reviewing Chebyshev interpolation of a function $f(t)$ on the
interval $t \in [-1, 1]$.  The function is discretized on a grid consisting of
the so-called \emph{Chebyshev points}, which for given $N$ read
\begin{align}
   \label{eq:cheb_points}
      t_i
   &= \cos \Bigl( \frac{i \pi}{N} \ms \Bigr)
      \qquad
      \text{with}
      \quad
      i = 0, \ldots, N
      \,.
\end{align}%
These are the points where the $N^{\mathrm{th}}$ Chebyshev polynomial of the
first
kind, $T_N(t)$, assumes its maxima $+1$ and minima $-1$. They form a descending
series from $t_0 = 1$ to $t_N = -1$ and satisfy the symmetry property $t_{N-i} =
- t_i$.  We call the set of Chebyshev points a \emph{Chebyshev grid}.

A sufficiently smooth function $f(t)$ on $t \in [-1, 1]$ can be approximated by
the \emph{Chebyshev interpolant} $p_N(t)$.  This is a series of Chebyshev
polynomials $T_j(t)$ with $j = 0, \ldots, N$ whose expansion coefficients are
such that one has $p_N(t_i) = f(t_i)$ at the Chebyshev points.  For details see
equations (2.6) to (2.8) in \refcite{Diehl:2021gvs}.

%-------------------------------------------------------------------------------
\paragraph{Barycentric formula.}
A simple and efficient way to compute the Chebyshev interpolant without having
to evaluate Chebyshev polynomials is given by the \emph{barycentric formula}
\begin{align}
   \label{eq:barycentric}
      p_N(t)
   &= \sum_{i=0}^{N} f(t_i) \, b_i(t)
\end{align}%
with the barycentric basis functions
\begin{align}
   \label{eq:bary_basis}
      b_i(t)
   &= \beta_i \, \frac{(-1)^{i}}{t - t_i}
      \Bigg/ \sum_{j=0}^{N} \beta_j \, \frac{(-1)^{j}}{t - t_j}
      \,,
\end{align}%
where $\beta_0 = \beta_N = 1/2$ and $\beta_i = 1$ otherwise.
Even though it is not directly evident from \eq{bary_basis}, the $b_i(t)$ are
polynomials of order $N$. The number of operations for evaluating the
barycentric formula scales linearly with $N$.  The barycentric formula is found
to be numerically stable in the interpolation interval (however, this is
\emph{not} the case for extrapolating the function $f(t)$ outside this interval,
see chapter 5 of \refcite{Trefethen}).

%-------------------------------------------------------------------------------
\paragraph{Accuracy estimate.}
A computationally inexpensive estimate for the accuracy of Chebyshev
interpolation is obtained by interpolating $f(t)$ on the Chebyshev grid
\emph{without} the end points $t_0$ and $t_{N}$.  The resulting interpolant
$q_{N-2}(t)$ is a polynomial of order $N - 2$ and can be computed with an
adapted form of the barycentric formula, see equations (2.23) to (2.27) in
\refcite{Diehl:2021gvs}.  Comparing $p_{N}(t)$ with $q_{N-2}(t)$ yields an
estimate for the interpolation accuracy, i.e.\ for the difference between
$p_{N}(t)$ and $f(t)$.  This estimate was found to work well for PDF
interpolation \cite{Diehl:2021gvs}, and we will investigate in later chapters
how reliable it is in the case of DPDs.

%-------------------------------------------------------------------------------
\paragraph{Integration.}
Using the expansion of $f$ in terms of Chebyshev polynomials $T_i(t)$ and the
analytic form for integrals of these polynomials, one readily obtains the
integration rule
\begin{align}
   \label{eq:CC-rule}
      \int_{-1}^1 \dd t \, f(t)
   &  \approx \sum_{i=0}^{N}\, w_i\ms f(t_i)\;
\end{align}%
%%%
with weights
\begin{align}
   \label{eq:CC-weights}
      w_i
   &= \frac{4 \beta_i}{N}\,
      \sum_{\genfrac{}{}{0pt}{}{j=0}{\text{even}}}^{N}\,
      \beta_j\, \frac{\cos (j \ms \theta_i)}{1-j^2}
      \,.
\end{align}%
This integration rule is known as \emph{Clenshaw-Curtis quadrature}. For a
detailed discussion of its accuracy (and comparison with Gauss quadrature), we
refer to chapter 19 of \refcite{Trefethen} and to section 2 of
\refcite{Diehl:2021gvs}.

An estimate for the integration accuracy can be obtained by using a quadrature
rule on the Chebyshev grid without its end points, in full analogy to the
procedure described above for interpolation.  The result is known as
\emph{Fej{\'e}r's second rule}. Details are given in equations (2.31) to (2.33)
of \refcite{Diehl:2021gvs}.

%===============================================================================
\subsection{Interpolation strategy}
\label{subsec:interpolation_strategy}
%===============================================================================
%
In this subsection, we consider the scales $\mu_1$ and $\mu_2$ to be fixed and
omit them as arguments for brevity.  We discuss the discretization of DPDs in the
three variables $x_1$, $x_2$, and $y$, which we generically denote by $z$.
Depending on the typical dependence of a DPD on $z$, different variable
transformations $z \to u$ are used to map an interval $[z_{\min}, z_{\max}]$ onto
a finite interval $[u_{\min}, u_{\max}]$.  For the momentum fractions $x_1$ and
$x_2$ we always use the transformation
\begin{align}
   \label{eq:trafo-x}
      u(x_i)
   &= \ln x_i
   &&
   \text{for } i=1,2,
\end{align}
which corresponds to the choice for interpolating PDFs in
\refcite{Diehl:2021gvs}.  For both $x_1$ and $x_2$, the interpolation interval
has a fixed upper limit $x_{\max} = 1$, whereas the lower limits may be equal or
different (but must be nonzero).  For the $y$ dependence (where $y_{\max}$ may
be infinite or finite) we use a number of different transformations, which are
presented in \subsec{variable-trafo-y}.

For reasons given below, the $z$ intervals are usually split into a few
subintervals.  On each subinterval, we perform a linear transformation from $u$
to a variable $t \in [-1,1]$, which is used for Chebyshev interpolation as
described in \subsec{chebyshev}.

Consider a particular subgrid with $N + 1$ points $u_0, \ldots, u_N$, which is
mapped by a linear transform onto the Chebyshev grid $t_0, \ldots, t_N$ given in
\eq{cheb_points}.  The corresponding grid points in the physical variable are
then given by the inverse variable transformation $z_i = z(u_i)$. Similarly to
the PDF case, it is advantageous to interpolate DPDs scaled by their momentum
fractions,
\begin{align}
   \label{eq:dpd-rescaled}
      \widetilde{F}(x_1, x_2, y)
   &= x_1 \ms x_2 \ms F(x_1, x_2, y)
   \,,
\end{align}
because this leads to less steep functions in $x_1$ and $x_2$.  The
interpolation in one variable is given by the barycentric formula
\begin{align}
   \label{eq:barycentric_dpd}
      \widetilde{F}(z, \ldots)
   &  \approx \sum_{i = 0}^N \widetilde{F}_i(\ldots) \; b_i(u(z))
   &  \text{for~} z_0 \le z \le z_N
      \,,
\end{align}%
where the ellipsis denotes the other variables and
\begin{align}
   \label{eq:bary_basis_u}
      \widetilde{F}_i(\ldots)
   &= \widetilde{F}(z_i, \ldots)
      \,,
   &  b_i(u)
   &= \beta_i \, \frac{(-1)^{i}}{u - u_i}
     \Bigg/ \sum_{j=0}^{N} \beta_j \, \frac{(-1)^{j}}{u - u_j}
     \,.
\end{align}%
The interpolation of the full DPD is obtained by discretization in both momentum
fractions and the interparton distance and reads
\begin{align}
   \label{eq:interpolation-dpd}
      \widetilde{F}(x_1, x_2, y)
   &  \approx
      \sum_{i=0}^{N_{x\bs, 1}} \sum_{j=0}^{N_{x\bs, 2}} \sum_{k=0}^{N_y}
      \widetilde{F}_{i j k} \;
      b_i\bigl( \ln(x_1) \bigr) \, b_j\bigl( \ln(x_2) \bigr) \,
      b_k\bigl( u(y) \bigr)
      \,,
\end{align}%
where $N_{x\bs, 1}$, $N_{x\bs, 2}$, and $N_y$ specify the polynomial order of
interpolation in each variable, and
\begin{align}
   \label{eq:Fjkl}
      \widetilde{F}_{i j k}
   &= \widetilde{F}(x_{1, i},\ms x_{2, j},\ms y_k)
\end{align}%
denotes the values of $\widetilde{F}$ on the interpolation grid.
\Eq{bary_basis_u} reflects that the form of the barycentric basis functions
\eqref{eq:bary_basis} remains unchanged under a linear transform of the
interpolation variable.  Analogous formulae can be used to interpolate functions
derived from DPDs, such as the Mellin convolutions of DPDs with an integral
kernel (see \subsec{mellin_convolution}).

Note that we interpolate $\widetilde{F}(x_1, x_2, y)$ on a full rectangle in the
$(x_1, x_2)$ plane, given by $x_1 \in [x_{1, \min}, 1]$ and $x_2 \in [x_{2,
\min}, 1]$.  This includes the region $x_1 + x_2 > 1$ where the DPD is zero.
In this way, we can use the same interpolation grid in $x_1$ for all values of
$x_2$ and vice versa, which in turn permits a simple implementation of DGLAP
evolution for the two partons.  The implication of this choice on the
interpolation accuracy is discussed in \sec{x1-x2-interpolation}.

Chebyshev interpolation is ``global'' in the sense that the interpolant at a
given point $z$ depends on the function values at \emph{all} points $z_i$ in the
interpolation interval.  For grids with many points, this can lead to an
accumulation of rounding errors, especially in regions of $z$ where the function
is much smaller than elsewhere in the interval.  To limit this effect, it is
advantageous to split the full $z$ domain into several subintervals and to use
the barycentric formula \eqref{eq:interpolation-dpd} separately on each
subinterval.  In \refcite{Diehl:2021gvs} this was found to substantially improve
the interpolation accuracy for PDFs, provided that the number of points in each
subinterval does not become too small.  We find that the same is true for the
interpolation of DPDs in $x_1$ and $x_2$, as well as in $y$.

To specify such a composite grid with $k$ individual subgrids, we use the
notation
\begin{align}
   \label{eq:composite-grid}
      [z_0,\ms z_1,\ms\ldots,\ms z_{k}]_{(p_1,\ms p_2,\ms \ldots,\ms p_k)}
      \,,
\end{align}%
where $z_i$ are the subinterval boundaries and $p_i = N_i + 1$ is the number of
Chebyshev points in subgrid number $i$. We refer to this as a $(p_1, p_2,
\ldots, p_k)$-point grid. Note that neighboring subgrids share their end points,
such that the total number of grid points is $p_{\text{total}} = \sum_{i = 1}^k
p_i - (k - 1)$.

Splitting the interpolation interval in $y$ has the additional benefit that one
can use different variable transformations $u(y)$ for regions with different
characteristic $y$ dependence, such as the power law behavior from perturbative
splitting and an exponential decrease at large distances.  Details will be
discussed in \subsec{variable-trafo-y}.

%===============================================================================
\subsection{Mellin convolution}
\label{subsec:mellin_convolution}
%===============================================================================
%
Consider now the convolution of a DPD with an integral kernel, which may be a
DGLAP evolution kernel, a flavor matching kernel, or a hard-scattering
coefficient.  The general expression to be computed reads
\begin{align}
   \label{eq:mellin_convolution}
      \bigl( K \conv{} \widetilde{F} \,\bigr) (x)
   &= \int_x^1 \frac{\dd z}{z} \, K(z) \,
      \widetilde{F} \left( \frac{x}{z} \right)
      \,,
\end{align}%
where $x$ stands for $x_1$ or $x_2$ and the dependence of the rescaled DPD on
all other variables is omitted for now.  The kernel $K(z)$ is appropriately
rescaled, given that $g_1 = g_2 \otimes g_3$ implies $h_1 = h_2 \otimes h_3$
with $h_k(z) = z \ms g_k(z)$.

To discretize \eq{mellin_convolution}, we consider for simplicity a single
Chebyshev grid with $p_x$ points for $x \in [x_{\min}, 1]$. The result of the
convolution $(K \otimes \widetilde{F})(x)$ is a function defined in the same $x$
interval as $\widetilde{F}(x)$ and can hence be interpolated on the same grid. It
is thus sufficient to evaluate \eq{mellin_convolution} at the grid points
$x_{i}$,
\begin{align}
   \label{eq:mellin_convolution_chebpts}
      \bigl( K \conv{} \widetilde{F} \,\bigr) (x_{i})
   &= \int_{x_{i}}^1 \! \frac{\dd{z}}{z} \, K(z) \,
         \widetilde{F} \left( \frac{x_{i}}{z} \right)
      \approx \int_{x_{i}}^1 \frac{\dd{z}}{z} \, K(z) \,
      \sum_{i' = 0}^{p_x - 1}
      \widetilde{F}(x_{i'}) \;
      b_{i'}\!\left( \ln \frac{x_{i}}{z} \right)
      \,,
\end{align}%
where in the second step we have approximated $\widetilde{F} (x_i/z)$ by its
interpolated form.  Here and in the following, we use the number of grid points
$p_x$ rather than the polynomial order $N_x = p_x - 1$ to specify summation
ranges. At the grid points, the convolution is thus obtained by a matrix
multiplication
\begin{align}
   \label{eq:mellin_product}
      \bigl( K \conv{} \widetilde{F} \,\bigr) (x_i)
   &  \approx \sum_{i' = 0}^{p_x - 1}
      K_{i i'} \, \widetilde{F}(x_{i'})
   \,,
\end{align}
where the \emph{kernel matrix}
\begin{align}
   \label{eq:kernel_matrix_def}
      K_{i i'}
   &= \int_{x_{i}}^1 \frac{\dd{z}}{z} \, K(z) \;
      b_{i'}\!\left( \ln \frac{x_{i}}{z} \right)
\end{align}%
can be computed using standard numerical integration techniques (we use an
adaptive Gauss-Kronrod routine).  If $K(z)$ contains plus- or $\delta$
distributions, this requires rewriting \eq{kernel_matrix_def} as specified in
equations (4.5) to (4.10) of \refcite{Diehl:2021gvs}.

It is straightforward to generalize the preceding discussion to the case where
the interval $[x_{\min}, 1]$ is split into several subintervals with
corresponding subgrids.  The relation \eqref{eq:mellin_product} remains valid,
with an appropriately generalized definition of the kernel matrix and with $p_x$
being the \emph{total} number of grid points in $x$.

The kernel matrix method can be readily used for flavor matching of DPDs,
discretizing the Mellin convolutions with the matching kernels in
\eq{dpd-flavor-matching}.  It is also a crucial part in DPD evolution, which is
presented next.

%===============================================================================
\subsection{DGLAP evolution}
\label{subsec:dglap_evolution}
%===============================================================================

We now discuss the evolution of DPDs, starting with evolution in $\mu_1$ at a
fixed value of $\mu_2$.  The evolution equation for the rescaled DPD reads
\begin{align}
   \label{eq:dglap-dpd}
      \frac{\dd}{\dd \ln \mu_1^2} \, \widetilde{F}(x_1,x_2,y; \mu_1,\mu_2)
   &= \Bigl[
         \widetilde{P}(\mu_1) \conv{1} \widetilde{F}(x_2,y; \mu_1,\mu_2)
      \Bigr](x_1)
      \,,
\end{align}
where $\widetilde{P}(z; \mu_1) = z \ms P(z; \mu_1)$ is the rescaled splitting
function and we omit parton labels and the associated sums for the time being.
After discretization on one or several subgrids with a total number of
$p_{x\bs,1}$, $p_{x\bs,2}$, $p_y$ points in $x_1$, $x_2$, $y$, the
integro-differential equation \eqref{eq:dglap-dpd} turns into a linear system of
ordinary differential equations in $\mu_1$, %
\begin{align}
   \label{eq:dglap-dpd-discretized}
      \frac{\dd}{\dd \ln \mu_1^2} \,  \widetilde{F}_{i j k}(\mu_1, \mu_2)
   &= \sum_{i'=0}^{p_{x\bs, 1} - 1} \,
      \widetilde{P}_{i i'}(\mu_1) \,
      \widetilde{F}_{i'\bs j k}^{}(\mu_1, \mu_2)
      \,,
\end{align}%
where $\widetilde{F}_{i j k}(\mu_1, \mu_2)$ is given in \eq{Fjkl} and
$\widetilde{P}_{i i'}$ is the matrix for the kernel $\widetilde{P}(z)$ as defined
in \eq{kernel_matrix_def} or its generalization to several subgrids. At NNLO
accuracy, we have
\begin{align}
   \widetilde{P}_{i i'}(\mu)
   &= \sum_{k=0}^{2} \biggl[ \frac{\alpha_s(\mu)}{4\pi} \biggr]^{k+1} \;
      \widetilde{P}_{i i'}^{(k)}
   \,,
\end{align}
and after a change of variables from the scales $\mu_i$ to the evolution times
\begin{align}
   \label{eq:mu-to-t}
      t_i
   &= - \ln \bigl( \alpha_s(\mu_i) \bigr)
   &&
   \text{for } i=1,2,
\end{align}%
we obtain
\begin{align}
   \label{eq:dglap-dpd-t-nnlo}
   \frac{\dd}{\dd t_1} \, \widetilde{F}_{i j k}(t_1, t_2)
   &=
   \sum_{i'=0}^{p_{x\bs, 1} - 1} \, K_{i i'}(t_1) \,
   \widetilde{F}_{i'\bs j k}(t_1, t_2)
\end{align}
with a new kernel matrix given by
\begin{align}
   \label{eq:dglap-dpd-t-kernel}
   K_{i i'}(t)
   &=
   \biggl[ \ms
         \beta_0 + \frac{e^{-t}}{4\pi} \, \beta_1
         + \frac{e^{-2 t}}{(4\pi)^2} \, \beta_2
   \, \biggr]^{-1} \;
   \biggl[ \ms
         \widetilde{P}_{i i'}^{(0)}
         + \frac{e^{-t}}{4\pi} \, \widetilde{P}_{i i'}^{(1)}
         + \frac{e^{-2 t}}{(4\pi)^2} \, \widetilde{P}_{i i'}^{(2)}
   \, \biggr]
   \,,
\end{align}
where $\beta_k$ are the expansion coefficients of the QCD $\beta$ function (see
equation~(5.4) in \cite{Diehl:2021gvs} for our normalization conventions).

To solve \eq{dglap-dpd-t-nnlo} we use a Runge-Kutta algorithm with a fixed
maximal step size in~$t_1$.  Notice that the $t_1$ dependence of $K_{i i'}(t_1)$
starts only at NLO, which leads to a more uniform accuracy of the algorithm
compared with the direct solution of \eq{dglap-dpd-discretized}, where
$\widetilde{P}_{i i'}(\mu_1)$ changes with $\ln \mu_1$ already at LO.  As
documented in appendix A of \refcite{Diehl:2021gvs}, we find that the use of a
higher-order Runge-Kutta algorithm --- in particular the $8^{\text{th}}$ order
algorithm of Dormand and Prince \cite{PRINCE198167} --- greatly improves the
accuracy of evolution for a given number $N_{\text{RK}}$ of intermediate time
values at which the r.h.s.\ of the differential equation is evaluated for
evolution from one value of $t_1$ to another.

The Runge-Kutta algorithm is also used to compute the transformation
\eqref{eq:mu-to-t}, solving the differential equation for the running of
$\alpha_s^{-1}$ as a function of $\ln\mu$.

Flavor labels and sums have to be added in \eq{dglap-dpd-t-nnlo} as in the
original version \eqref{eq:double-dglap}.  The mixing between parton flavors
under evolution is the same as for PDFs, and correspondingly we solve the
evolution equations in the same basis that is used for PDFs in
\refcite{Diehl:2021gvs}.  For each of the two partons, we thus form the linear
combinations
\begin{align}
   \label{eq:evol_diagonal_basis}
   \Sigma^{\pm} = \sum_i q_i^\pm ,
   && u^\pm    - d^\pm   ,
   && d^\pm    - s^\pm   ,
   && s^\pm    - c^\pm   ,
   && c^\pm    - b^\pm   ,
   && b^\pm    - t^\pm
      \,,
\end{align}%
where $q^\pm = q \pm \qbar$.  The combination $\Sigma^-$ and all flavor
differences then evolve by themselves, and mixing is reduced to the combination
$\Sigma^+$ and the gluon.

The system \eqref{eq:dglap-dpd-t-nnlo} needs to be solved separately for each
combination of the indices $j, k$ that correspond to the ``inactive'' variables
$x_2$ and $y$ during evolution in $t_1$, and for each of the $2 \nf{2} + 1$
active flavors of the second parton.  This gives a large number of $p_{x\bs, 2}
\; p_y \, (2 \nf{2} + 1)$ combinations --- typically several thousand.  It is
much more efficient to apply the Runge-Kutta algorithm not to each of these
combinations, but to a set of $p_{x\bs, 1}$ basis vectors $e_{i}^{j}$ that
generates all possible initial conditions on the grid for $x_1$.  With the choice
of basis $e_{i}^{\ms j} = \delta_{i j}^{}$, evolution from $t_{0 1}$ to $t_1$
requires solving the system
\begin{align}
   \label{eq:evolution-matrix}
      \frac{\dd}{\dd t_{1}} \, U_{i j}(t_{1}, t_{0 1})
   &= \sum_{i' = 0}^{p_{x\bs, 1} - 1} K_{i i'}(t_{1}) \,
      U_{i' j}(t_{1}, t_{0 1})
      \,,
\end{align}%
with the boundary condition $U_{i j}(t_{0 1}, t_{0 1}) =
\delta_{i j}$.  The solution of \eq{dglap-dpd-t-nnlo} is then obtained by a
matrix multiplication
\begin{align}
   \label{eq:dglap-greens-discrete}
      \widetilde{F}_{i j k}(t_{1}, t_{2})
   &= \sum_{i' = 0}^{p_{x\bs, 1} - 1}
      U_{i i'}(t_{1}, t_{0 1}) \,
      \widetilde{F}_{i' j k}^{}(t_{0 1}, t_2)
      \,,
\end{align}%
where the evolution matrix $U_{i i'}(t_{1}, t_{0 1})$ acts as the discretized
version of the Green function for the original evolution equation
\eqref{eq:dglap-dpd}. With flavor labels reinstated, \eq{evolution-matrix} is
solved separately for the evolution kernels of the different channels (i.e.\ for
$q_i^{+} - q_j^{+}$, $q_i^{-} - q_j^{-}$, $\Sigma^{-}$, and the coupled system of
$\Sigma^{+}$ and the gluon).

This approach is also attractive in the PDF case when a large set of different
PDFs is evolved simultaneously, and it is for instance used in the PDF evolution
codes \texttt{APFEL} \cite{Bertone:2013vaa} and \texttt{EKO}
\cite{Candido:2022tld}, with $U_{i i'}(t_{1}, t_{0 1})$ being called
an evolution operator and an evolution kernel operator, respectively.  In
\chili, the evolution of a DPD in one scale is implemented in the same way as
the evolution of a set of PDFs.

The number of operations for evaluating \eq{dglap-greens-discrete} for all
indices $i, j, k$ and flavors for the second parton scales like $(p_{x\bs,
1})^2 \; p_{x\bs, 2} \; p_y \, (2 \nf{2} + 1)$, and the number of operations for
solving the evolution equation \eqref{eq:evolution-matrix} scales like
$(p_{x\bs, 1})^3 \, N_{\text{RK}}$ with $N_{\text{RK}}$ introduced below
\eq{dglap-dpd-t-kernel}.  The number of grid points for the interpolation in the
momentum fractions is therefore a major factor determining the computational
speed of DPD evolution.

The preceding discussion is trivially extended to evolution in the scale $\mu_2$
at fixed $\mu_1$.  To evolve from $(\mu_{0 1}, \mu_{0 2})$
to $(\mu_{1}, \mu_{2})$, we first evolve from $(\mu_{0 1}, \mu_{0 2})$
to $(\mu_{1}, \mu_{0 2})$  and from there to $(\mu_{1},
\mu_{2})$.  We verified that evolving to $(\mu_{0 1}, \mu_{2})$ in the
first step leads to the same final result with very high numerical accuracy (see
also \subsec{path-indep-evol-match}).

\rev{Let us briefly sketch how the initial conditions for DPD evolution are set
up in \chili.  In a first step, the values of $F_{a_1 a_2}\bigl(x_1, x_2, y;
\mu(y), \mu(y)\bigr)$ for a selected number $\nf{0}$ of active flavors are
computed on the grid points in $x_1$, $x_2$ and $y$, either from pre-defined
functions for the splitting form \eqref{eq:F-spl}, or from functions implementing
a product ansatz like the one in \eqref{eq:F-int}, or from functions entirely
provided by the user.  As is adequate for the physics, the input scale $\mu(y)$
may depend on $y$ as specified by the user.  For each grid point~$y_k$, the
discretized DPD is then evolved to a pair of scales $(\mu_{0}, \mu_{0})$ that is
independent of $y$ and stored as an initial condition.  After this, all evolution
calls can use the same set of evolution matrices for all grid points $y_k$, which
significantly reduces computing time.  One may also add two DPDs (after evolution to a
common scale pair), which is for instance needed for calculating $F^{\text{int}}
+ F^{\text{spl}}$.}

\rev{In the procedure just described, the flavor number $\nf{0}$ is equal for
both partons.  DPDs with larger values of $\nf{1}$ or $\nf{2}$ are then obtained
by flavor matching, with initial conditions at the pair of scales $(\mu_1,
\mu_2)$ where the matching is carried out.  More detail is given in
\subsec{path-indep-evol-match}.}

%%%%%%%%%%%%%%%%%%%%%%%%%%%%%%%%%%%%%%%%%%%%%%%%%%%%%%%%%%%%%%%%%%%%%%%%%%%%%%%%
\section{Interpolation of DPDs in both momentum fractions}
\label{sec:x1-x2-interpolation}
%%%%%%%%%%%%%%%%%%%%%%%%%%%%%%%%%%%%%%%%%%%%%%%%%%%%%%%%%%%%%%%%%%%%%%%%%%%%%%%%

As shown in the previous section, our approach to interpolating DPDs in $x_1$
and $x_2$ is a straightforward two-dimensional generalization of the method
presented in \refcite{Diehl:2021gvs} for the interpolation of PDFs in $x$.  In
the present section, we study the performance of this approach.  Let us point
out some particularities of the DPD case:
\begin{enumerate}
\item To limit computation time, interpolation grids should not have too many
points, while still giving numerically reliable results.  For PDFs one can
typically afford denser grids, and the demands on interpolation accuracy are
often higher than for DPDs.
\item The DPD is a non-analytic function of $x_1$ at the point $x_1 = 1 - x_2$
because it vanishes identically above that value.  Since we interpolate on the
same grid in $x_1$ for all values of $x_2$, this singular point is generally in
the middle of an interpolation interval.  Polynomial interpolation is not very
accurate in the vicinity of such a point, but we will see that this effect is in
general relatively mild.
\item The perturbative splitting mechanism gives a dependence on $x_1$ that is
qualitatively different from the $x$ dependence of a PDF.  In some channels, the
splitting contribution \eqref{eq:F-spl} goes to zero for $x_1 \ll x_2$ and thus
results in a \emph{strong} rise of the DPD as a function of $x_1$.
\end{enumerate}
Of course, the last two points also apply to interpolation in $x_2$ at a given
$x_1$.

In the following, we will study the interpolation accuracy for DPDs that have a
known analytic expression.  To this end, we simplify the forms \eqref{eq:F-int}
and \eqref{eq:F-spl} to
\begin{align}
   \label{eq:F-int-no-y}
      \Fint{a_1 a_2}{(r)}(x_1,x_2)
   &= \frac{(1 - x_1 - x_2)^r}{(1 - x_1)^r (1 - x_2)^r} \;
      f_{a_1}(x_1)\, f_{a_2}(x_2)
\intertext{and}
   \label{eq:F-spl-no-y}
      \Fspl{a_1 a_2}(x_1,x_2)
   &= V^{(1)}_{a_1 a_2, a_0} \biggl( \frac{x_1}{x_1 + x_2} \biggr) \,
      \frac{f_{a_0}(x_1 + x_2)}{x_1 + x_2}
   \,,
\end{align}%
respectively.  For the PDFs, we take the simple parameterizations of the Les
Houches benchmark PDFs given in section 1.32 of \refcite{Giele:2002hx} and
section 4.4 of \refcite{Dittmar:2005ed}.

Corresponding to the typical use case, we take the same interpolation grids for
$x_1$ and $x_2$ (a situation in which different grids are useful is encountered
in \sec{dove}).  We consider three composite grids of the form
\begin{align}
   \label{eq:x-grids}
   [10^{-5}, 5 \times 10^{-3}, 0.5, 1]_{(p,\, p,\, p)}
   & &
   \text{with }
   p =
   \begin{cases}
      12 & \text{(coarse $x$ grid)} \\
      16 & \text{(medium $x$ grid)} \\
      24 & \text{(fine $x$ grid)}
   \end{cases}
\end{align}
where the shorthand notation from \eq{composite-grid} has been used.  We find
that for the given $x$ range, three subgrids with the above interval limits
generally give the best accuracy for a given total number of points.
As a measure for the quality of interpolation, we use the relative accuracy
defined as
\begin{align}
   \label{eq:rel-accuracy}
      \text{relative accuracy}
   &= \bigl| \text{interpolated result} / \text{exact result} - 1 \bigr|
   \,.
\end{align}%
In the following, this interpolation accuracy will be shown as a function of
$x_1$ for fixed values of $x_2$.  The latter are \emph{not} grid points, so that
interpolation is performed for both momentum fractions.

%===============================================================================
\subsection{Interpolation of the product form}
\label{subsec:interpolation-x1-x2-product}
%===============================================================================

We start with the product form \eqref{eq:F-int-no-y} for the three values $r =
0, 1, 2$ of the power in the phase space factor.  The case $r = 0$ corresponds
to a ``naive'' product form with a discontinuity at $x_1 + x_2 = 1$ that is not
very plausible on physics grounds.  In fact, evolution to higher scales quickly
changes such a discontinuity to a steep but continuous decrease.  We
nevertheless include the case $r = 0$ here, in order to see how our method
performs in this limiting case.  The following plots show the $u \ubar$
distribution; we verified that for other parton combinations the interpolation
accuracy is not significantly better or worse.

\begin{figure}
   \centering
   \subfigure[
      \label{fig:product-PSF-0-small-x2-uubar}
      $r = 0, ~ x_2 = 10^{-4}$
   ]{%
   \includegraphics[height=\HeightTwoSubfigs]{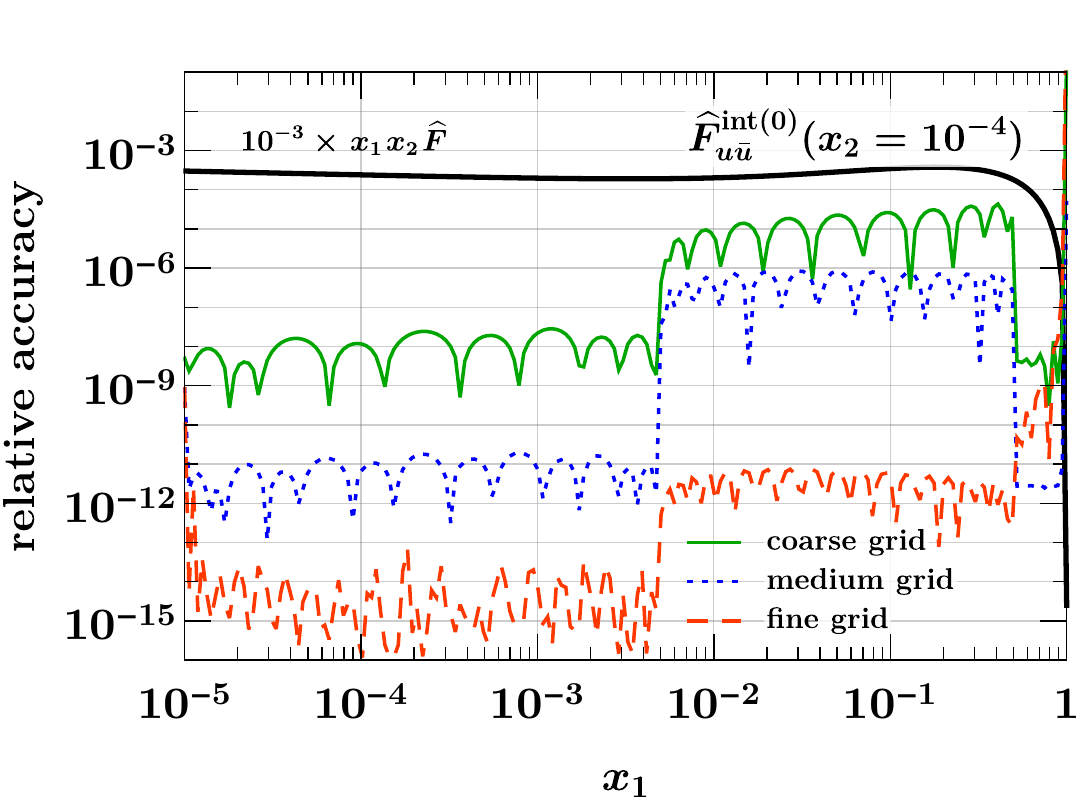}%
   }%
   \hfill
   \subfigure[%
      \label{fig:product-PSF-0-large-x2-uubar}
      $r = 0, ~ x_2 = 0.3$
   ]{%
   \includegraphics[height=\HeightTwoSubfigs]{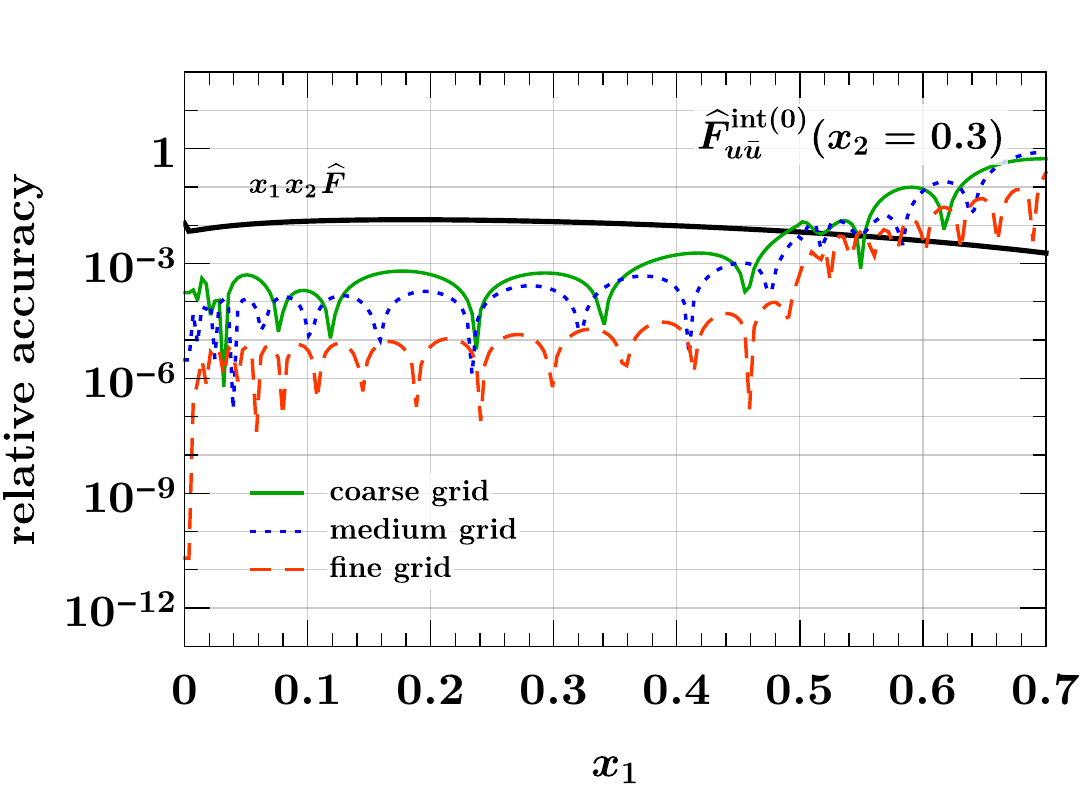}%
   }%
   \\
   \subfigure[
      \label{fig:product-PSF-1-small-x2-uubar}
      $r = 1, ~ x_2 = 10^{-4}$
   ]{%
   \includegraphics[height=\HeightTwoSubfigs]{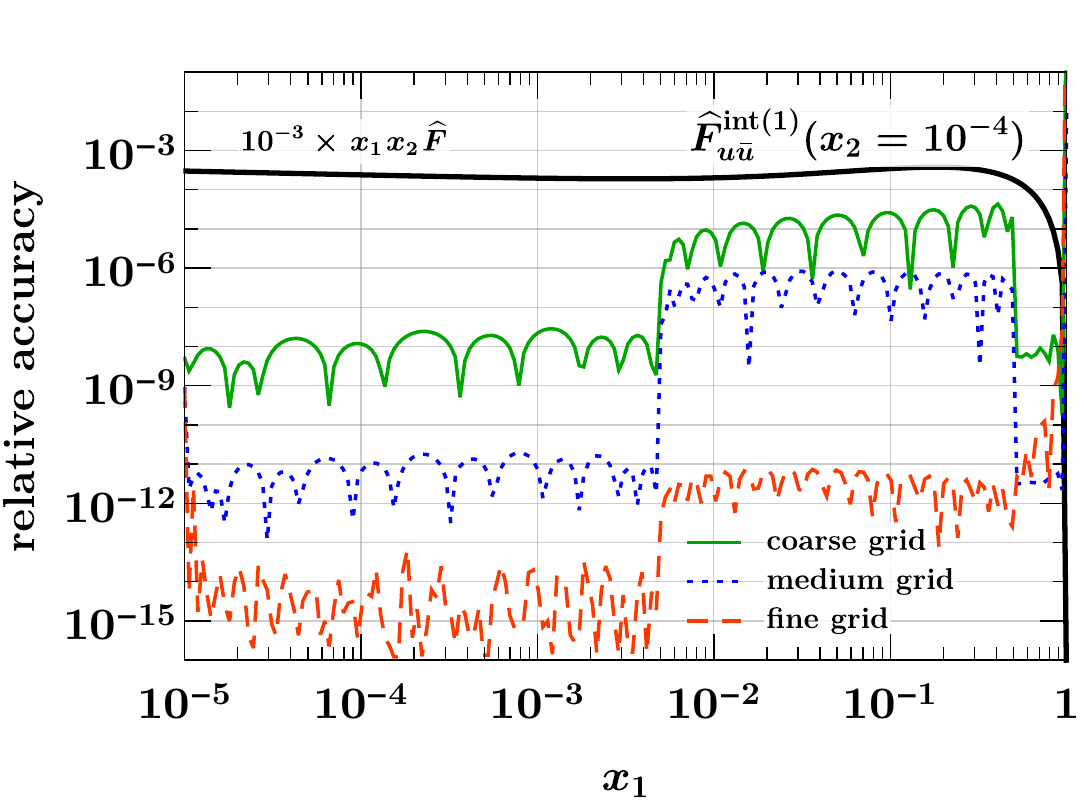}%
   }%
   \hfill
   \subfigure[
      \label{fig:product-PSF-1-large-x2-uubar}
      $r = 1, ~ x_2 = 0.3$
   ]{%
   \includegraphics[height=\HeightTwoSubfigs]{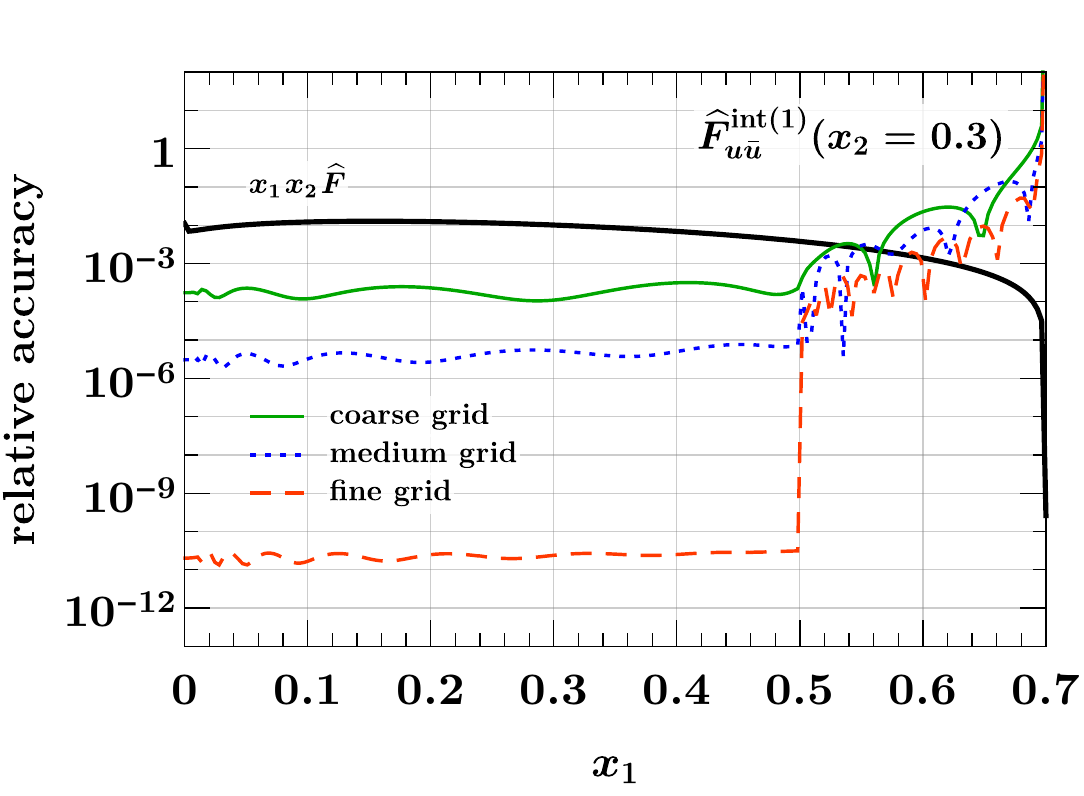}%
   }%
   \\
   \subfigure[
      \label{fig:product-PSF-2-small-x2-uubar}
      $r = 2, ~ x_2 = 10^{-4}$
   ]{%
   \includegraphics[height=\HeightTwoSubfigs]{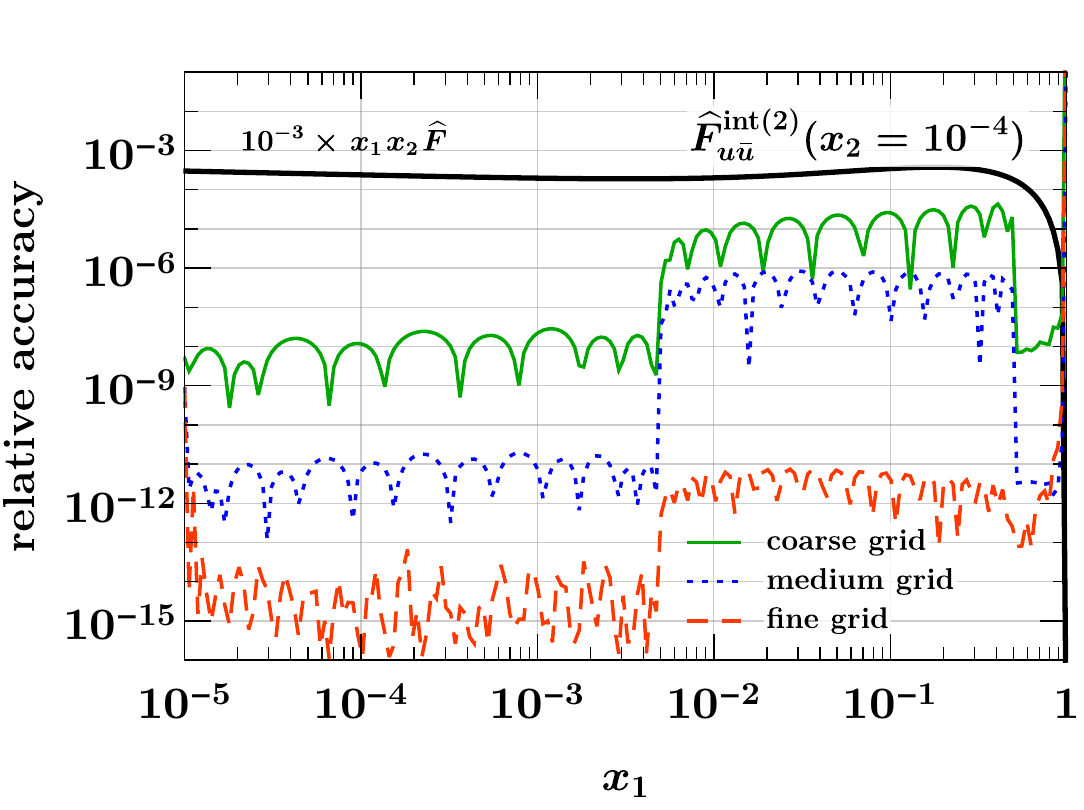}%
   }%
   \hfill
   \subfigure[
      \label{fig:product-PSF-2-large-x2-uubar}
      $r = 2, ~ x_2 = 0.3$
   ]{%
   \includegraphics[height=\HeightTwoSubfigs]{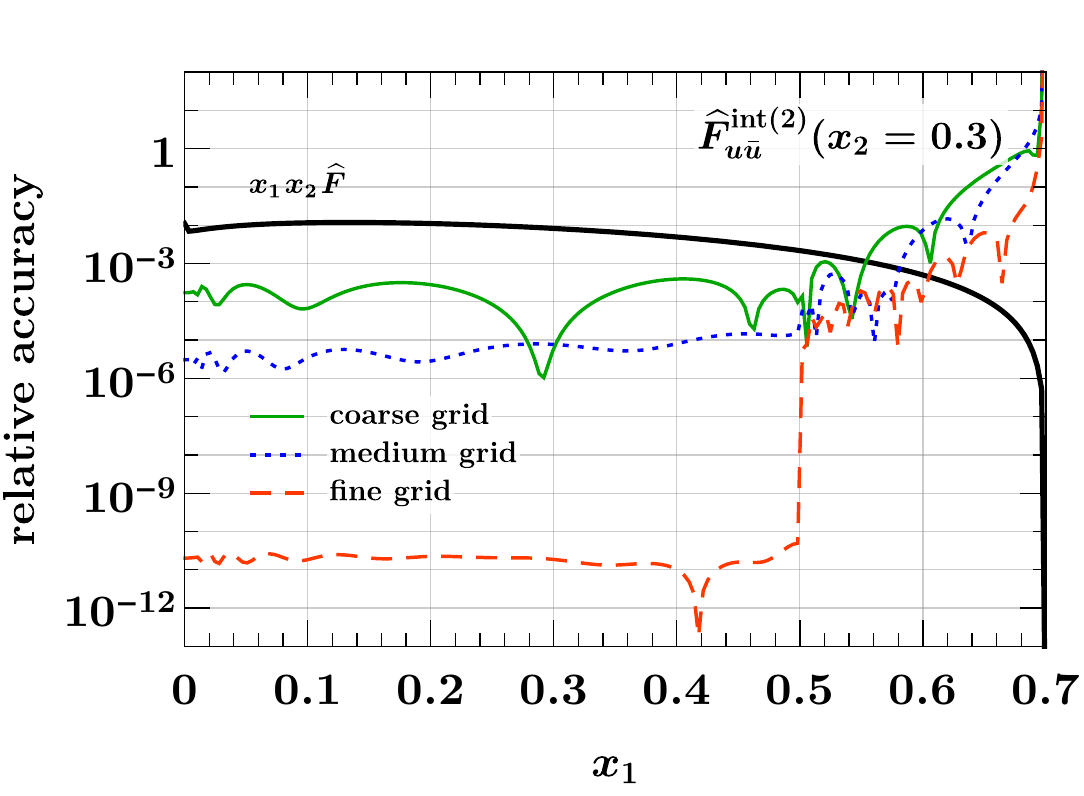}%
   }%
   \caption{\label{fig:interpol-x1x2-product}Relative accuracy
\protect\eqref{eq:rel-accuracy} for interpolation in $x_1$ and $x_2$, evaluated
for the product form $\Fint{u \ubar}{(r)}(x_1,x_2)$ of \eq{F-int-no-y} with $r =
0,1,2$.  The interpolation grids are specified in \eq{x-grids}.  Here and in
similar plots, the exact result that is being interpolated is shown in black.}
\end{figure}

In the left panels of \fig{interpol-x1x2-product} we see that all three grids
yield an excellent interpolation accuracy at small momentum fractions.  Not
surprisingly, there is little dependence on the power $r$ in this region, given
that the phase space factor is very close to $1$ for $x_1, x_2 \ll 1$.  In the
right panels of the figure, we see that a degradation of the relative accuracy
sets in as $x_1$ approaches its kinematic limit $1 - x_2 = 0.7$, and that the
degradation is stronger for smaller $r$.  Nevertheless, the accuracy is better
than 1\% as long as the distance $1 - (x_1 + x_2)$ from the kinematic threshold
is above $0.15$ for $r=1,2$ and above $0.2$ for $r=0$.

\begin{figure}
   \hspace{0.65em}
   \subfigure[
      \label{fig:kinematic-boundary-product-PSF-0-large-x2-uubar}
      $r = 0, ~ x_2 = 0.3$
   ]{%
      \includegraphics[height=\HeightTwoSubfigs]{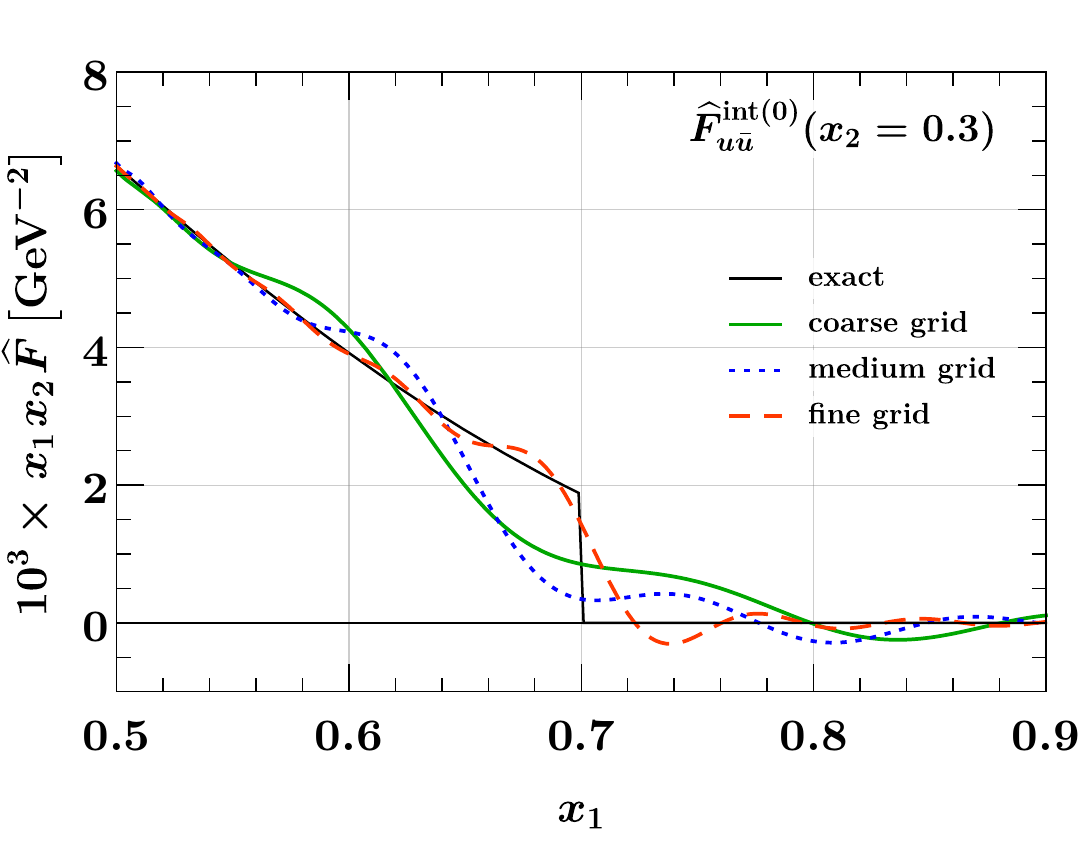}%
   }%
   \hfill
   \subfigure[
      \label{fig:kinematic-boundary-product-PSF-0-large-x2-uubar-abs}
      $r = 0, ~ x_2 = 0.3$
   ]{%
      \includegraphics[height=\HeightTwoSubfigs]{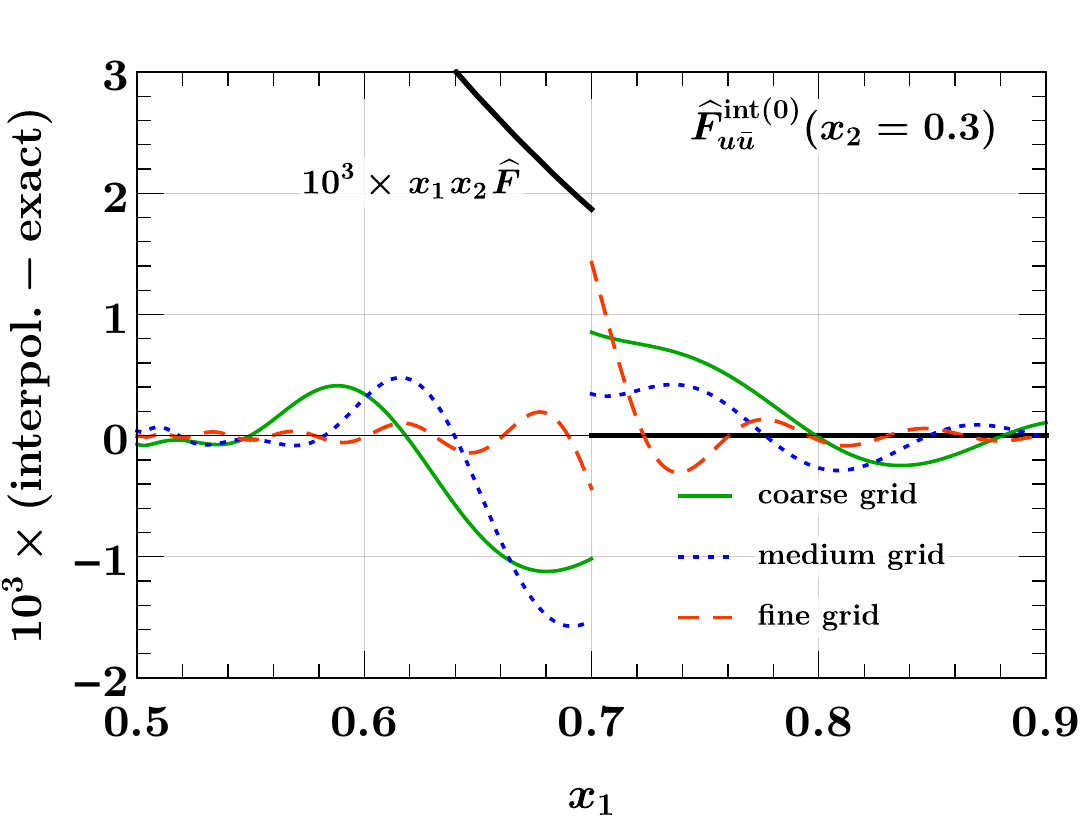}%
   }%
   \\
   \subfigure[
      \label{fig:kinematic-boundary-product-PSF-1-large-x2-uubar}
      $r = 1, ~ x_2 = 0.3$
   ]{%
      \includegraphics[height=\HeightTwoSubfigs]{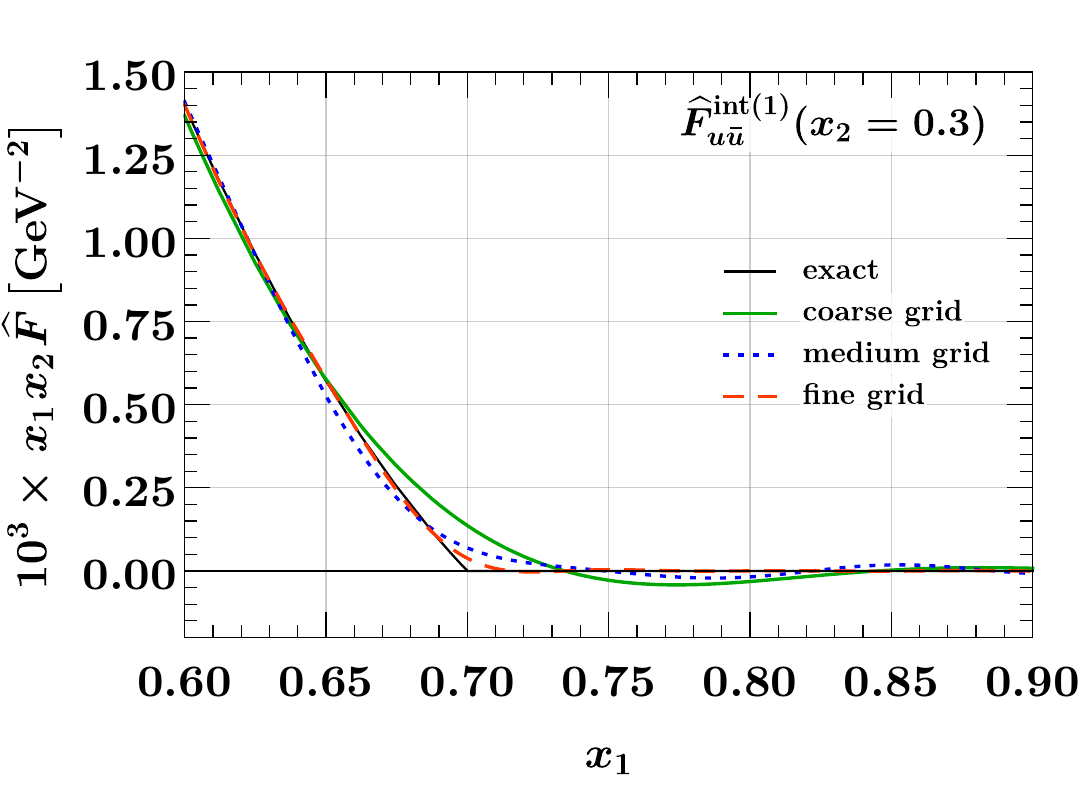}%
   }%
   \hfill
   \subfigure[
      \label{fig:kinematic-boundary-product-PSF-1-large-x2-uubar-abs}
      $r = 1, ~ x_2 = 0.3$
   ]{%
      \includegraphics[height=\HeightTwoSubfigs]{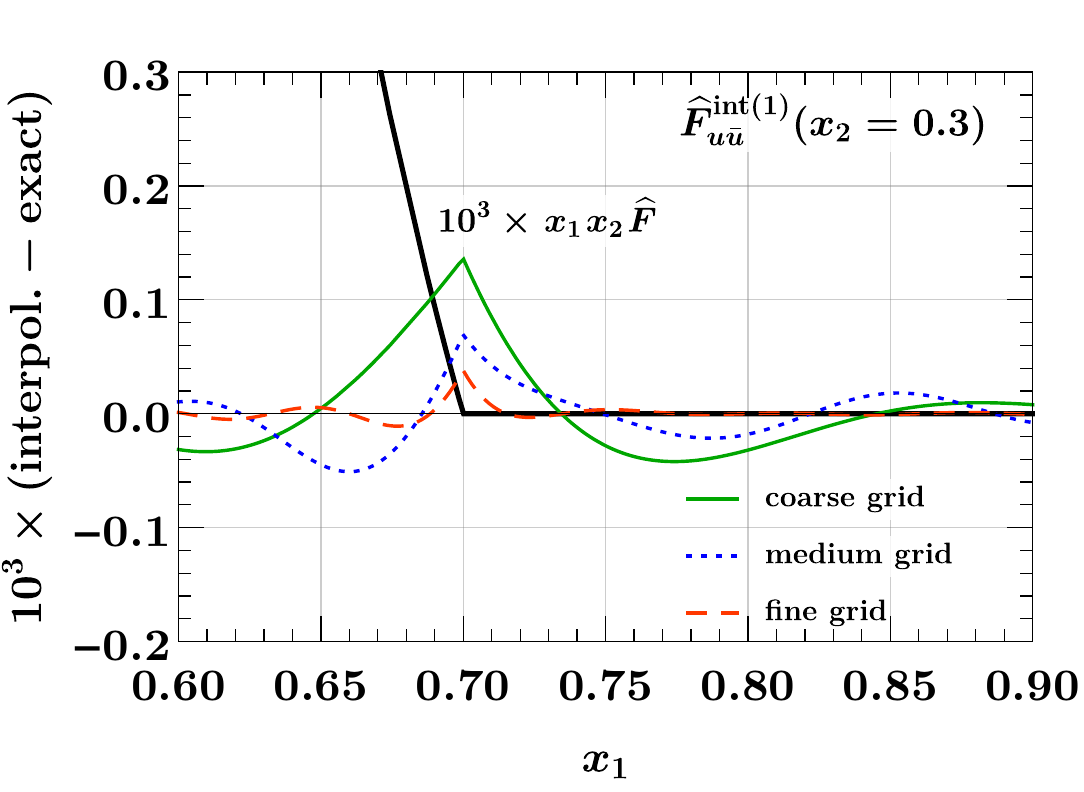}%
   }%
   \\
   \subfigure[
      \label{fig:kinematic-boundary-product-PSF-2-large-x2-uubar}
      $r = 2, ~ x_2 = 0.3$
   ]{%
      \includegraphics[height=\HeightTwoSubfigs]{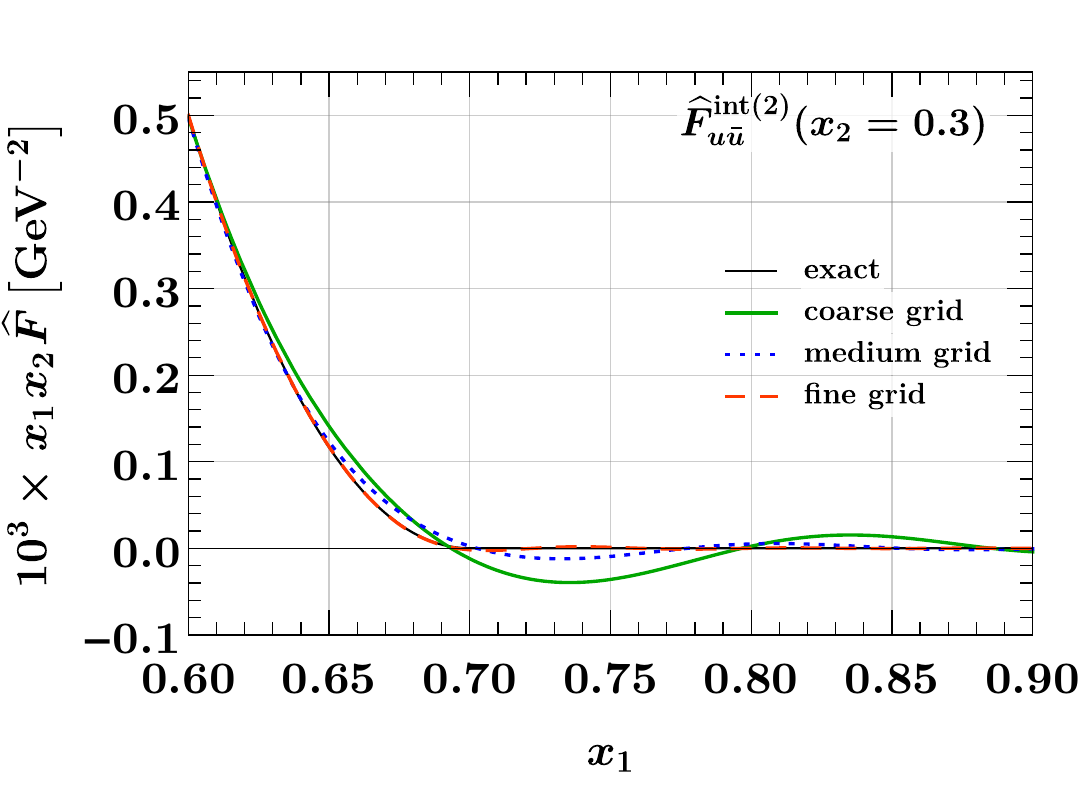}%
   }%
   \hfill
   \subfigure[
      \label{fig:kinematic-boundary-product-PSF-2-large-x2-uubar-abs}
      $r = 2, ~ x_2 = 0.3$
   ]{%
      \includegraphics[height=0.34\textwidth]{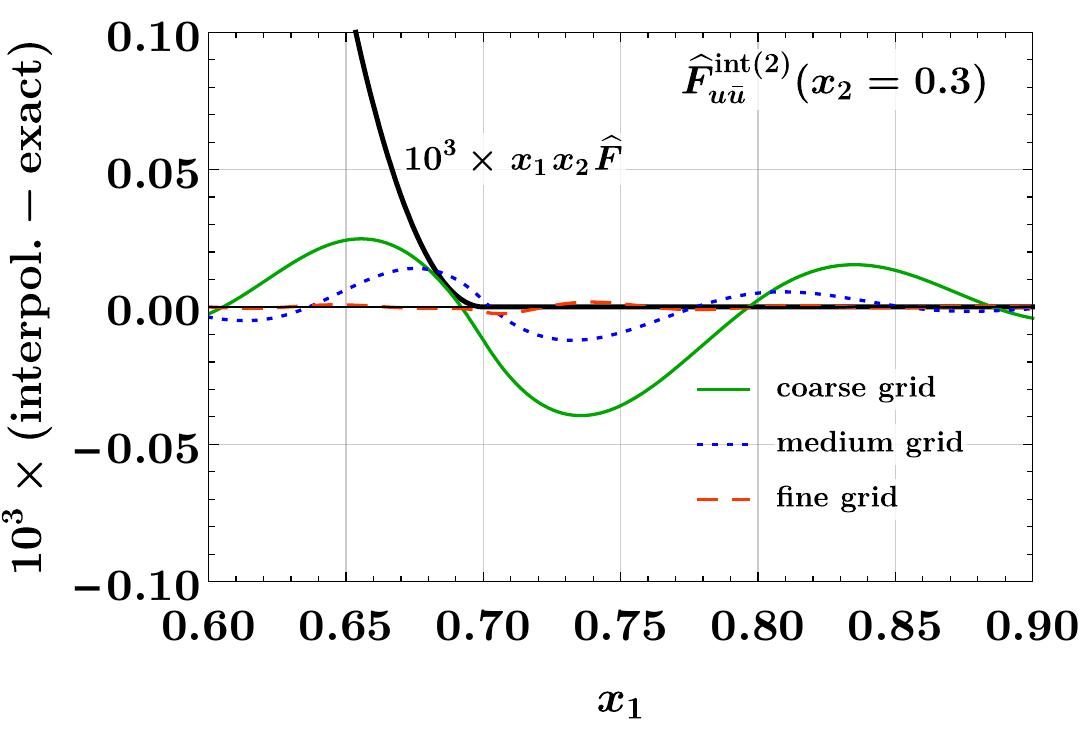}%
   }%
   \caption{\label{fig:interpol-x1x2-product-boundary}Interpolation of DPDs at
large momentum fractions.  The DPDs considered are the same as in the right
panels of \fig{interpol-x1x2-product}.  The plots on the left compare the exact
and interpolated forms of the DPDs.  The plots on the right show the difference
between the interpolated and the exact values, along with the exact form. Notice
that the plots in the top row have a different $x_1$ range than the remaining
ones.}
\end{figure}

The plots in \fig{interpol-x1x2-product-boundary} show the $x_1$ region around
the point where the DPD has a non-analytic behavior, which is at $x_1 = 0.7$ in
this example.  In these plots, we include the region $x_1 > 1 - x_2$, where the
interpolated DPD oscillates around its true value zero.  Note that this region is
included when one computes a Mellin convolution for the DPD with the kernel
matrix method described in \subsec{mellin_convolution}.  However, the effect of
oscillations around the true value will partially cancel out in convolution
integrals.

We see that for $r=0$, one obtains a rather poor approximation of the DPD in the
range $x_1 \in [0.6, 0.8]$ with the coarse or medium grid.  In situations where
this region is deemed important, one should take a dense grid and check the
reliability of the results by comparing with an even denser grid.  Of course,
this is only needed in the relevant subinterval (i.e.\ for $x_1 \in [0.5, 1]$ in
our example) and hence entails only a moderate increase in the total number of
grid points.  For $r=1$, where the DPD is continuous but has a discontinuous
first derivative at $x_1 = 0.7$, we see that the fine grid performs very well,
and for $r=2$ already the medium grid yields a good approximation of the DPD.

%===============================================================================
\subsection{Interpolation of the splitting form}
\label{subsec:interpolation-x1-x2-splitting}
%===============================================================================

We now turn to the interpolation of the splitting form.  We see in
\eq{F-spl-no-y} that it decreases almost as fast as a PDF for $x_1 + x_2 \to 1$,
so that according to the results of the previous subsection one does not expect
particular problems with interpolation in that region.  This is confirmed by our
numerical studies.

\begin{figure}[t]
   \subfigure[
      \label{fig:splitting-small-x2-uubar}
      $u\ubar, ~ x_2 = 10^{-4}$
   ]{%
      \includegraphics[height=\HeightTwoSubfigs]{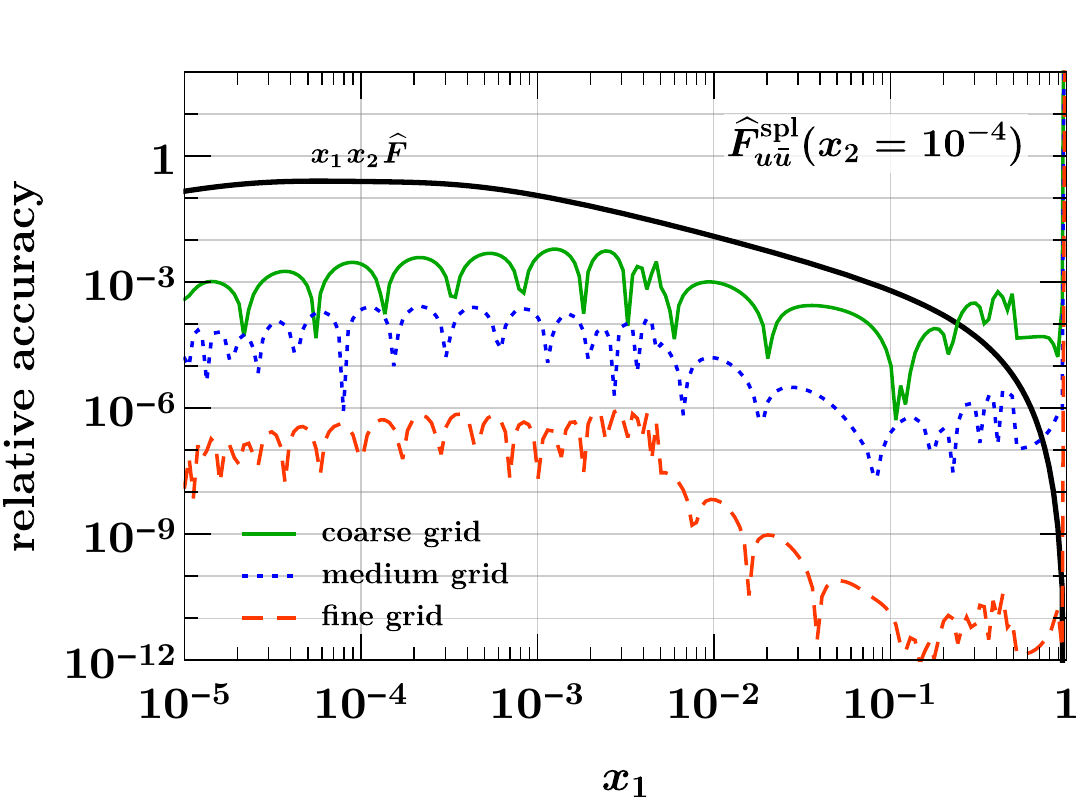}%
   }%
   \hfill
   \subfigure[
      \label{fig:splitting-large-x2-uubar}
      $u\ubar, ~ x_2 = 0.3$
   ]{%
      \includegraphics[height=\HeightTwoSubfigs]{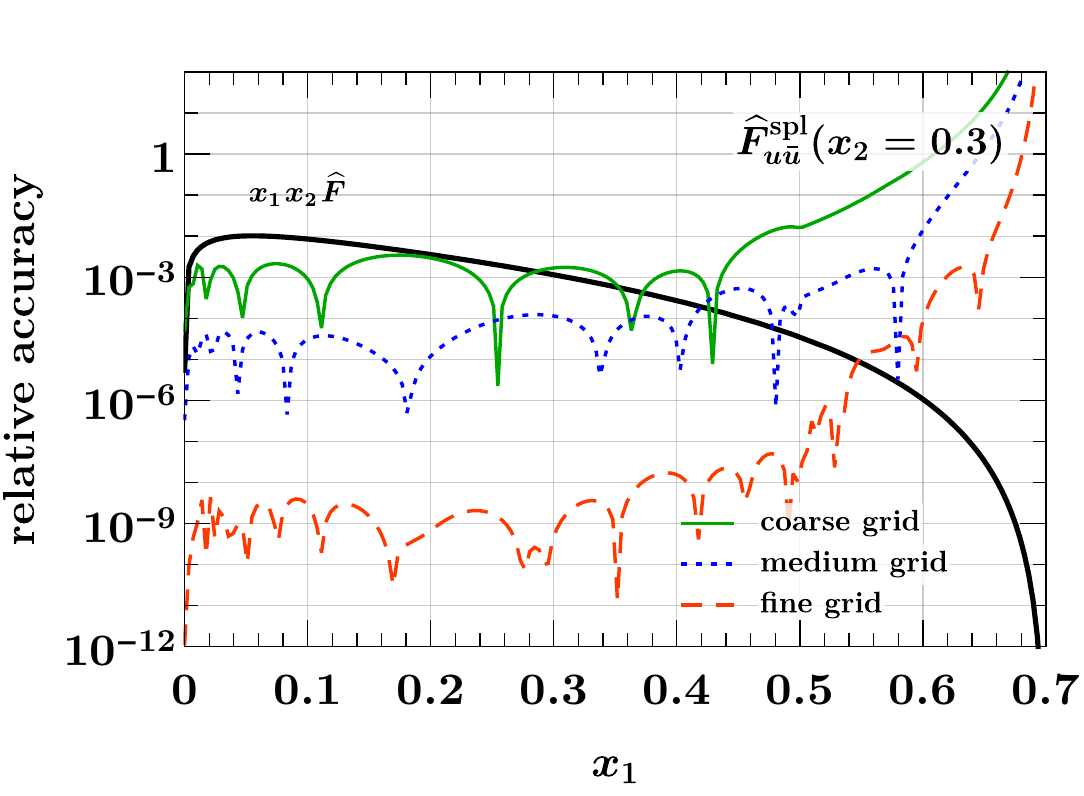}%
   }%
   \\
   \subfigure[
      \label{fig:splitting-small-x2-dgdg}
      $\delta g \delta g, ~ x_2 = 10^{-4}$
   ]{%
      \includegraphics[height=\HeightTwoSubfigs]{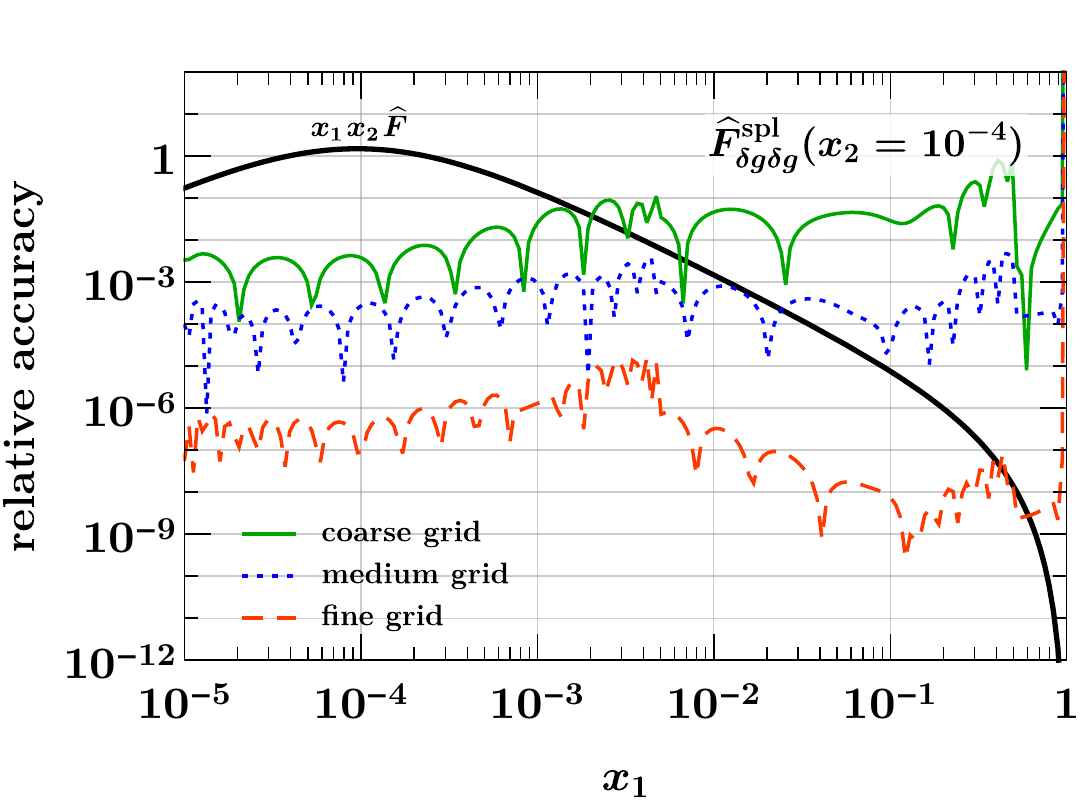}%
   }%
   \hfill
   \subfigure[
      \label{fig:splitting-large-x2-dgdg}
      $\delta g \delta g, ~ x_2 = 0.3$
   ]{%
      \includegraphics[height=\HeightTwoSubfigs]{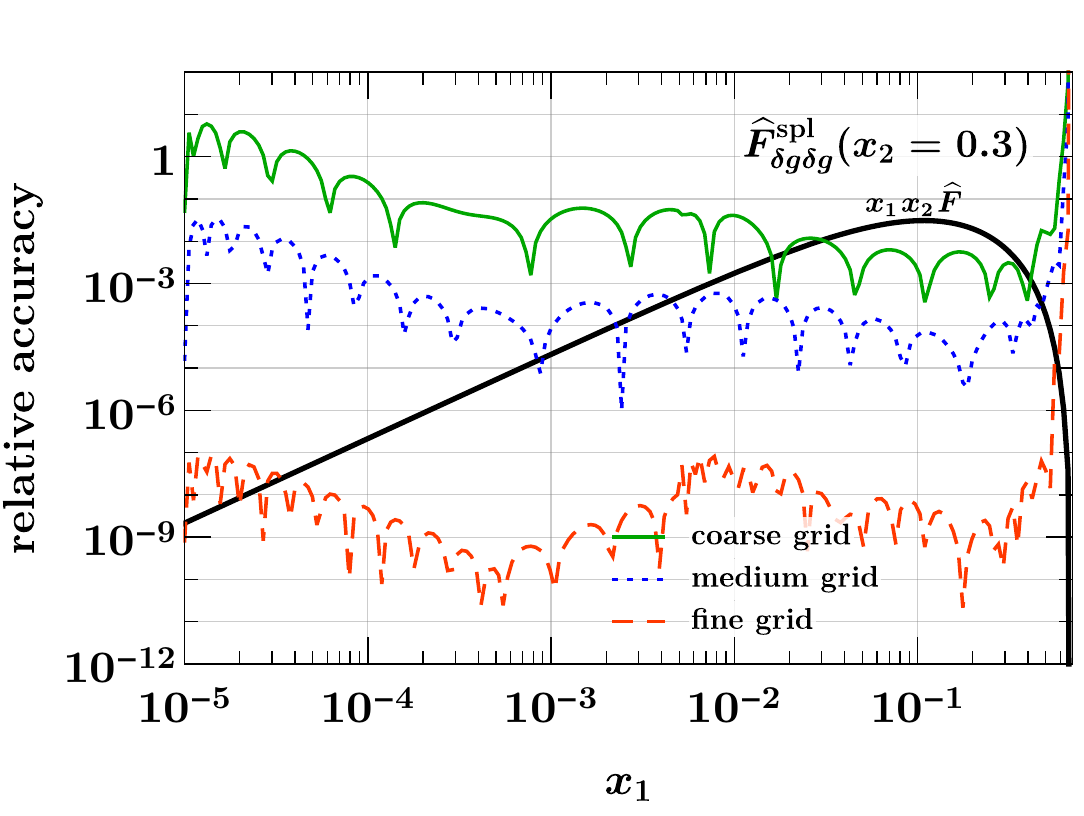}%
   }%
   \caption{\label{fig:interpol-x1x2-split}Relative accuracy
\protect\eqref{eq:rel-accuracy} for interpolation in $x_1$ and $x_2$, evaluated
for the splitting form $\Fspl{a_1 a_2}(x_1,x_2)$ of \eq{F-spl-no-y}.  The
interpolation grids are specified in \eq{x-grids}.  The plots in the top row are
for an unpolarized $u \ubar$ pair, and those in the bottom row for a pair of
linearly polarized gluons.  Notice that in panel (d) a logarithmic scale for $x_1$
is used to plot the DPD at $x_2 = 0.3$.}
\end{figure}%

The shape of the splitting form is strongly influenced by the end-point behavior
of the splitting kernel $V^{(1)}_{a_1 a_2, a_0}$, and in the following we show
results for two qualitatively different cases:
\begin{align}
   \label{eq:Vqqbarg-Vdgdgg}
      V^{(1)}_{u \ubar, g}(z)
   &= T_F \bigl[ z^2 + (1 - z)^2 \bigr]
   \,,
   &
      V^{(1)}_{\delta g \delta g, g}(z)
   &= 2 C_A \, z (1 - z)
   \,.
\end{align}%
Whereas the first kernel is finite for $z \to 0$ and $z \to 1$, the second one
goes to zero in both limits.  This implies a strong decrease of $\Fspl{\delta g
\delta g}(x_1, x_2)$ for both $x_1 \ll x_2$ and $x_1 \gg x_2$.  Other splitting
kernels diverge at the end points --- namely when an unpolarized gluon produced
by the splitting becomes soft --- and we verified that the interpolation
accuracy for these channels is not worse than the one for $u \ubar$.

We see in the top row of \fig{interpol-x1x2-split} that already the medium grid
yields a very good interpolation as long as the momentum fractions are not too
close to the kinematic limit.  At small $x_1$ and $x_2$, the accuracy is
somewhat less good than for the product form in \fig{interpol-x1x2-product},
which can be understood because the splitting form has more structure in that
region.  This is also seen for the interpolation of $\Fspl{\delta g \delta g}$
at small momentum fractions in panel (c), where the accuracy of the medium grid
is at the permille level.  In panel (d) we see that the linear decrease of the
distribution for $x_1 \ll x_2$ requires the fine grid for a satisfactory
interpolation at the smallest $x_1$, where the DPD is many orders of magnitude
smaller than its maximum value.  On the other hand, the medium grid still
performs well for $x_1 > 10^{-4}$.  We note that the coarse grid does not give a
good accuracy in any of the cases considered here, and it is entirely unreliable
in the case of panel (d).  We conclude that at at least 16 points per subgrid
should be taken, unless one considers only unpolarized DPDs and has only very
moderate accuracy requirements.

%===============================================================================
\subsection{Estimating the interpolating accuracy}
\label{subsec:interpolation-x1-x2-error-estimate}
%===============================================================================

Let us finally study the reliability of the accuracy estimate for interpolation
described in \subsec{chebyshev}, given by
\begin{align}
   \label{eq:accuracy-estimate}
   \text{relative accuracy estimate}
   &=
   \biggl| \,
      \frac{\text{value interpolated without end points}}
           {\text{value interpolated with end points}} - 1
   \, \biggr|
   \,.
\end{align}
\begin{figure}
   \subfigure[
      \label{fig:error-estimates-product-PSF-0-large-x2-uubar}
      $\Fint{u \ubar}{(0)}$ at $x_2 = 0.3$
   ]{%
      \includegraphics[height=\HeightTwoSubfigs]{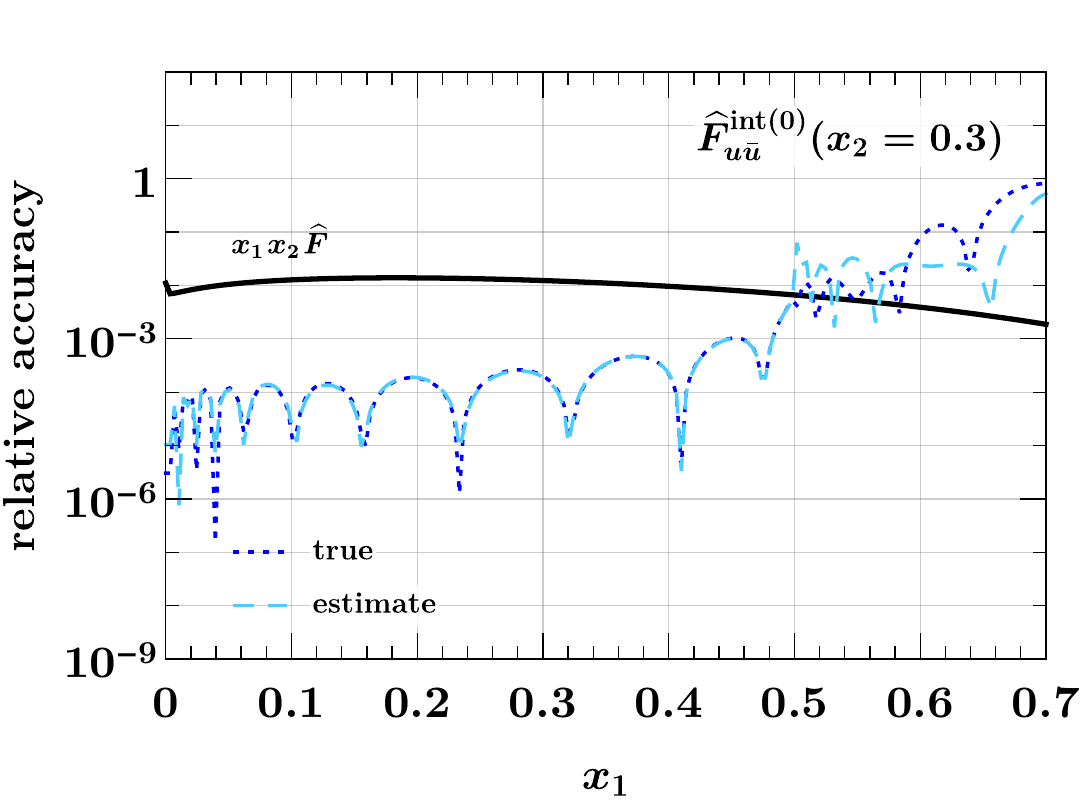}%
   }%
   \hfill
   \subfigure[
      \label{fig:error-estimates-product-PSF-1-large-x2-uubar}
      $\Fint{u \ubar}{(1)}$ at $x_2 = 0.3$
   ]{%
      \includegraphics[height=\HeightTwoSubfigs]{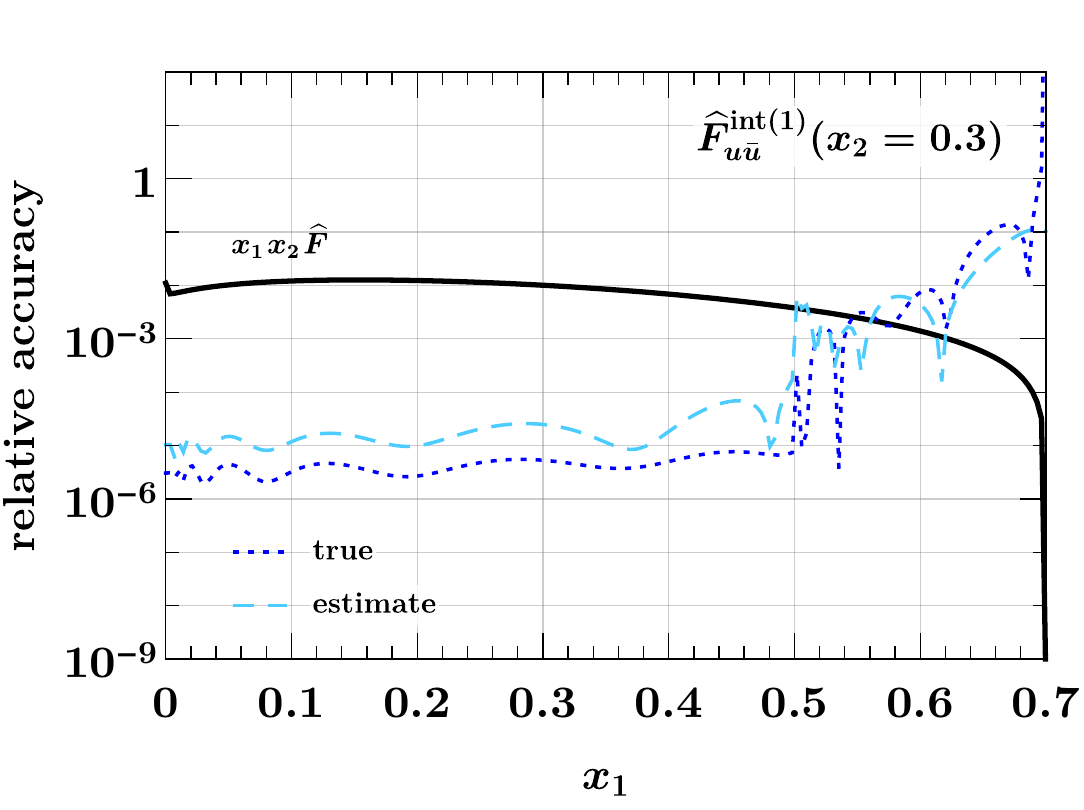}%
   }%
   \\
   \subfigure[
      \label{fig:error-estimates-product-PSF-2-large-x2-uubar}
      $\Fint{u \ubar}{(2)}$ at $x_2 = 0.3$
   ]{%
      \includegraphics[height=\HeightTwoSubfigs]{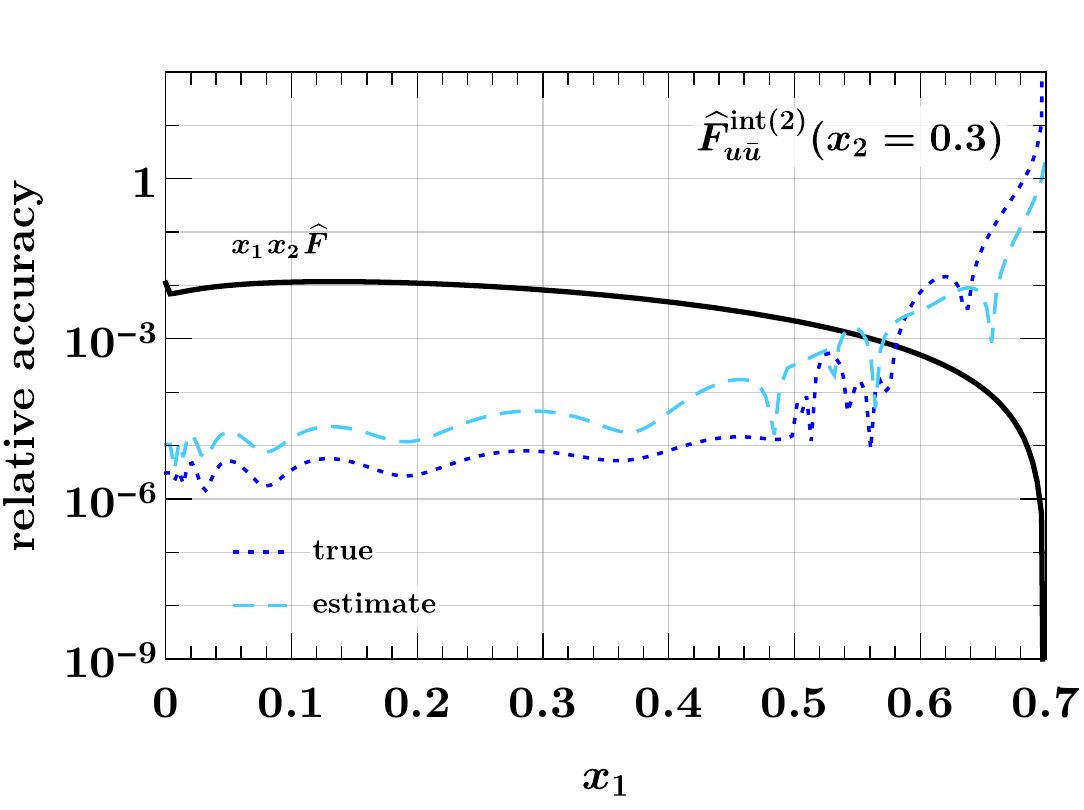}%
   }%
   \hfill
   \subfigure[
      \label{fig:error-estimates-splitting-small-x2-uubar}
      $\Fspl{u \ubar}$ at $x_2 = 10^{-4}$
   ]{%
      \includegraphics[height=\HeightTwoSubfigs]{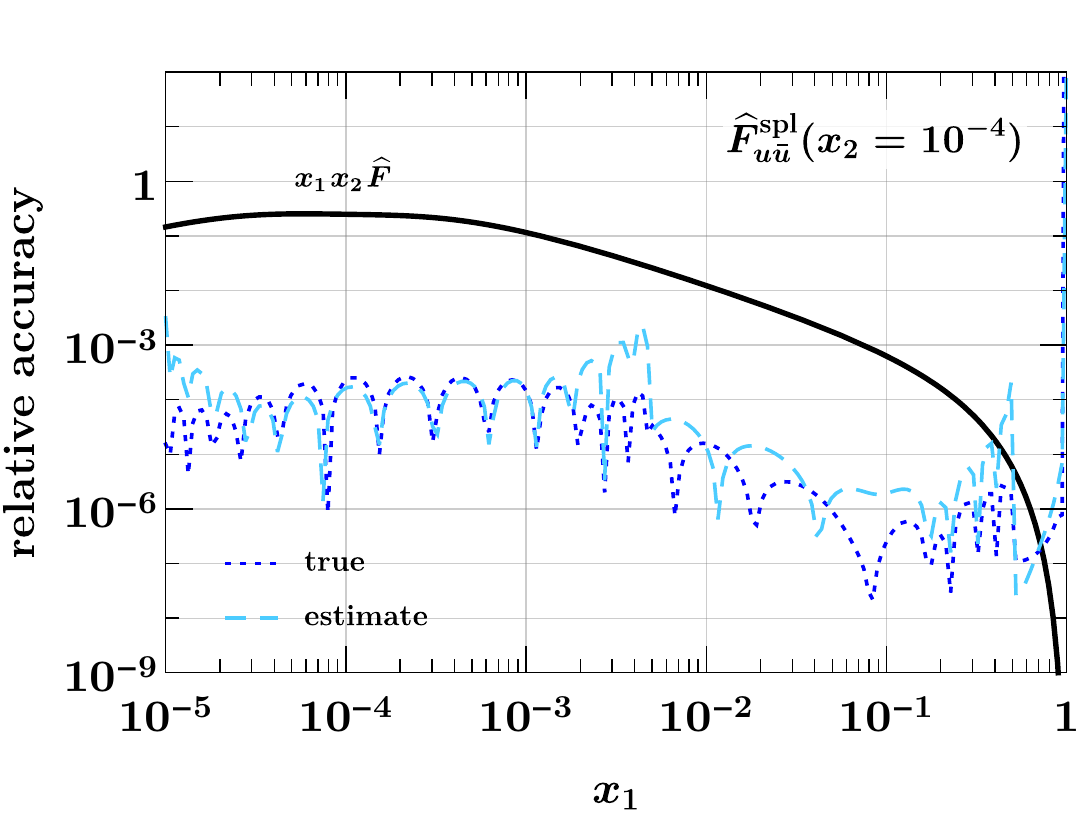}%
   }%
   \caption{\label{fig:interpol-x1x2-err-est}True interpolation accuracy in
$x_1$ and $x_2$ compared with its estimate \eqref{eq:accuracy-estimate}.  All
values refer to the medium $x$ grid in \eq{x-grids}.  The DPDs used are
specified in \eqs{F-int-no-y}{F-spl-no-y}.}
\end{figure}%
In \fig{interpol-x1x2-err-est}, we compare this estimate with the true relative
accuracy \eqref{eq:rel-accuracy} for the medium $x$ grid in \eq{x-grids}.  We
see that the estimate is typically not more than two orders of magnitude away
from the true accuracy (except around points where the error estimate or the
true error has a zero crossing).  Somewhat surprisingly, the estimate almost
coincides with the true accuracy in some cases, as seen in figure
\ref{fig:error-estimates-product-PSF-0-large-x2-uubar}.  Closer inspection
reveals that the estimate tends to be above the true accuracy close to
subinterval end points; this is expected because in that region interpolation
without the end points is particularly bad by construction.  We find a
qualitatively similar picture for the accuracy estimate in other kinematic
settings, and for the other two grids in \eq{x-grids}.  Although not perfectly
accurate in all regions, the method based on \eq{accuracy-estimate} provides a
good and easy-to-compute indicator for the quality of interpolation.

%%%%%%%%%%%%%%%%%%%%%%%%%%%%%%%%%%%%%%%%%%%%%%%%%%%%%%%%%%%%%%%%%%%%%%%%%%%%%%%%
\section{Evolution and flavor matching of DPDs}
\label{sec:dpd-evolution}
%%%%%%%%%%%%%%%%%%%%%%%%%%%%%%%%%%%%%%%%%%%%%%%%%%%%%%%%%%%%%%%%%%%%%%%%%%%%%%%%

In this section, we demonstrate the accuracy of DGLAP evolution and flavor
matching for DPDs with the methods presented in
\subsecs{mellin_convolution}{dglap_evolution}.  As a default setting for DPD
evolution, we take a maximal step size of $h = 0.22$ for the evolution time
\eqref{eq:mu-to-t} in the Runge-Kutta method.  We use the $8^{\text{th}}$ order
algorithm of Dormand and Prince \cite{PRINCE198167}, where each Runge-Kutta step
is divided into 13 substeps for which the r.h.s.\ of the evolution equation is
evaluated.

As initial conditions we take the splitting form \eqref{eq:F-spl} evaluated
with the Les Houches benchmark PDFs of \refscite{Giele:2002hx, Dittmar:2005ed}.
Following the settings in these references, we take
\begin{align}
   \label{eq:mu0-benchmark}
   \mu_0 &= \sqrt{2} \GeV
\end{align}
as initial scale with $\nf{} = 3$ active flavors for both partons.  The evolution
accuracy is independent of $y$, and we fix this variable to the value
$y = b_0 / \mu_0 \approx 0.794 \GeV^{-1}$.  Unless
specified otherwise, flavor matching is carried out at the canonical scales
$\mu_q = m_q$, with the quark masses $m_c = \sqrt{2} \GeV$, $m_b = 4.5 \GeV$, and
$m_t = 175 \GeV$.  The strong coupling at the initial scale is taken to be
$\alpha_s(\mu_0) = 0.35$.
Throughout this section, we consider unpolarized partons and use evolution and
flavor matching at NNLO.  We specify the scales and the number of active flavors
for the two partons in the form $(\mu_1, \mu_2)$ and $(\nf{1}, \nf{2})$.

Guided by the studies in \sec{x1-x2-interpolation}, we interpolate both $x_1$
and $x_2$ on the grid
\begin{align}
   \label{eq:x-grid-final}
      [10^{-5}, 5 \times 10^{-3}, 0.5, 1]_{(16, 16, 24)}
      \,.
\end{align}%
This corresponds to the medium grid in \eq{x-grids}, except for the third
subinterval, where we afford a finer grid for improved interpolation
accuracy at large momentum fractions.

%===============================================================================
\subsection{Path independence of evolution and flavor matching}
\label{subsec:path-indep-evol-match}
%===============================================================================

For a given heavy quark flavor $q$, we apply the matching equations
\eqref{eq:dpd-flavor-matching} at a fixed scale~$\mu_{q}$, so that in the
$(\mu_1, \mu_2)$ plane we match at the line $(\mu_{q}, \mu_2)$ for the first
parton and at the line $(\mu_1, \mu_{q})$ for the second one.  We show in
\app{path-indep} that, under these conditions, the result of evolution and flavor
matching between two points in the $(\mu_1, \mu_2)$ plane is independent of the
evolution path\rev{, including the order in which the different flavor
matching steps are carried out}.

\begin{figure}
   \centering
   \includegraphics[width=0.68\textwidth]{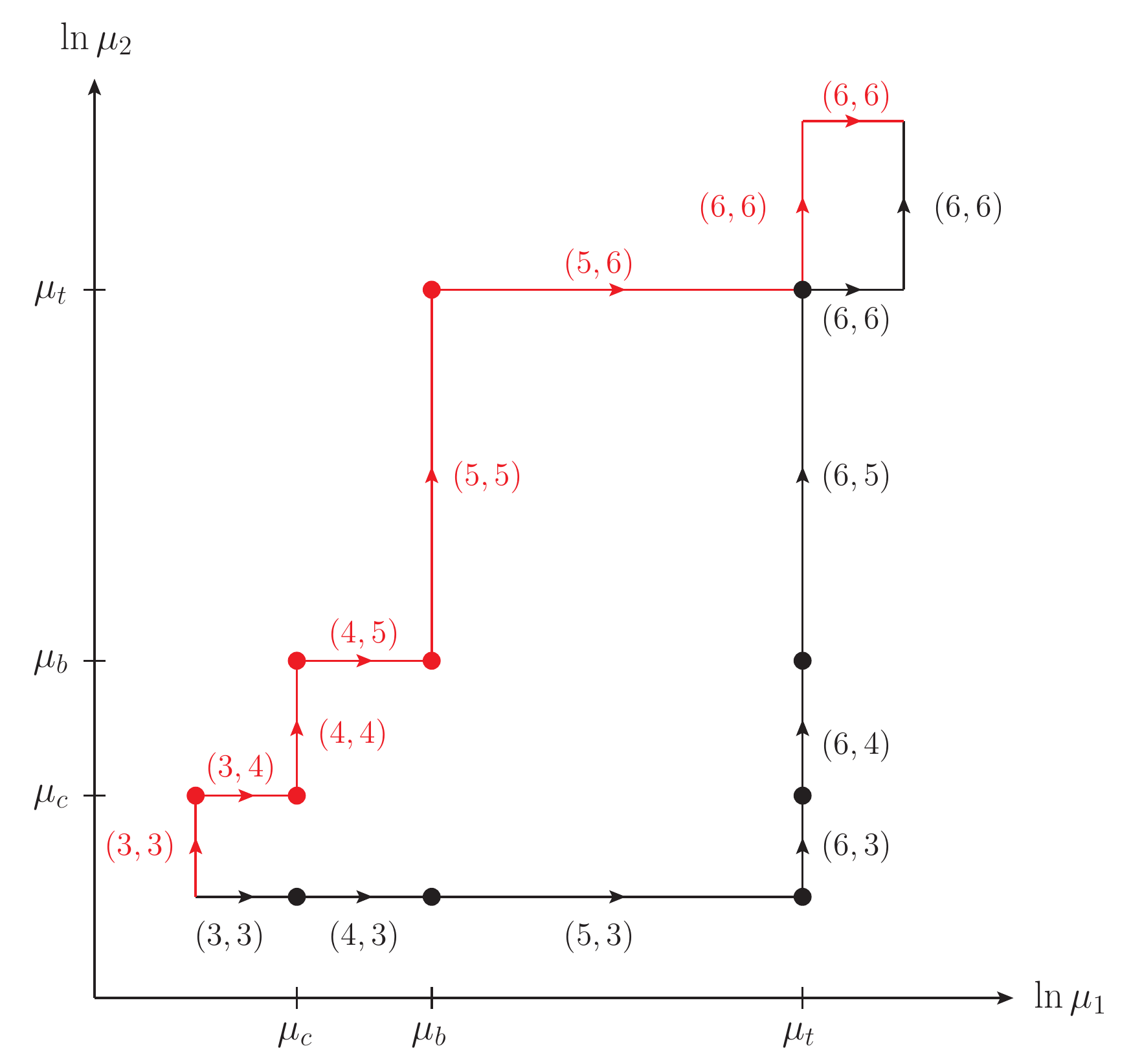}%
   \caption{\label{fig:evolution-path}Two paths for evolution and flavor
matching in the $(\mu_1, \mu_2)$ plane.  Next to each line segment, we specify
the numbers $(\nf{1}, \nf{2})$ of active flavors for which the DPD is evolved.
Flavor matching is performed at the points marked by dots.  The path in black is
the \chili\ default for matching from $(3,3)$ to $(6,6)$ and for subsequent
evolution up to the final scale pair.}
\end{figure}%

We verify this numerically by comparing the results of evolving and matching from
$(1 \GeV, 1 \GeV)$ with $(3, 3)$ flavors to $(500 \GeV, 1000 \GeV)$ with $(6, 6)$
flavors along the two paths shown in \fig{evolution-path}.  The path in black is
the default taken in \chili\ when matching from 3 to 6 flavors for both partons
and then evolving to the final pair of scales.  The staircase-like path in red
can be selected by subsequent evolution calls for each line segment.
\rev{Each point marked by a dot in the figure corresponds to the initial
condition of the DPD with given flavor numbers $(\nf{1}, \nf{2})$ in \chili.
For any scale pair $(\mu_1, \mu_2)$, the DPD with these flavor numbers is
computed by evolution from this initial condition.}

\begin{figure}
   \subfigure[
      \label{fig:path-indep-evol-match-small-x2-ttbar}
      $t\tbar, ~ x_2 = 10^{-4}$
   ]{%
      \includegraphics[height=\HeightTwoSubfigs]{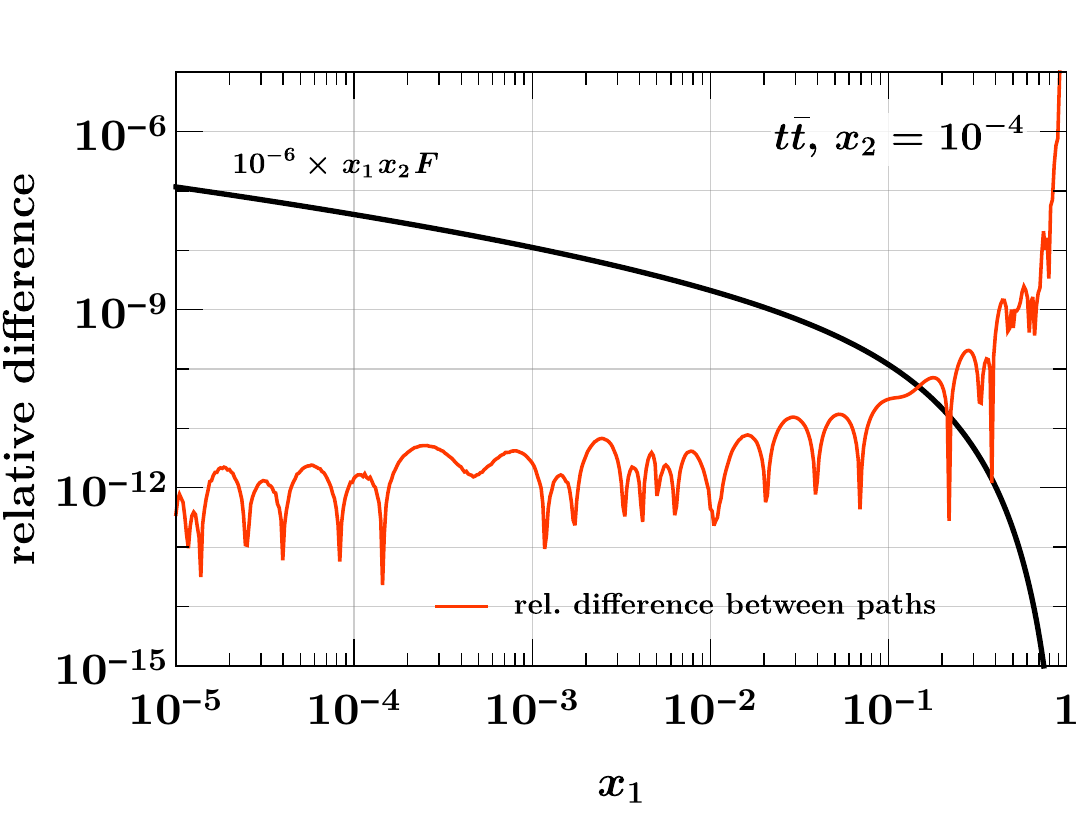}%
   }%
   \hfill
   \subfigure[
      \label{fig:path-indep-evol-match-large-x2-ttbar}
      $t\tbar, ~ x_2 = 0.3$
   ]{%
      \includegraphics[height=\HeightTwoSubfigs]{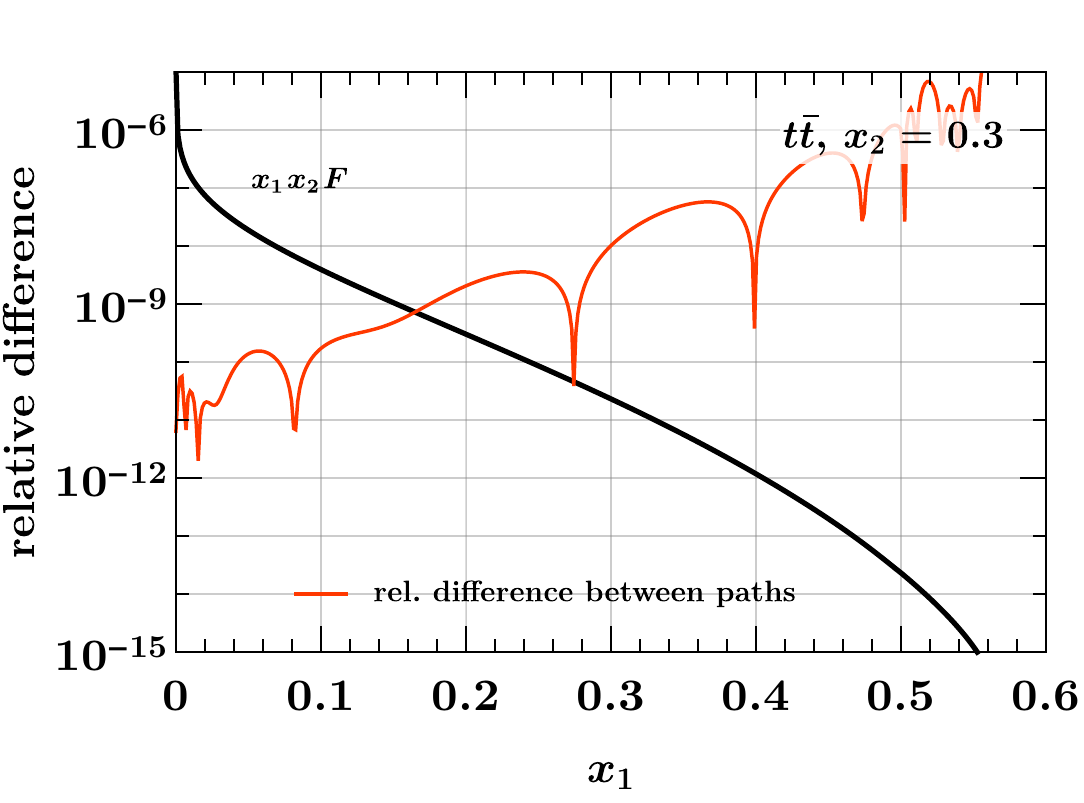}%
   }%
   \caption{\label{fig:path-indep-evol-match}Relative difference between the $t
\tbar$ distribution obtained by evolving and matching along one of the two paths
in \fig{evolution-path}.  Both paths lead from $(1 \GeV, 1 \GeV)$ with $(3, 3)$
flavors to $(500 \GeV, 1000 \GeV)$ with $(6, 6)$ flavors.  The distributions at
$(1 \GeV, 1 \GeV)$ are obtained by evolving backwards from the starting
conditions at $(\mu_0, \mu_0)$ specified at the beginning of the section.}
\end{figure}

The results \rev{of evolving and matching along the two specified paths}
are in excellent agreement, as illustrated in \fig{path-indep-evol-match} for the
$t \tbar$ distribution, for which the differences are largest among all flavor
combinations.  The agreement deteriorates close to the kinematic limit $x_1 = 1 -
x_2$ but remains small compared with the interpolation accuracy in that region.
The choice for a particular evolution path in our implementation is therefore of
no consequence for the accuracy.

%===============================================================================
\subsection{Evolution and flavor matching accuracy}
\label{subsec:accuracy-evol-match}
%===============================================================================

We now study evolution and flavor matching to higher scales, using the initial
conditions specified at the beginning of this section. To assess the accuracy, we
use three settings:
\begin{enumerate}
\item our default setting with the grid
$[10^{-5}, 5 \times 10^{-3}, 0.5, 1]_{(16, 16, 24)}$ and a maximal step size $h
= 0.22$ for the Runge-Kutta algorithm,
\item a finer grid $[10^{-5}, 5 \times 10^{-3}, 0.5, 1]_{(40, 40, 40)}$ with the
same maximal step size $h = 0.22$,
\item the same finer grid with a smaller maximal step size $h = 0.0088$.
\end{enumerate}
The difference between settings 1 and 2 is taken as an estimate of the error
from discretization in $x_1$ and $x_2$ with our default grid, and the difference
between settings 2 and 3 is taken as an estimate of the error due to the
Runge-Kutta algorithm.

\begin{figure}
   \subfigure[
      $u \ubar, ~ x_2 = 10^{-4}$
   ]{%
      \includegraphics[width=\WidthTwoSubfigs]{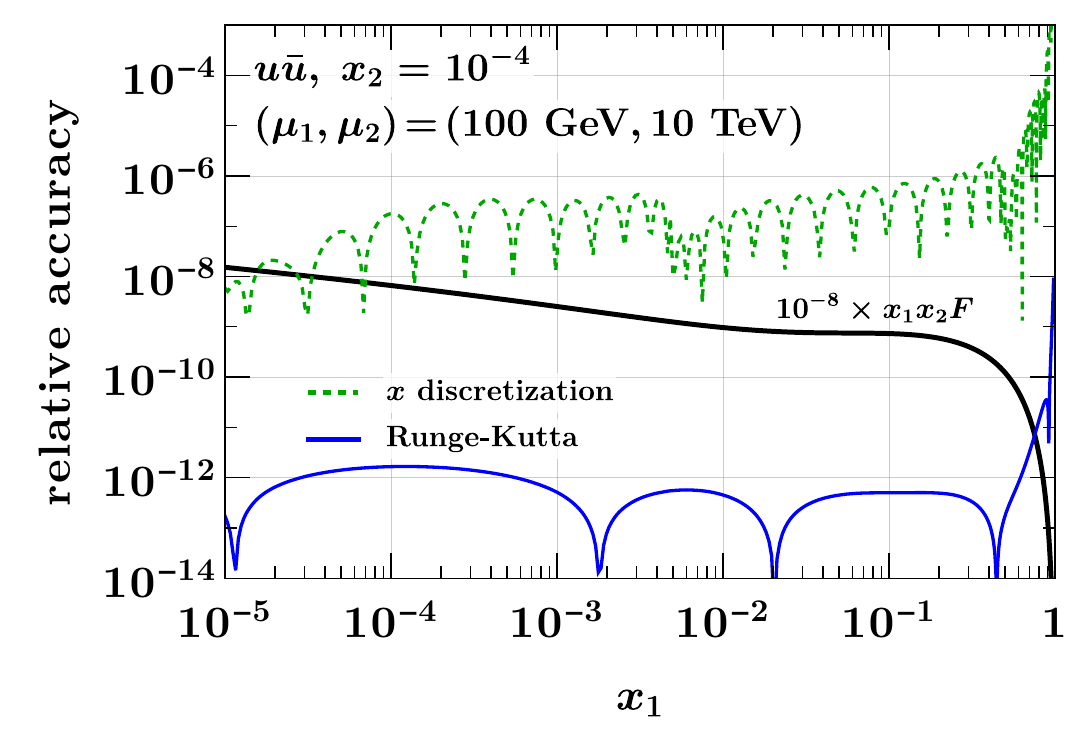}%
   }%
   \hfill
   \subfigure[
      $d g, ~ x_2 = 10^{-4}$
   ]{%
      \includegraphics[width=\WidthTwoSubfigs]{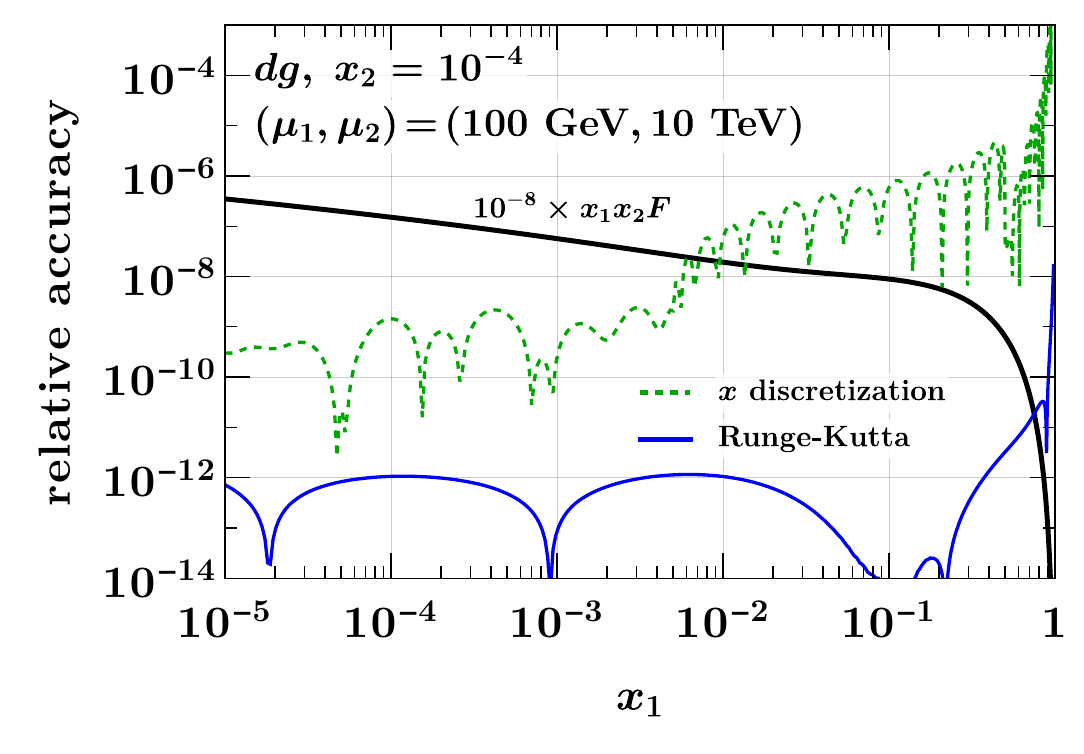}%
   }%
   \\
   \subfigure[
      $d \sbar, ~ x_2 = 0.3$
   ]{%
      \includegraphics[width=\WidthTwoSubfigs]{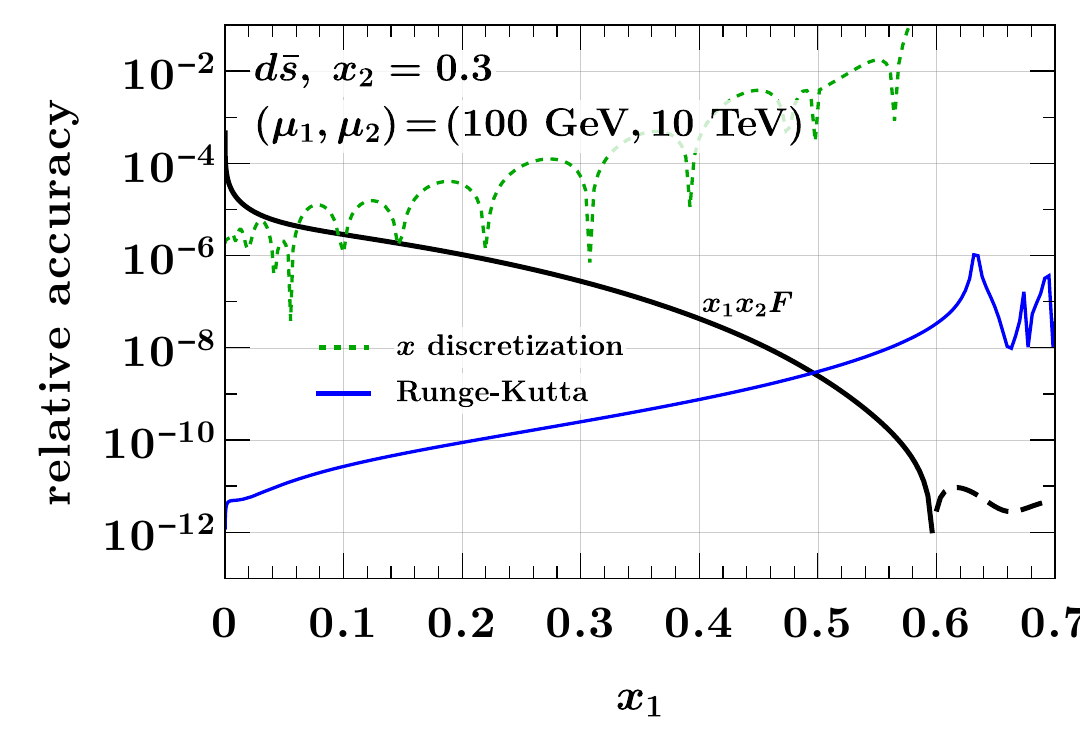}%
   }%
   \hfill
   \subfigure[
      $g \ubar, ~ x_2 = 0.3$
   ]{%
      \includegraphics[width=\WidthTwoSubfigs]{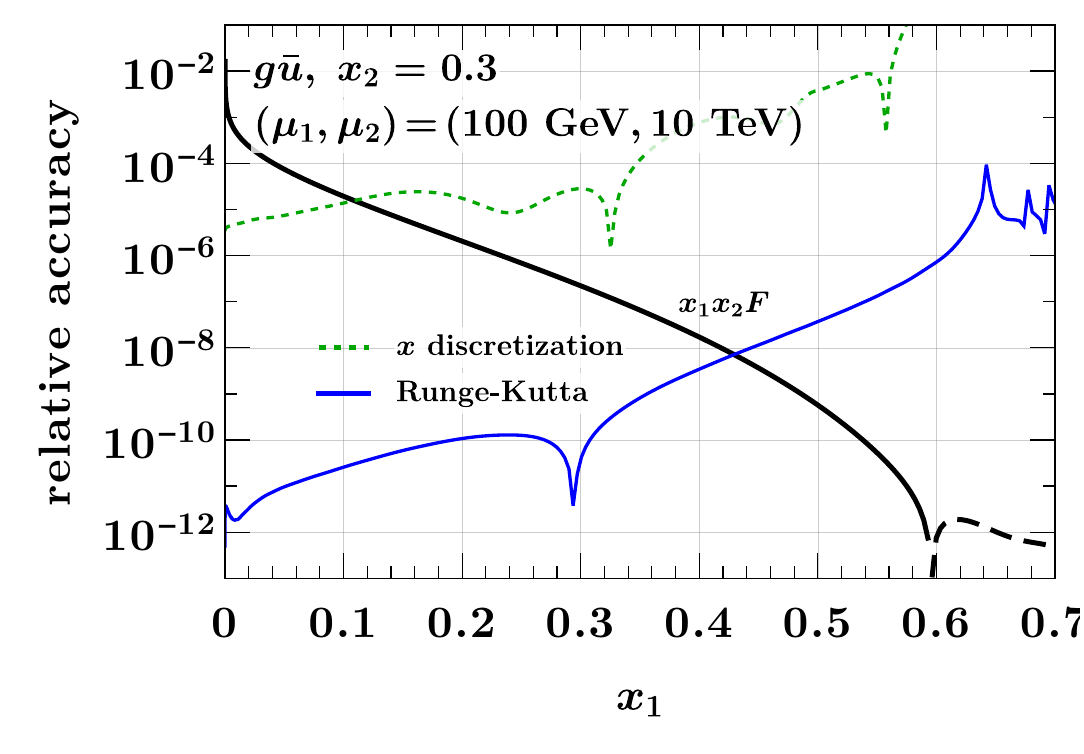}%
   }%
   \caption{\label{fig:evol-match-accuracy-hi}Relative Runge-Kutta and
discretization errors for DPD evolution and flavor matching from scales $(\mu_0,
\mu_0)$ with $(3, 3)$ flavors to $(100\GeV, 10\TeV)$ with $(5, 6)$ flavors.  The
initial conditions are specified at the beginning of the section.}
\end{figure}

These errors are shown in \fig{evol-match-accuracy-hi} for evolution and
matching to $(100\GeV, 10\TeV)$ with $(5, 6)$ active flavors.  The plots are for
selected parton combinations; we verified that the errors for the other
combinations are of similar size or even smaller.
We see that the Runge-Kutta errors are always negligible compared with the
errors from discretization.  The latter are of similar size as (or even smaller
than) the interpolation errors for the initial conditions, which are shown for
$u \ubar$ in \fig{interpol-x1x2-split}.  Evolution to high scales does therefore
\emph{not} degrade the overall accuracy of DPDs with our method.  Only close to
the kinematic limit does the relative error from discretization become large.
With the chosen grid, it remains below 1\% for $1 - (x_1 + x_2)$ above $0.2$.

\begin{figure}
   \subfigure[
      $\cbar b, ~ x_2 = 10^{-4}$
   ]{%
      \includegraphics[width=\WidthTwoSubfigs]{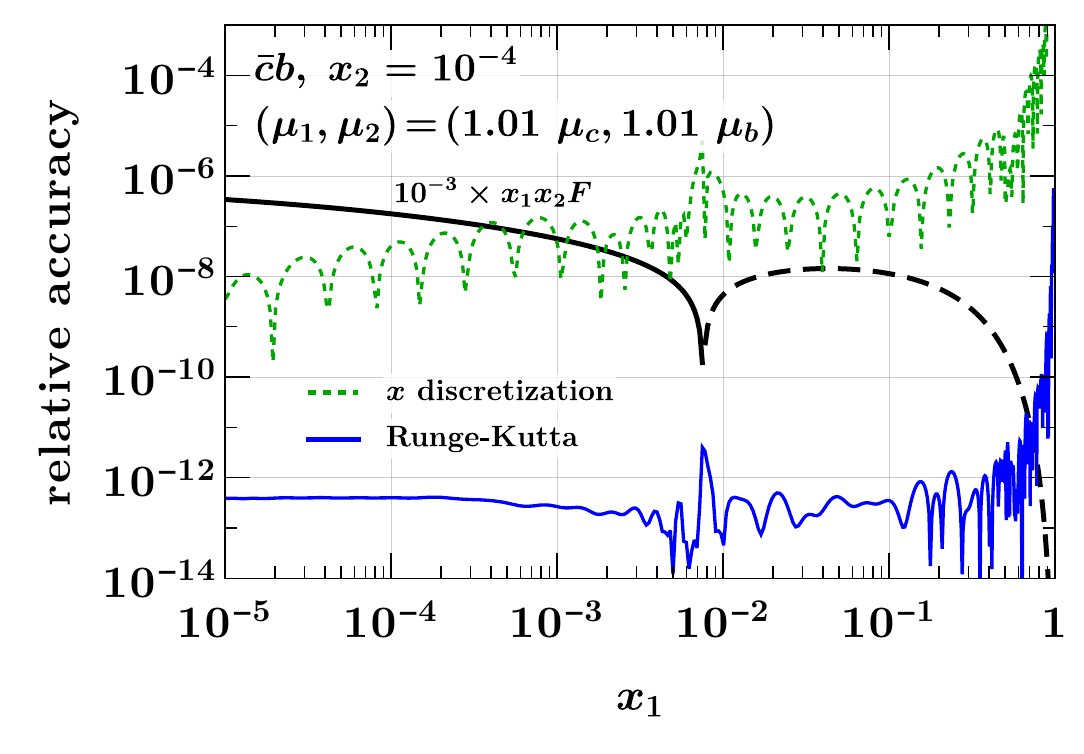}%
   }%
   \hfill
   \subfigure[
      $s \sbar, ~ x_2 = 10^{-4}$
   ]{%
      \includegraphics[width=\WidthTwoSubfigs]{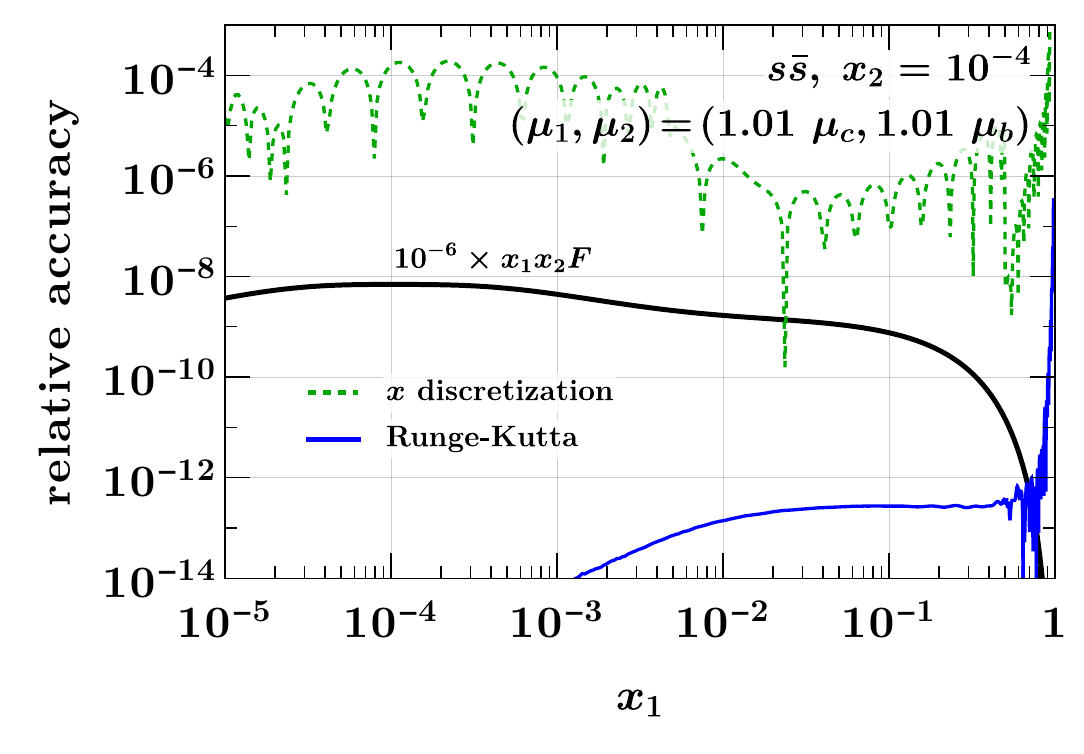}%
   }%
   \\
   \subfigure[
      $g g, ~ x_2 = 0.3$
   ]{%
      \includegraphics[width=\WidthTwoSubfigs]{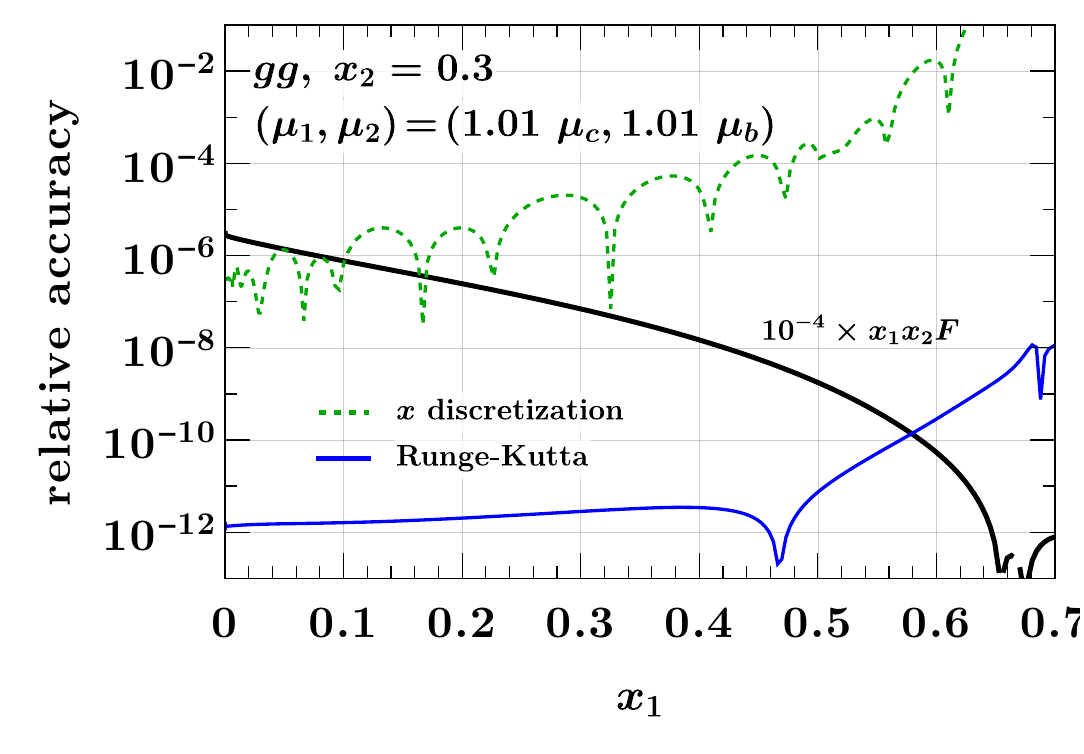}%
   }%
   \hfill
   \subfigure[
      $\ubar b, ~ x_2 = 0.3$
   ]{%
      \includegraphics[width=\WidthTwoSubfigs]{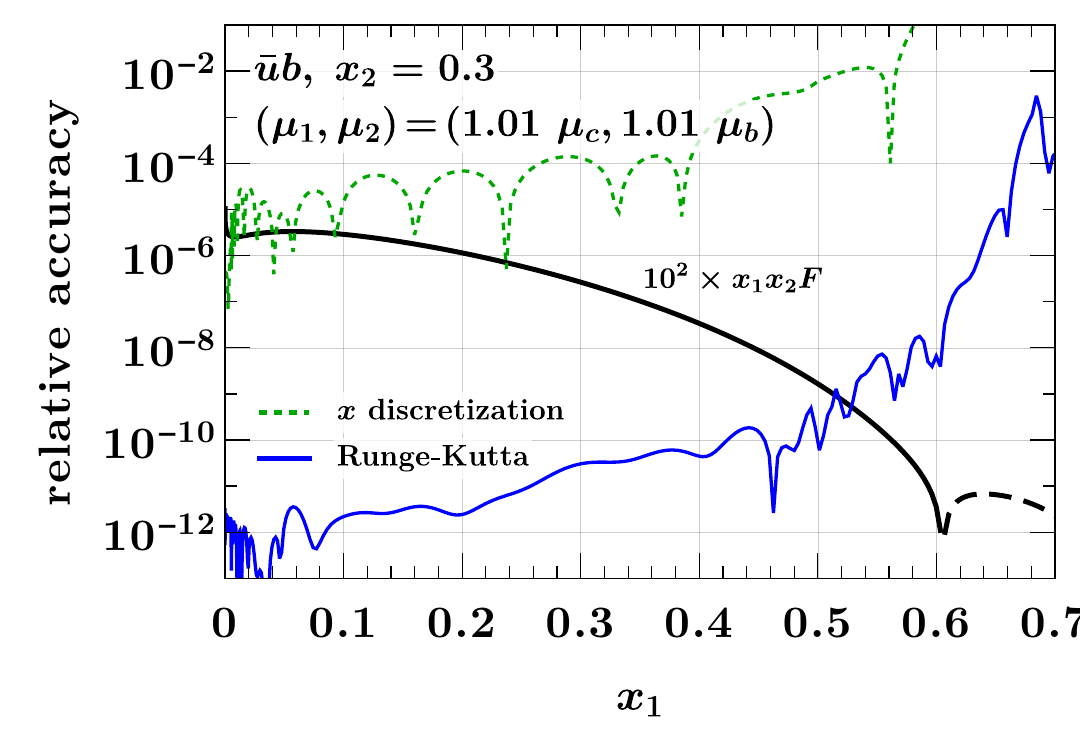}%
   }%
   \caption{\label{fig:evol-match-accuracy-lo}As \fig{evol-match-accuracy-hi},
but for different flavor combinations evolved and matched to scales $(1.01 \ms
\mu_c, 1.01 \ms \mu_b)$ with $(4, 5)$ flavors.  A dashed black line indicates
that the interpolated function is negative.}
\end{figure}

In \fig{evol-match-accuracy-lo}, we show the errors for evolution and matching
to $(1.01 \ms \mu_c, 1.01 \ms \mu_b)$ with $(4, 5)$ flavors, i.e.\ to scales
just slightly above the matching scales for the final flavor numbers.  The
accuracy remains excellent, even for DPDs with charm as first or bottom as
second parton, which are still tiny at the chosen scales.

%===============================================================================
\subsection{Backward evolution accuracy}
\label{subsec:accuracy-back-evol}
%===============================================================================

A stringent way to test the accuracy of an evolution algorithm is backward
evolution from large to small scales.  Whilst small inaccuracies of
distributions at a given scale are diminished when evolving to higher scales,
they are amplified by backward evolution.  In the following, we adapt the
backward evolution test performed for PDFs in \refcite{Diehl:2021gvs} to the
case of DPDs.

We start with the three-flavor input DPD specified at the beginning of this
section, match to $4$ flavors at the input scales $(\mu_0, \mu_0)$, and then
evolve to $(\muLow, \muLow)$ with $\muLow = m_b/2 = 2.25 \GeV$.  At this point,
we match from $4$ to $5$ flavors for both partons.  Since the matching scale
differs from $m_b$, all DPDs are nonzero at that point, including the ones with
bottom quarks or antiquarks.  We take these distributions as starting conditions
for evolution with $5$ flavors, evolving
\begin{enumerate}
\item from $(\muLow, \muLow)$ to $(\muHi, \muHi)$ with $\muHi = 1 \TeV$ in step
1,
\item from $(\muHi, \muHi)$ back to $(\muLow, \muLow)$ in step 2.
\end{enumerate}
We use our default value of $h = 0.22$ for the maximal Runge-Kutta step size.

\begin{figure}
   \subfigure[
      \label{fig:back-evol-accuracy-small-x2}
      $x_2 = 10^{-4}$
   ]{%
      \includegraphics[width=0.498\textwidth]{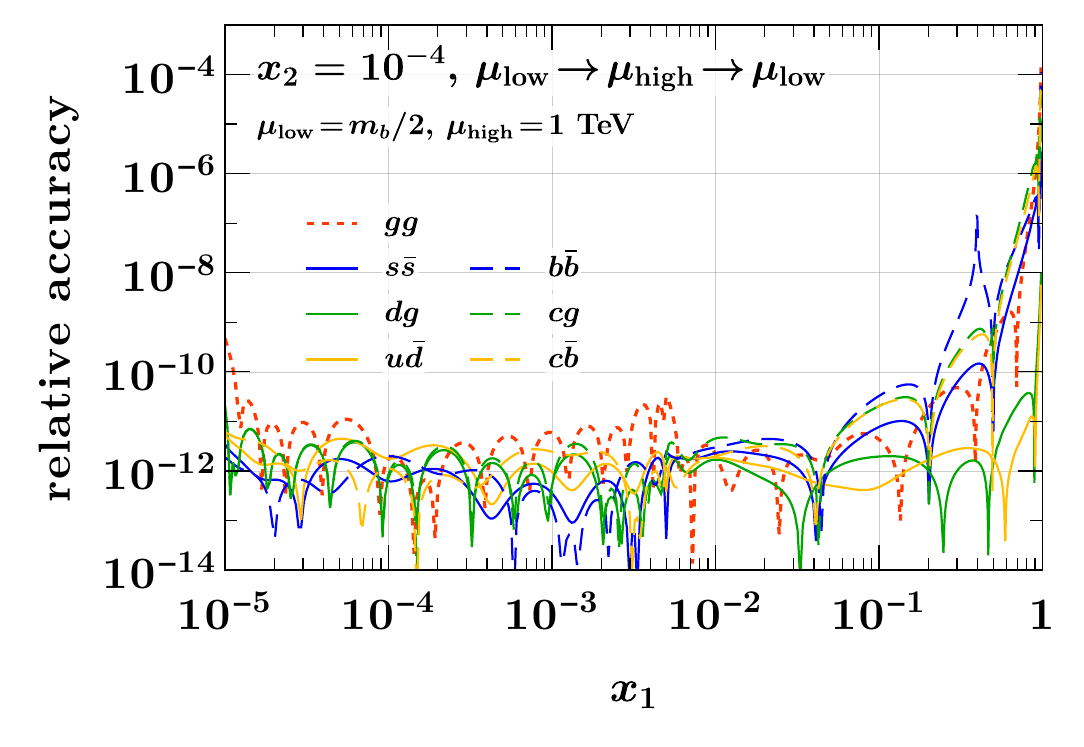}%
   }%
   \hfill
   \subfigure[
      \label{fig:back-evol-accuracy-large-x2}
      $x_2 = 0.3$
   ]{%
      \includegraphics[width=0.498\textwidth]{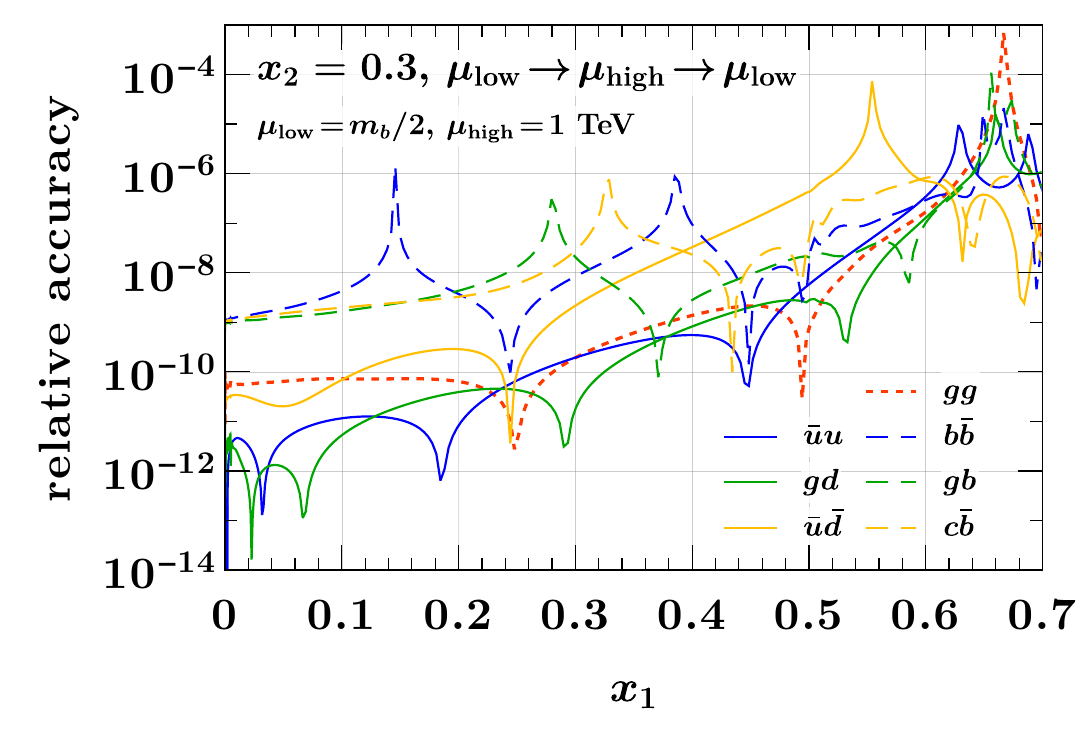}%
   }%
   \caption{\label{fig:back-evol-accuracy}Relative accuracy of evolving DPDs from
$\muLow = m_b /2$ to $\muHi = 1\TeV$ and back to $\muLow$ for both partons.
Details are given in the text.}
\end{figure}

The relative difference between the input to the first step and the output of the
second step is sensitive to the accuracy of forward \emph{and} backward
evolution.  In \fig{back-evol-accuracy}, this quantity is shown for selected
parton combinations; the accuracy is not worse for other combinations.  We find
that backward evolution is very precise, with the relative errors in our exercise
being below $10^{-6}$ over most of the phase space.  This also holds for
distributions containing bottom quarks or antiquarks, which have zero crossings
at the scale $\muLow$ where the ratio shown in the figure is evaluated.

We note that backward evolution is inherently more demanding for two scales than
for a single scale.  Indeed, step 2 of our study corresponds to backward
evolution first from $(\muHi, \muHi)$ to $(\muLow, \muHi)$ and then from
$(\muLow, \muHi)$ to $(\muLow, \muLow)$.  Inaccuracies accumulated in the first
substep are amplified in the second one.  We repeated the above exercise for
evolution in only one scale, evolving from $(\muLow, \muLow)$ up to $(\muHi,
\muLow)$ and then down to $(\muLow, \muLow)$.  We find that indeed the accuracy
is even higher in that case, by one or two orders of magnitude in some kinematic
regions.

%%%%%%%%%%%%%%%%%%%%%%%%%%%%%%%%%%%%%%%%%%%%%%%%%%%%%%%%%%%%%%%%%%%%%%%%%%%%%%%%
\section{Interpolation of DPDs in the distance between partons}
\label{sec:y-interpolation}
%%%%%%%%%%%%%%%%%%%%%%%%%%%%%%%%%%%%%%%%%%%%%%%%%%%%%%%%%%%%%%%%%%%%%%%%%%%%%%%%

In the discussion so far, the transverse distance $y$ between partons did not
play an active role, given that evolution and flavor matching of DPDs are
performed for each discretized value of $y$ individually, without any cross talk
between different values of $y$.  Once a DPD is computed for selected
scales $(\mu_1, \mu_2)$ and flavor numbers $(\nf{1}, \nf{2})$, two different
tasks are concerned with the $y$ dependence:
\begin{enumerate}
\item interpolation of the DPD for specific values of $y$, in order to study how
evolution and matching have modified the $y$ dependence of the initial
conditions.  In the study described in \refcite{Diehl:2014vaa}, visible effects
of evolution on the large-$y$ dependence of DPDs were reported, and in
\refcite{Diehl:2017kgu} it was found that evolution can substantially flatten
the $y^{-2}$ behavior generated by the perturbative splitting mechanism at small
distances.  Numerically reliable interpolation of DPDs over a wide range of $y$
is hence important for understanding the dynamics of evolution.
\item integration over $y$ from a minimum value $y_{\min}$ up to infinity.
First and foremost, this is needed for computing the double parton luminosities
\eqref{eq:DPD-lumi-def} that enter in cross sections of double parton scattering
processes.  It is also needed for evaluating sum rules (see \sec{dove}), which
DPDs must satisfy and which provide valuable constraints for their modeling.
Note that one integrates the product of two DPDs when computing double parton
luminosities and single DPDs when evaluating the sum rules.
\end{enumerate}
Both tasks can be carried out using Chebyshev grids, as described in
\subsec{chebyshev}.  To this end, one first needs to find variable
transformations that map the characteristic $y$ dependence of DPDs to a
dependence in the variable $u$ that is well suited for interpolation by
high-order polynomials.  In the next two subsections, we show how this can be
achieved.  After that, we will demonstrate the resulting accuracy of
interpolation and integration.

%===============================================================================
\subsection{Characteristic \texorpdfstring{$y$}{y} dependence and associated
variable transformations}
\label{subsec:variable-trafo-y}
%===============================================================================

It is natural to distinguish the region of small $y$, where perturbative
mechanisms are at work from the domain of large $y$.

As mentioned earlier and discussed at length in \refcite{Diehl:2017kgu}, DPDs at
small $y$ can be represented as the superposition of a part due to perturbative
splitting, with a power behavior like $y^{-2}$, and an ``intrinsic'' part that
behaves like $y^{0}$ and is related to twist-four distributions.  Both power laws
are modified by logarithms from loop corrections.  The relative weight of the two
parts strongly depends on the parton combination and on the momentum fractions
$x_1$ and $x_2$ in the DPD.  As already mentioned, the $y^{-2}$ behavior can be
substantially flattened by evolution to high scales. In the small-$y$ region, the
transformation $u(y)$ should therefore be suited to handle the interpolation of a
power behavior $y^{-q}$ with $0 \le q \le 2$.

A complication arises from the fact that for certain parton combinations, such
as $u u$ or $u \dbar$, the leading-order splitting contribution \eqref{eq:F-spl}
is zero.  Quark-gluon mixing results in nonzero values for these combinations
when evolving from the scale $\mu_y \sim 1/y$ at which the splitting formula is
evaluated to the final scale $\mu$.  Hence, the splitting part of the DPD at
scale $\mu$ has a zero around $y \sim 1/\mu$.  The position of this zero is
shifted when the intrinsic part is added.

At large distances, one expects a gradual decrease of DPDs governed by some
characteristic nonperturbative mass scale.  This is often modeled by a
Gaussian, $F \sim \exp(- M^2 y^2)$, but one may also argue in favor of a
Yukawa-like decrease, $F \sim \exp(- M y)$.  The characteristic mass scale $M$
may depend on the parton combination and on $x_1$ and $x_2$ \cite{Diehl:2014vaa,
Corke:2011yy}, and it can slowly change under evolution.

We now discuss variable transformations that we found to work well for the types
of $y$ dependence just discussed.  An overview of the transformations and their
properties is given in \tabs{y-grid-info-1}{y-grid-info-2}.  By convention, all
transformations are such that $u(y)$ has a positive derivative in the domain
where it is used.  \rev{In the case of interpolation, the function $f(y)$ in
\tab{y-grid-info-2} represents a DPD at fixed momentum fractions or scales. For
sum rule integrals $f(y)$ is $2 \pi y$ times a DPD (see \eq{msbar-dpd-def}), and
for double parton luminosities $f(y)$ is $2 \pi y$ times the product of two DPDs
(see \eq{DPD-lumis}).  The overall normalization of $f(y)$ is irrelevant for the
interpolation or integration accuracy and therefore set to an arbitrary value
here.}

\begin{table}[t]
   \centering
   \begin{tabular}{@{}c|cccc@{}}
      \toprule
      no. &
         ${}-u(y)$ &
         $y(u)$ &
         $y(-1)$ &
         $\dd y / \dd u$
         \\ \midrule
      1  &
         $y^{-\alpha}$ &
         $|u|^{- 1 / \alpha}$ &
         $1$ &
         $\frac{1}{\alpha} \, |u|^{\ms -(1+\alpha)/\alpha}$
         \\[0.25em]
      2  &
         $\exp\Bigl[- \frac{1}{4} \ms (m^2 y^2 + m y) \Bigr]$ &
         $\frac{1}{2 m} \, \bigl(\sqrt{16 L(u) + 1} - 1\bigr)$ &
         $0$ &
         $\frac{4}{m} \, |u|^{-1} \, \bigl(16 L(u) + 1\bigr)^{-1/2}$
         \\[0.25em]
      3  &
         $\exp\Bigl[-\frac{1}{4} \ms m y \Bigr]$ &
         $\frac{4}{m} \ms L(u)$ &
         $0$ &
         $\frac{4}{m} \ms |u|^{-1}$
         \\[0.25em]
      4  &
         $\exp\Bigl[1 - \sqrt{1 + m y / 2}\Bigr]$ &
         $\frac{2}{m} \, \bigl(L^2(u) + 2 L(u)\bigr)$ &
         $0$ &
         $\frac{4}{m} \, |u|^{-1} \, \bigl(L(u) + 1\bigr)$
         \\ \bottomrule
   \end{tabular}
   \caption{\label{tab:y-grid-info-1}Variable transformations suitable for
interpolation in $y$.  The parameters \rev{$\alpha$} and $m$ must be positive.
All
transformation satisfy $u \le 0$, $y(0) = \infty$, and $\dd y / \dd u > 0$ for
$u \in [-1,0]$.
   We abbreviate $L(u) = \ln\bigl( 1 / |u| \bigr)$.}
   \vspace{0.5cm}
   \begin{tabular}{@{}c|cccc@{}}
      \toprule
      no. &
         $f(y)$ &
         $f\bigl( y(u) \bigr)$ &
         $\int^\infty \dd y \, f(y)$ requires
         \\ \midrule
      1  &
         $y^{-\beta}$ &
         $|u|^{\rev{\ms \beta / \alpha}}$ &
         $\alpha < \beta - 1$
         \\[0.25em]
      2  &
         $\exp(-M^2 y^2)$ &
         $|u|^{\ms (2 M / m)^2} \, \exp\Bigl[ \frac{M^2}{m} \ms y(u) \Bigr]$ &
         $m < 2 M$
         \\[0.35em]
      3  &
         $\exp(-M^2 y^2)$ &
         $|u|^{\ms L(u) \; (4 M / m)^2}$ &
         no condition
         \\[0.25em]
      3  &
         $\exp(-M y)$ &
         $|u|^{\ms 4 M /m}$ &
         $m < 4 M$
         \\[0.25em]
      4  &
         $\exp(-M y)$ &
         $|u|^{\ms L(u) \; 2 M /m} \; |u|^{\ms 4 M /m}$ &
         no condition
         \\ \bottomrule
   \end{tabular}
   \caption{\label{tab:y-grid-info-2}Behavior of the variable transformations in
\tab{y-grid-info-1} for selected functions $f(y)$.  In the third column we used
$\exp\bigl[ -c \ms L(u) \bigr] = |u|^{c}$.  The condition in the last column
ensures that $\dd y/ \dd u \, f\bigl( y(u) \bigr) \to 0$ for $u\to0$.  \rev{It
is understood that the parameters $M$ and $m$ are always positive.}}
\end{table}

The last three transformations in the table are designed for functions with a
decrease like $\exp(- M^2 y^2)$ or $\exp(- M y)$ at large $y$, possibly modified
by a power law in $y$.  They map the physical interval $y \in [0, \infty]$ onto
$u \in [-1, 0]$ and depend on a mass parameter $m$, which is normalized such that
$\dd y / \dd u = 4/m$ at $y=0$.
The integral
\begin{align}
   \int_{y_{\min}}^{\infty} \!\! \dd y \; f(y)
   &= \int_{u_{\min}}^{0} \!\! \dd u \; \frac{\dd y}{\dd u} \, f\bigl(y(u)\bigr)
\end{align}
is computed with the Clenshaw-Curtis quadrature rule, which includes the
interval boundary $u=0$ where the Jacobian $\dd y / \dd u$ is infinite.  The
upper limit on $m$ specified in the last column of \tab{y-grid-info-2} ensures
that the product $\dd y / \dd u \, f\bigl(y(u)\bigr)$ is zero at the point $u =
0$, so that this point does not contribute to the Clenshaw-Curtis rule. We note
that the value of the mass scale $M$ in the integrand depends on whether one
integrates a product of two DPDs or a single DPD.
\begin{description}
\item[Inverse power law:] Transformation no.~1 in the table is $u(y) = -
y^{-\alpha}$.  We will show that with $\alpha \sim 0.2$ this is well suited for
the range of power laws $y^{-\beta}$ expected for DPDs at small distances.  The
more general form $u(y) = - (y + a)^{-\alpha}$ can be used down to $y=0$ if
$a>0$, but we will not need this here.
\item[Gaussian:] Transformation no.~2 in the table has a Gaussian falloff at
large $y$ and is designed for functions with a Gaussian behavior, $f(y) \sim
\exp(- M^2 y^2)$.  The transformed function $f\bigl( y(u) \bigr)$ approximately
behaves like a power of $|u|$ for $u\to 0$, and we find that good interpolation
and integration accuracy is obtained for $m \sim M$.
\item[Exponential:] Transformation no.~3 in the table is an exponential in $y$.
It is suitable for both $f(y) \sim \exp(- M y)$ and $f(y) \sim \exp(- M^2 y^2)$.
 In the first case, the transformed function $f\bigl( y(u) \bigr)$ behaves like
a power of $|u|$ for $u\to 0$, whereas in the second case it falls off faster
than any power of $|u|$.  In both cases, values $m \sim M$ yield good
interpolation and integration accuracy.
\item[Exponential with square root:] Transformation no.~4 in the table is
designed for functions $f(y) \sim \exp(- M y)$ with a Yukawa-type decrease.  The
transformed function $f\bigl( y(u) \bigr)$ decreases faster than any power of
$|u|$ for $u\to 0$, and good accuracy is obtained with $m \sim M$.
\end{description}

When handling DPDs with different flavor combinations and in different kinematic
regions, the mass scale $M$ of $f(y)$ in the preceding discussion does not have a
unique value. $M$ should instead be understood as an average, or in the case of
the bounds in the last column of \tab{y-grid-info-2} as the lower limit of the
relevant values.

%===============================================================================
\subsection{Composite grids}
\label{subsec:composite-y-grids}
%===============================================================================

Given the different behavior of DPDs at small and large $y$, it is natural to
use interpolation grids with different transformations $u(y)$.  At small $y$,
there is an additional physics reason for using several subgrids.  Not only the
scale $\mu_y \sim 1/y$ but also the number $\nf{}$ of active flavors for which
the splitting DPD is naturally initialized depends on $y$. Since the value
of a DPD in general changes with $\nf{}$, one should use a unique value of
$\nf{}$ in a subgrid, so as to avoid interpolating discontinuous functions.
As discussed at length in \refcite{Diehl:2022dia}, it is appropriate to evaluate
the perturbative splitting formula \eqref{eq:F-spl} with $\nf{}$ active flavors
for $\gamma \ms m_{n} < \mu_y < \gamma \ms m_{n+1}$ with $\gamma \sim 1$.  We
therefore place subgrid boundaries in $y$ at values around the inverse heavy
quark masses.

In the following studies, we will assume a $y$ dependence
\begin{align}
   \label{eq:y-shapes}
      F(y; \kappa)
   &= y^{-2 \kappa} \, \exp(-M^2 y^2)\,,
\end{align}%
for a single DPD, with $\kappa = 0$ for the intrinsic and $\kappa = 1$ for the
splitting part.  We take $M = 0.21 \GeV$, which roughly corresponds to the
Gaussian parameter in quark-gluon DPDs of the model used in
\refscite{Diehl:2017kgu, Diehl:2020xyg}.  \rev{As for the function $f(y)$ in
\tab{y-grid-info-2}, the overall normalization of $F(y; \kappa)$ is irrelevant
for the interpolation accuracy.}

The smallest value of $y$
needed in a given physics setting corresponds to the inverse of the largest hard
scale that will be considered. In the following, we will take $y_{\min} = 1 / (7
\TeV)$, which covers the production of very heavy states at the LHC.
For the interpolation of \eq{y-shapes} and for the computation of the associated
double parton luminosities, we use the following combination of grids and
transformations:
\begin{align}
   \label{eq:small-y-grid}
   &  \bigl[ \ms
         1 / (7 \TeV), 1/m_b, 1/m_c
      \, \bigr]_{p_1,\, p_2}
   && \text{
         inverse power law transformation with
         $\alpha = 0.2$,
      }
   \\
   \label{eq:large-y-grid}
   &  \bigl[ \ms
         1/m_c, \infty
      \, \bigr]_{p_3}
   && \text{
         Gaussian transformation with $m = 2 M$,
      }
\end{align}%
where the ``inverse power law'' and the ``Gaussian'' transformation are
respectively given in entries no.~1 and 2 of \tab{y-grid-info-1}.
We find that taking $m = 2 M$ results in a somewhat better overall accuracy than
the rule-of-thumb value $m \sim M$ given in the previous subsection.  We will
consider three different settings for the number of points in each subgrid:
\begin{align}
   \label{eq:y-grids}
   (p_1, p_2, p_3)
   &=
      \begin{cases}
      (12,  6, 12) & \text{(coarse $y$ grid)} \\
      (16,  8, 16) & \text{(medium $y$ grid)} \\
      (24, 12, 24) & \text{(fine $y$ grid)}
   \end{cases}
\end{align}
Notice the relatively small number of points in the central grid, which covers
the narrow $y$ region in which one would initialize the splitting part of the
DPD with $4$ active flavors.  For the quark masses, we use the values specified
at the beginning of \sec{dpd-evolution}, which results in subgrid boundaries at
$1/m_b \approx 0.22 \GeV^{-1}$ and $1/m_c \approx 0.71 \GeV^{-1}$.

%===============================================================================
\subsection{Accuracy of interpolation in \texorpdfstring{$y$}{y}}
\label{subsec:accuracy-y-interpolation}
%===============================================================================

\begin{figure}
   \subfigure[
      \label{fig:interpol-y-power-0}
      $\exp(-M^2 y^2)$
   ]{%
      \includegraphics[width=\WidthTwoSubfigs]{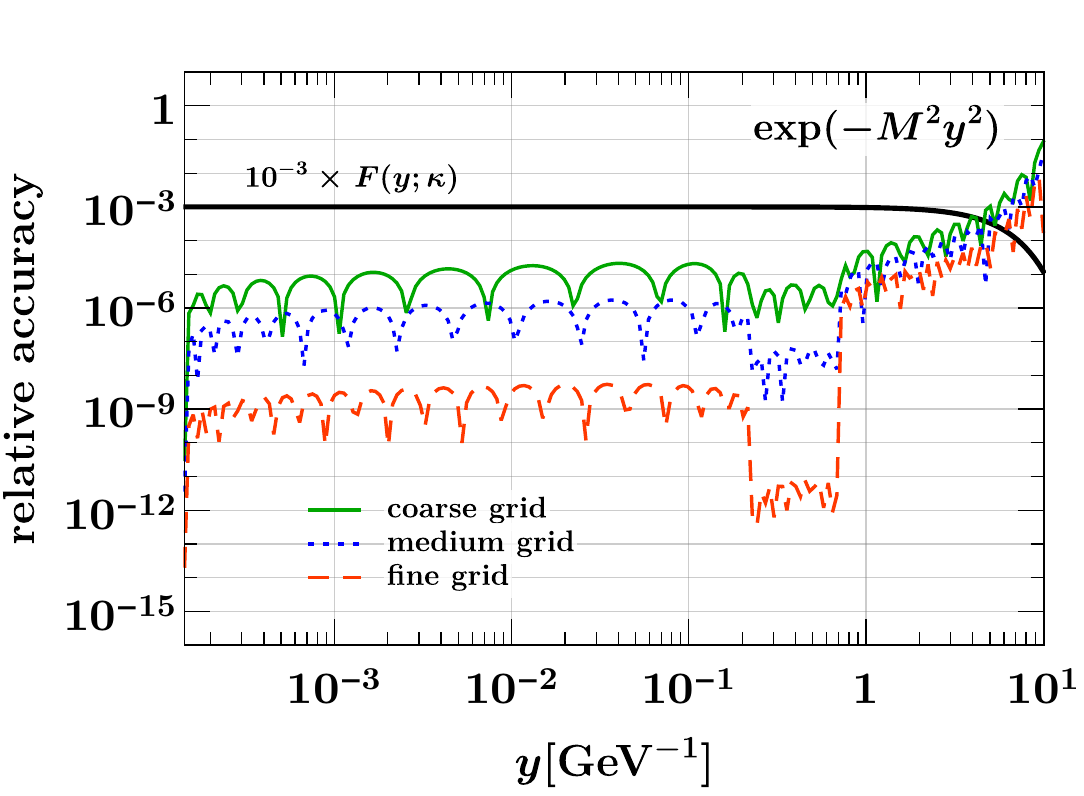}%
   }%
   \hfill
   \subfigure[
      \label{fig:interpol-y-power-1}
      $y^{-2} \exp(-M^2 y^2)$
   ]{%
      \includegraphics[width=\WidthTwoSubfigs]{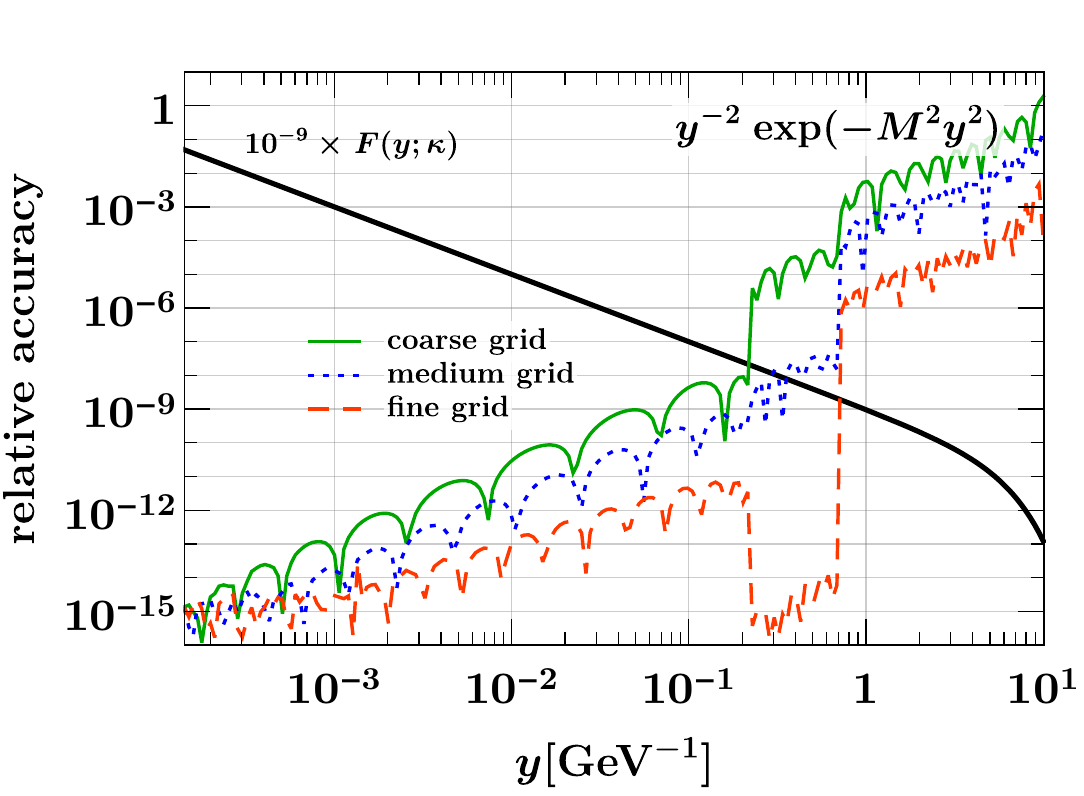}%
   }%
   \\
   \subfigure[
      \label{fig:interpol-y-lin-100}
      $F(y; y_0)$ with $y_0 = 10^{-2}$
   ]{%
      \includegraphics[width=\WidthTwoSubfigs]{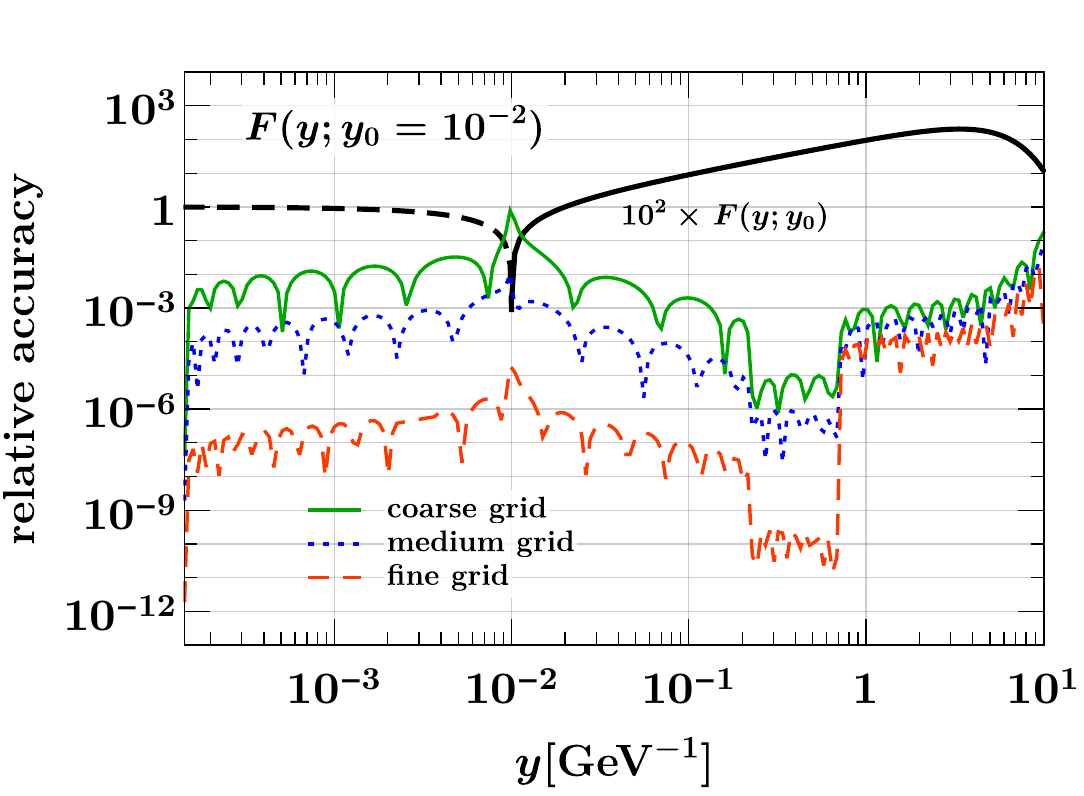}%
   }%
   \hfill
   \subfigure[
      \label{fig:interpol-y-lin-1000}
      $F(y; y_0)$ with $y_0 = 10^{-3}$
   ]{%
      \includegraphics[width=\WidthTwoSubfigs]{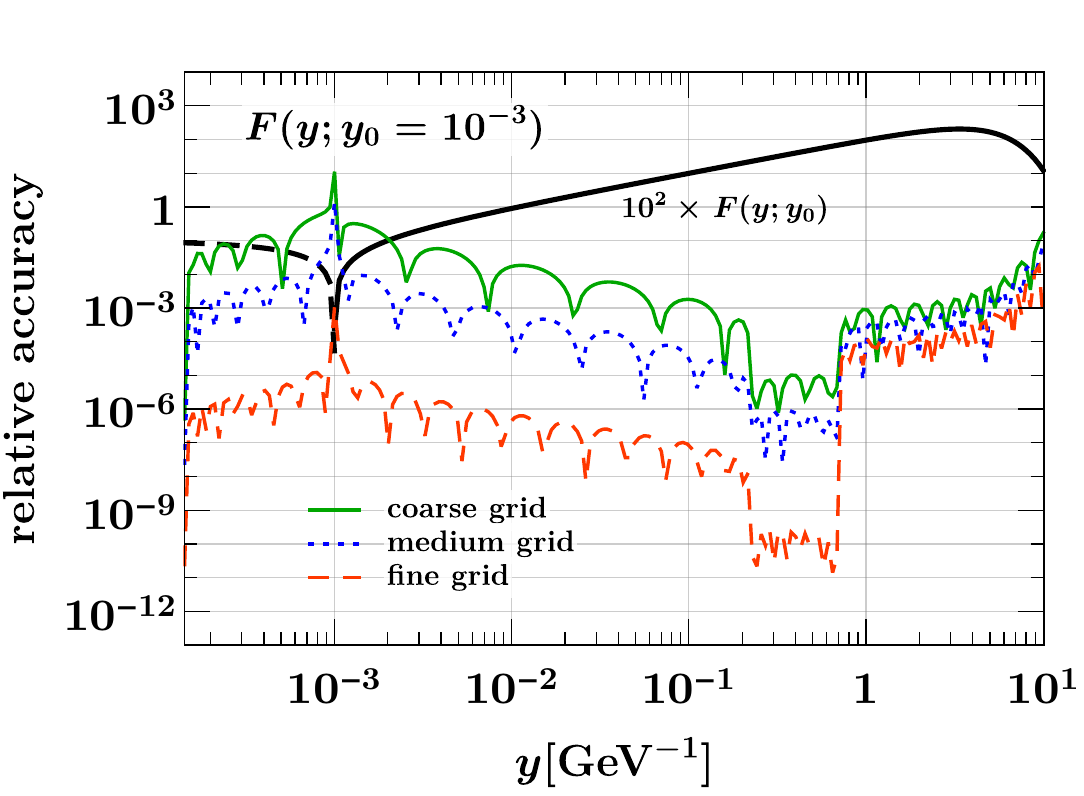}%
   }%
   \caption{\label{fig:interpol-y}Relative interpolation accuracy
\eqref{eq:rel-accuracy} on the grids specified in
\eqss{small-y-grid}{large-y-grid}{y-grids}.  The interpolated
functions are $F(y; \kappa)$ in the upper panels and $F(y; y_0)$ in the lower
ones, see \eqs{y-shapes}{y-shape-lin}.  A dashed black line indicates that the
interpolated function is negative.  Here and in all following plots, the mass
parameter in the functions is $M = 0.21 \GeV$.}
\end{figure}

In the top panels of \fig{interpol-y}, we show the interpolation accuracy for the
form \eqref{eq:y-shapes} with $\kappa = 0$ and $\kappa = 1$.  In the first and
second subgrid, a very accurate interpolation is obtained even with the coarse
grid setting.  In the third subgrid, the accuracy degrades rather quickly for
large distances, but for the medium grid it remains below 1\% for $y$ up to about
$4 \GeV^{-1} \approx 0.8 \operatorname{fm}$.  The fine grid yields an acceptable
accuracy at even larger distances.

Before discussing this issue in more detail, we consider the interpolation of a
DPD with a zero crossing in $y$, which arises for certain parton combinations as
explained above.  As an example for this case, we  take the simple form
\begin{align}
   \label{eq:y-shape-lin}
      F(y; y_0)
   &= (y - y_0) \exp(-M^2 y^2)
      \,,
\end{align}%
with different values of $y_0$ and the same mass parameter $M = 0.21 \GeV$ as
before.  The resulting accuracy is shown in the bottom panels of
\fig{interpol-y}.  We find that the coarse grid performs poorly, whereas with
the medium grid the relative accuracy stays below 1\% except in the immediate
vicinity of the zero crossing.  Very high accuracy is obtained with the fine
grid.

We note that for the intermediate $y$ grid from $1/m_b$ to $1/m_c$, the setting
with $p_2 = 8$ grid points is sufficient to obtain a relative interpolation
accuracy of $10^{-6}$ or better in all cases considered here.  In each of the
grids for smaller and at larger $y$, one may take either $16$ or $24$ points,
depending on how stringent the accuracy requirements are.  This gives
$40$, $48$, or $56$ for the total number of points.\footnote{%
Here we count the common end points of neighboring subgrids twice, since
different subgrids may contain DPDs
with different flavor numbers.}

\begin{figure}
   \subfigure[
      \label{fig:interpol-y-power-0-large-y}
      $\exp(-M^2 y^2)$
   ]{%
      \includegraphics[width=\WidthTwoSubfigs]{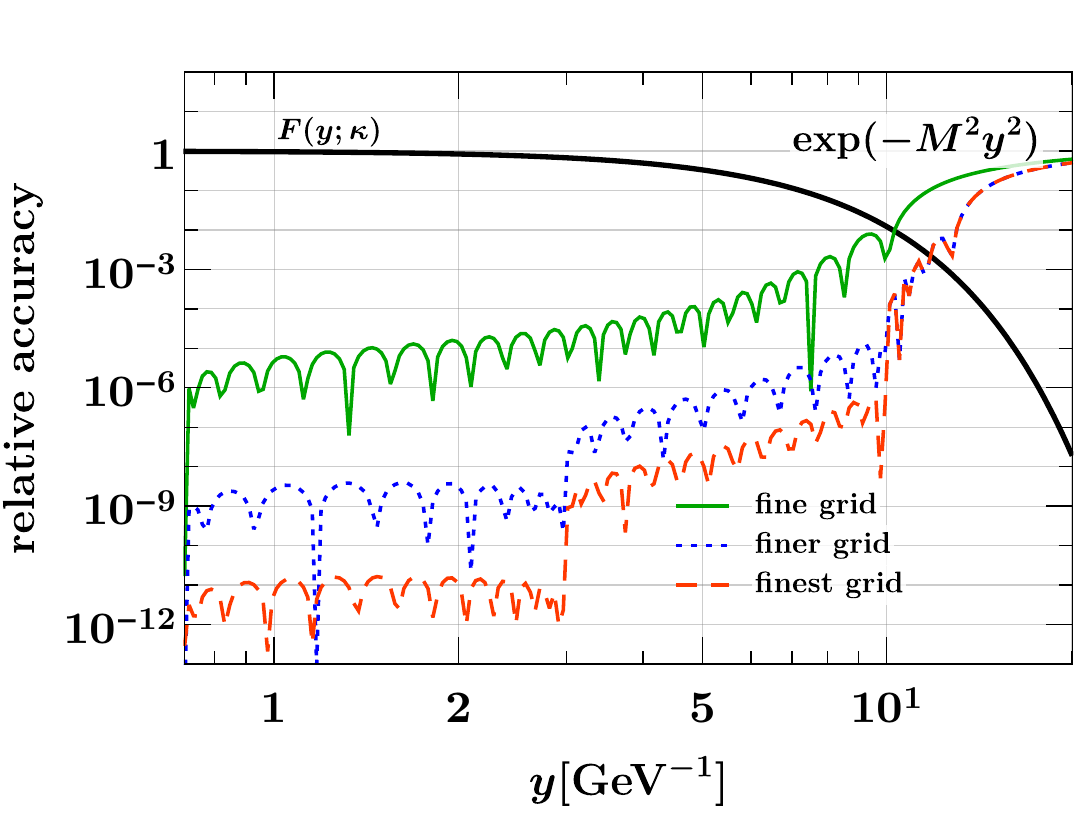}%
   }%
   \hfill
   \subfigure[
      \label{fig:interpol-y-power-1-large-y}
      $y^{-2} \exp(-M^2 y^2)$
   ]{%
      \includegraphics[width=\WidthTwoSubfigs]{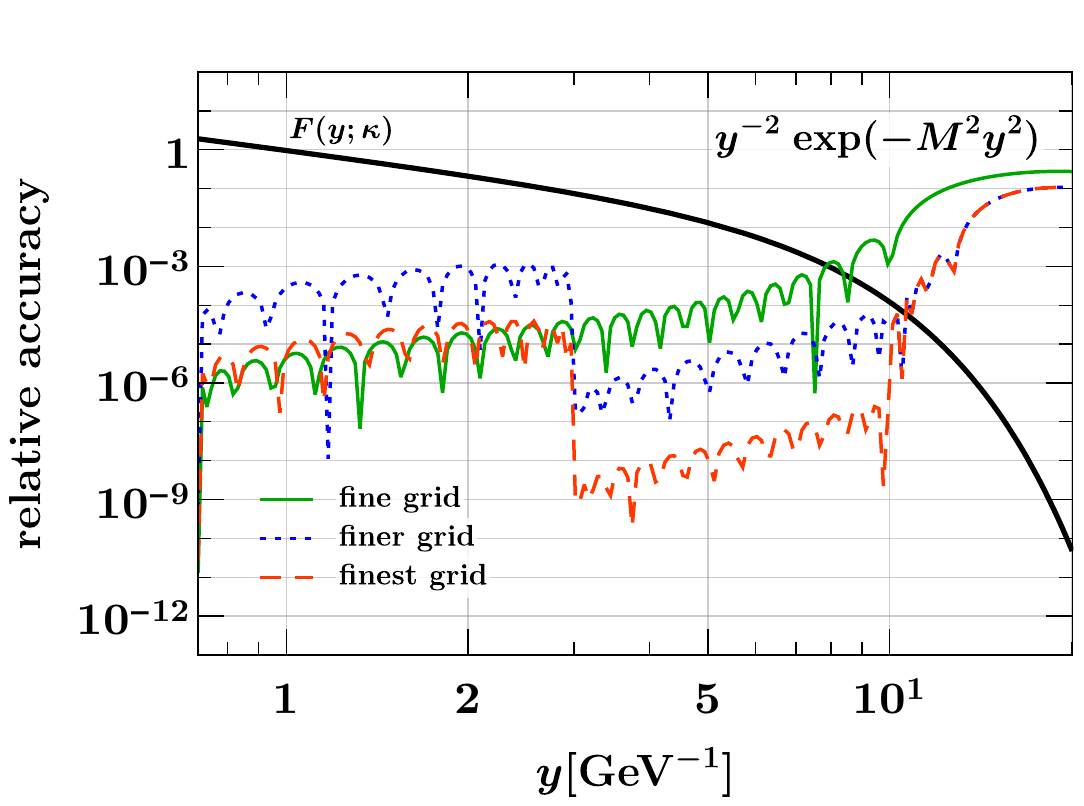}%
   }%
   \caption{\label{fig:interpol-y-large-y}Relative interpolation accuracy at
large $y$ for the grid in \eq{large-y-grid} with $p_3 = 24$ points (fine grid)
and for the composite grids specified by
\eqss{large-y-subgrid-low}{large-y-subgrid-hi}{y-grids-large-y}
(finer and finest grid).  The interpolated functions are the same as in
\figs{interpol-y-power-0}{interpol-y-power-1}, respectively.}
\end{figure}%

With the grids considered so far, the interpolation accuracy at very large $y$
becomes poor mainly because there are too few grid points in that region.  For
many purposes, the behavior of the distributions in the extreme tail region is
not of particular interest, and it contributes only little to integrals over
$y$.  Nevertheless, we wish to show that with an appropriate choice of subgrids
and grid transformations, one can also accurately describe this region.
To this end, we replace the large-$y$ grid in \eq{large-y-grid} with
\begin{align}
   \label{eq:large-y-subgrid-low}
   &  \bigl[
         1 / m_c, 3 \GeV^{-1}, 10 \GeV^{-1}
      \bigr]_{p_{a},\, p_{b}}
   && \text{
         Gaussian transformation with $m = \sqrt{2} M$,
      }
   \\
   \label{eq:large-y-subgrid-hi}
   &  \bigl[
         10 \GeV^{-1}, \infty
      \bigr]_{p_{c}}
   && \text{
         Gaussian transformation with $m = 2 M$,
      }
\end{align}%
where we consider two settings for the number of points:
\begin{align}
   \label{eq:y-grids-large-y}
   (p_{a}, p_{b}, p_{c})
   &=
   \begin{cases}
      (12, 12, 8) & \text{(finer $y$ grid)} \\
      (16, 16, 8) & \text{(finest $y$ grid})
   \end{cases}
\end{align}
In \fig{interpol-y-large-y}, the resulting interpolation accuracy is compared
with the previous setting of \eq{large-y-grid} with $p_3 = 24$ points.  We see
that both the finer and the finest grid extend the domain of very accurate
interpolation up to $y \approx 10 \GeV^{-1} \approx 2 \operatorname{fm}$, with
only a moderate increase in the number of grid points.

\begin{figure}
   \subfigure[
      \label{fig:error-estimates-y-power-1}$y^{-2} \exp(-M^2 y^2)$
   ]{%
      \includegraphics[width=\WidthTwoSubfigs]{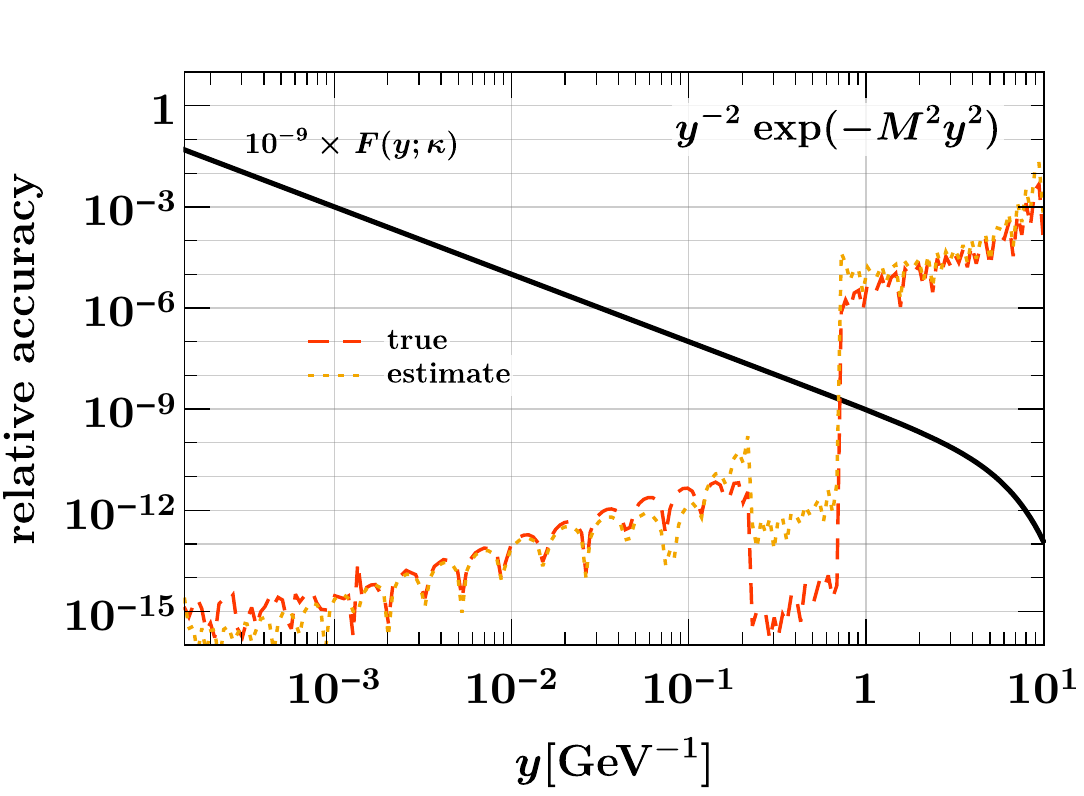}%
   }%
   \hfill
   \subfigure[
      \label{fig:error-estimates-y-lin-1000-error}
      $F(y; y_0)$ with $y_0 = 10^{-3}$
   ]{%
      \includegraphics[width=\WidthTwoSubfigs]{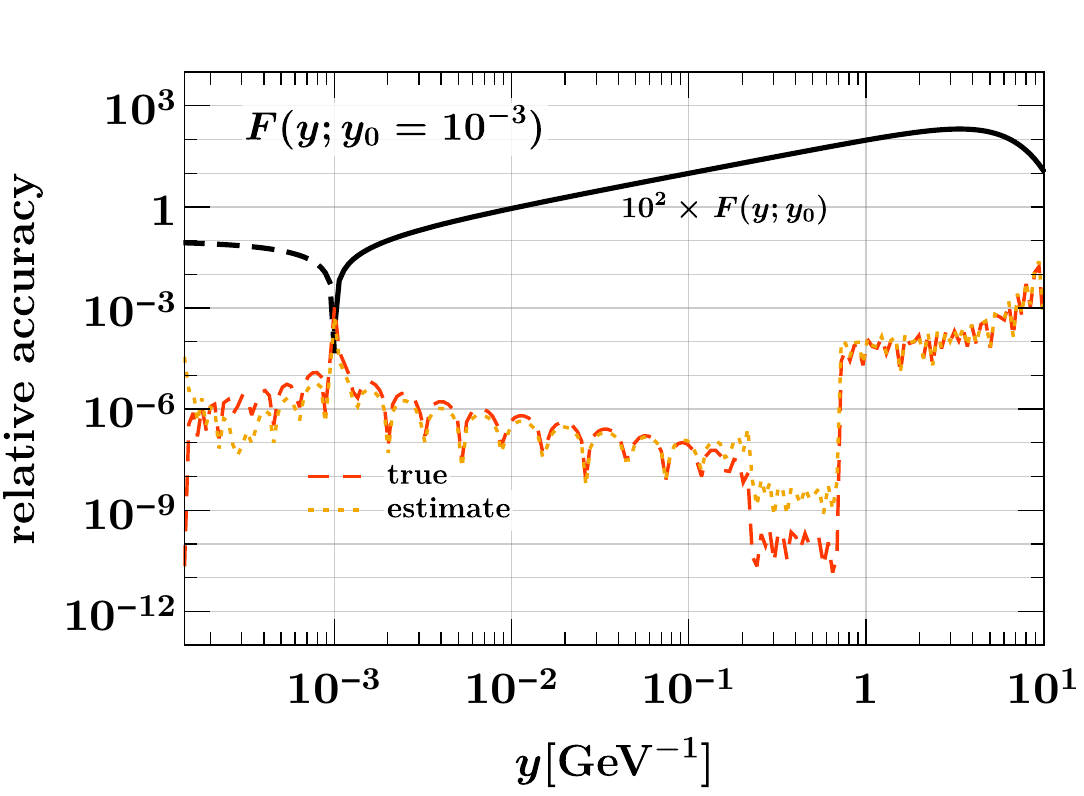}%
   }%
   \caption{\label{fig:interpol-y-error-estimates}True and estimated relative
interpolation accuracy in $y$, as defined in
\eqs{rel-accuracy}{accuracy-estimate}, respectively.  The curves are for the
fine grid setting in \eq{y-grids}.}
\end{figure}%

In analogy to \subsec{interpolation-x1-x2-error-estimate}, we can estimate the
interpolation accuracy in $y$ by comparing the results of interpolation with and
without subinterval end points.  We find that this method works very well, as
shown by the examples in \fig{interpol-y-error-estimates}.  The biggest
discrepancies between true and estimated errors appear near subgrid boundaries,
and in the central grid, where the omission of the end points decreases the
polynomial order of interpolation from 11 to 9.

%===============================================================================
\subsection{Accuracy of integration over \texorpdfstring{$y$}{y}}
\label{subsec:accuracy-y-integration}
%===============================================================================

Integration over $y$ from a lower cutoff $y_{\min}$ to infinity is needed to
evaluate the double parton luminosities \eqref{eq:DPD-lumi-def} and DPD sum
rules.  If $y_{\min}$ coincides with the end point of a subinterval of the
interpolation grid, one can directly use Clenshaw-Curtis quadrature to compute
the integral.  Otherwise, we perform a number of intermediate steps:
\begin{enumerate}
\item identify the subinterval $[y_a, y_b]$ that contains $y_{\min}$,
\item define a reduced subgrid on the interval $[y_{\min}, y_b]$ with the same
number of points and the same variable transformation as the original subgrid,
\item compute the required DPDs on the reduced subgrid by interpolation on
$[y_a, y_b]$.
\end{enumerate}
The integral is then computed using Clenshaw-Curtis quadrature on the reduced
subgrid and on the subgrids for $y \ge y_b$.

Let us investigate the accuracy of this procedure for the double parton
luminosities
\begin{align}
   \label{eq:DPD-lumis}
      \mathcal{L}(\ymin, \kappa_1 + \kappa_2)
   &= 2 \pi \!
      \int\limits_{\ymin}^{\infty} \! \dd y \; y \,
      F(y; \kappa_1) \, F(y; \kappa_2)
    = 2 \pi \!
      \int\limits_{\ymin}^{\infty} \!
      \frac{\dd y\; y}{y^{\ms 2 (\kappa_1 + \kappa_2)}} \,
      \exp(-2 M^2 y^2)
\end{align}%
that correspond to the simplified shapes of DPDs in \eq{y-shapes}.  We consider
the cases $\kappa_1 + \kappa_2 = 0, 1, 2$, which respectively correspond to
taking the intrinsic parts of both DPDs, the intrinsic part of one and the
splitting part of the other DPD, or the splitting parts of both DPDs.  Although
only the sum of intrinsic and splitting parts enters in physical cross sections,
it is useful to investigate these different combinations separately, as was for
instance done in the studies in \refscite{Diehl:2017kgu, Diehl:2022dia}.

\begin{figure}
   \subfigure[
      \label{fig:integration-y-power-0}
      $\exp(-2 M^2 y^2)$
   ]{%
      \includegraphics[height=\HeightTwoSubfigs]{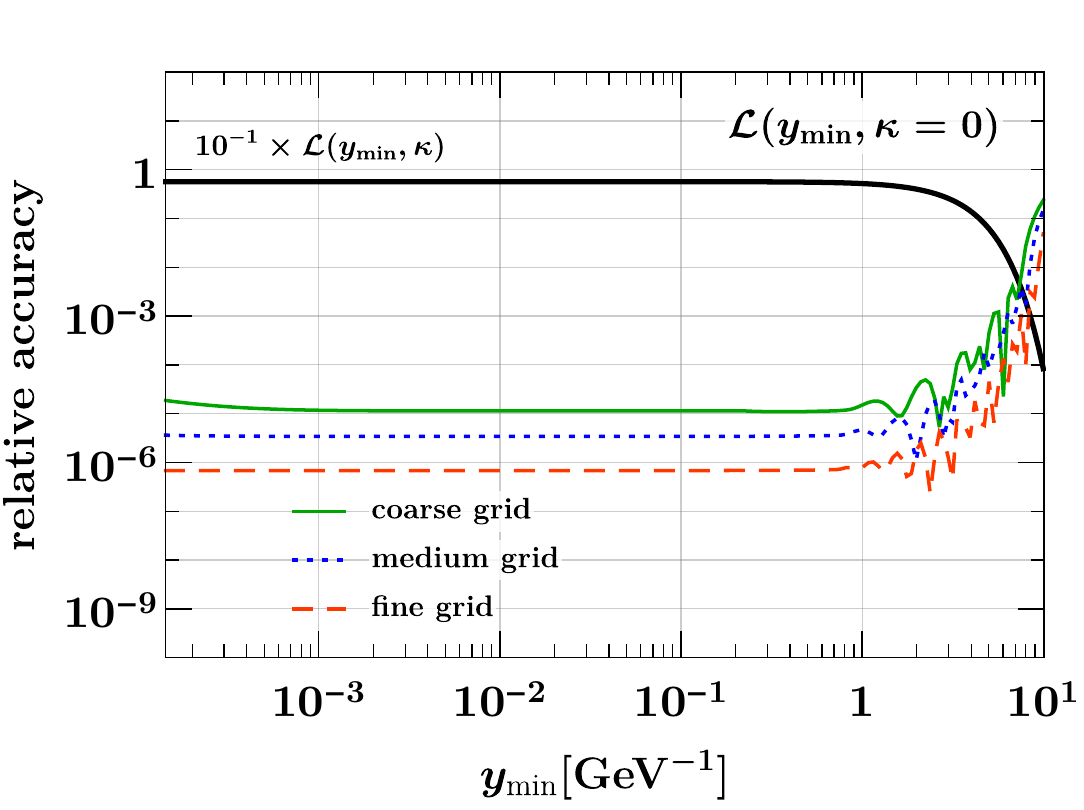}%
   }%
   \hfill
   \subfigure[
      \label{fig:integration-y-power-0-errors}
      $\exp(-2 M^2 y^2)$
   ]{%
      \includegraphics[height=\HeightTwoSubfigs]{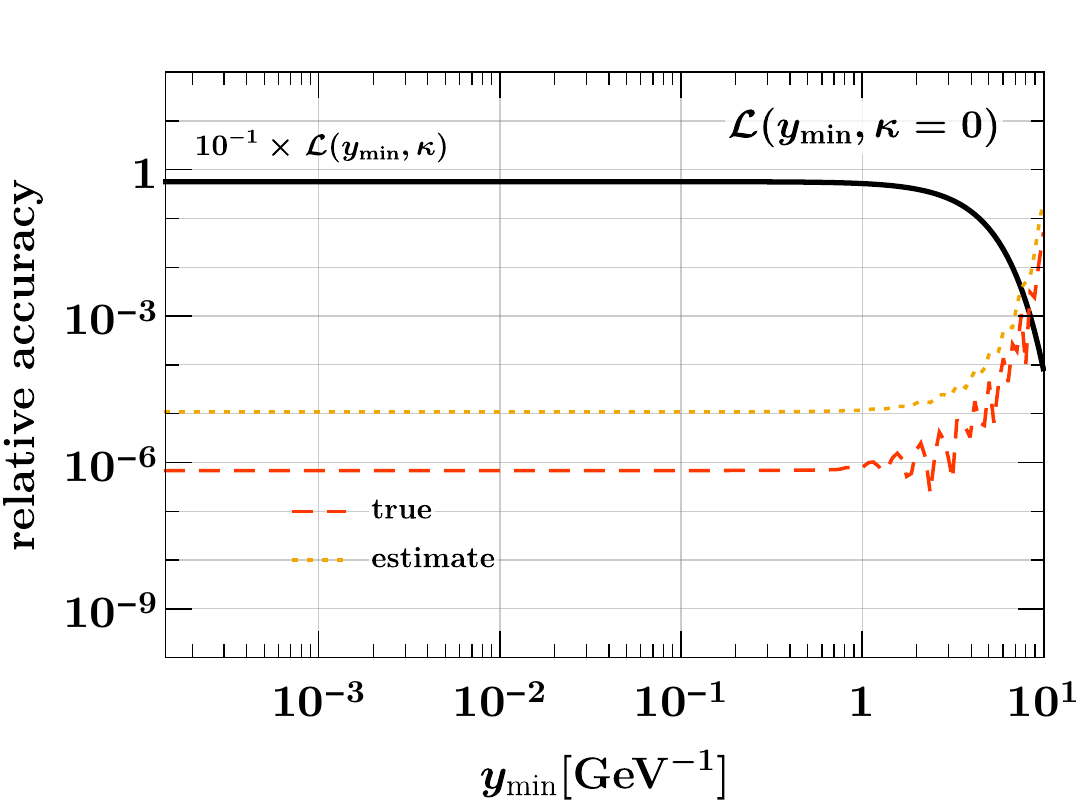}%
   }%
   \\
   \subfigure[
      \label{fig:integration-y-power-1}
      $y^{-2} \exp(-2 M^2 y^2)$
   ]{%
      \includegraphics[height=\HeightTwoSubfigs]{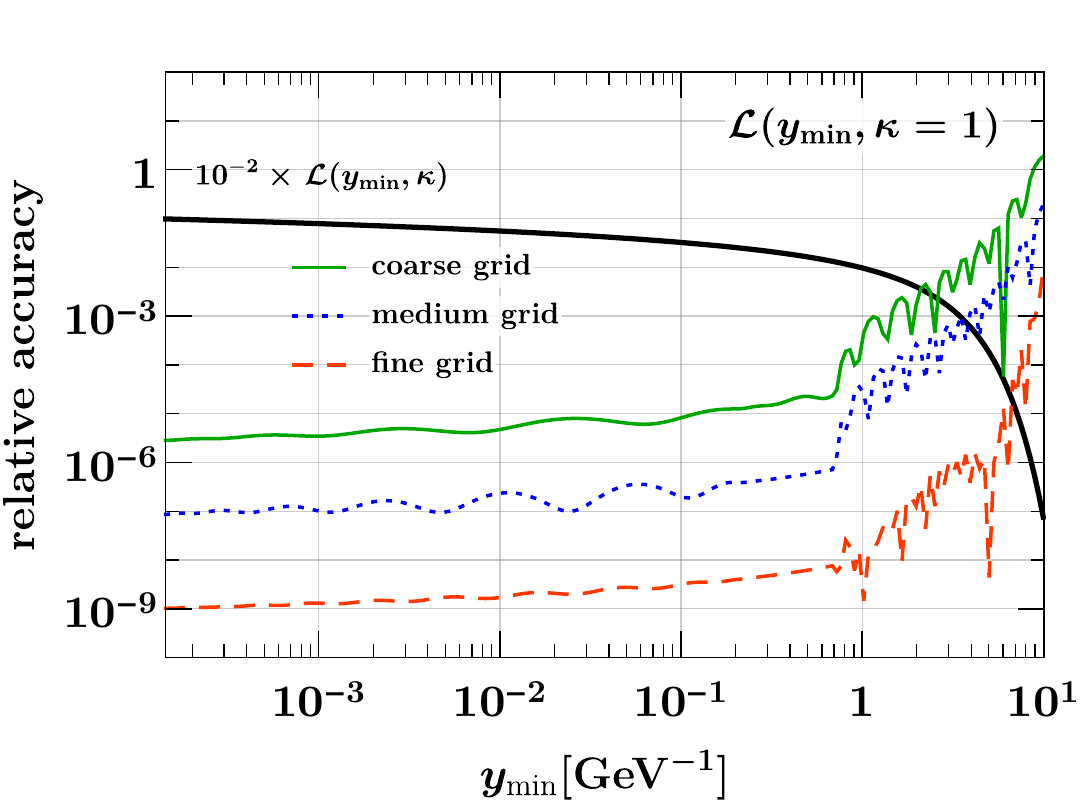}%
   }%
   \hfill
   \subfigure[
      \label{fig:integration-y-power-1-errors}
      $y^{-2} \exp(-2 M^2 y^2)$
   ]{%
      \includegraphics[height=\HeightTwoSubfigs]{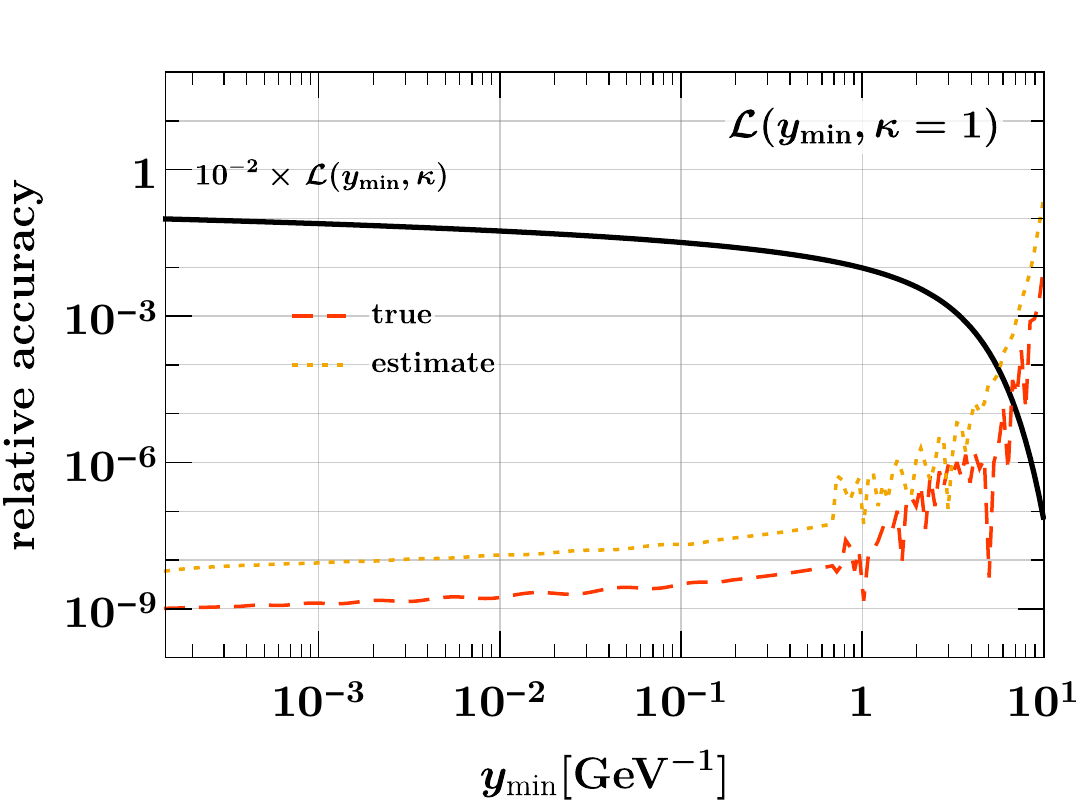}%
   }%
   \\
   \subfigure[
      \label{fig:integration-y-power-2}
      $y^{-4} \exp(-2 M^2 y^2)$
   ]{%
      \includegraphics[height=\HeightTwoSubfigs]{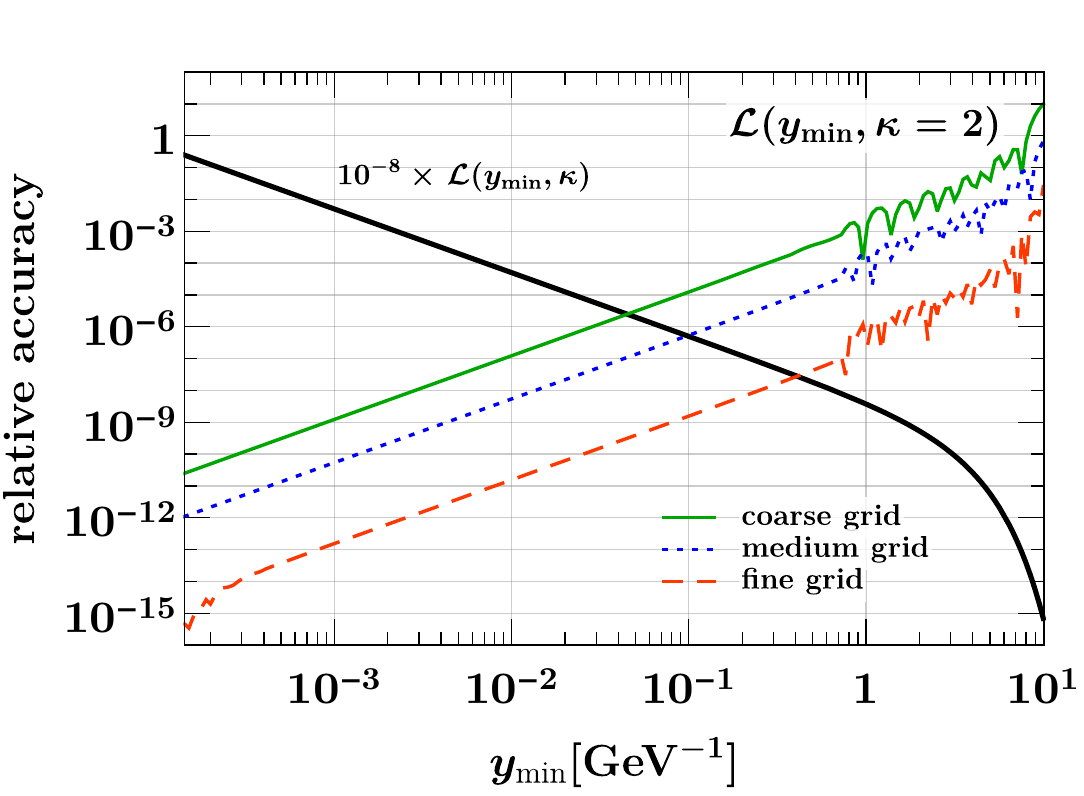}%
   }%
   \hfill
   \subfigure[
      \label{fig:integration-y-power-2-errors}
      $y^{-4} \exp(-2 M^2 y^2)$
   ]{%
      \includegraphics[height=\HeightTwoSubfigs]{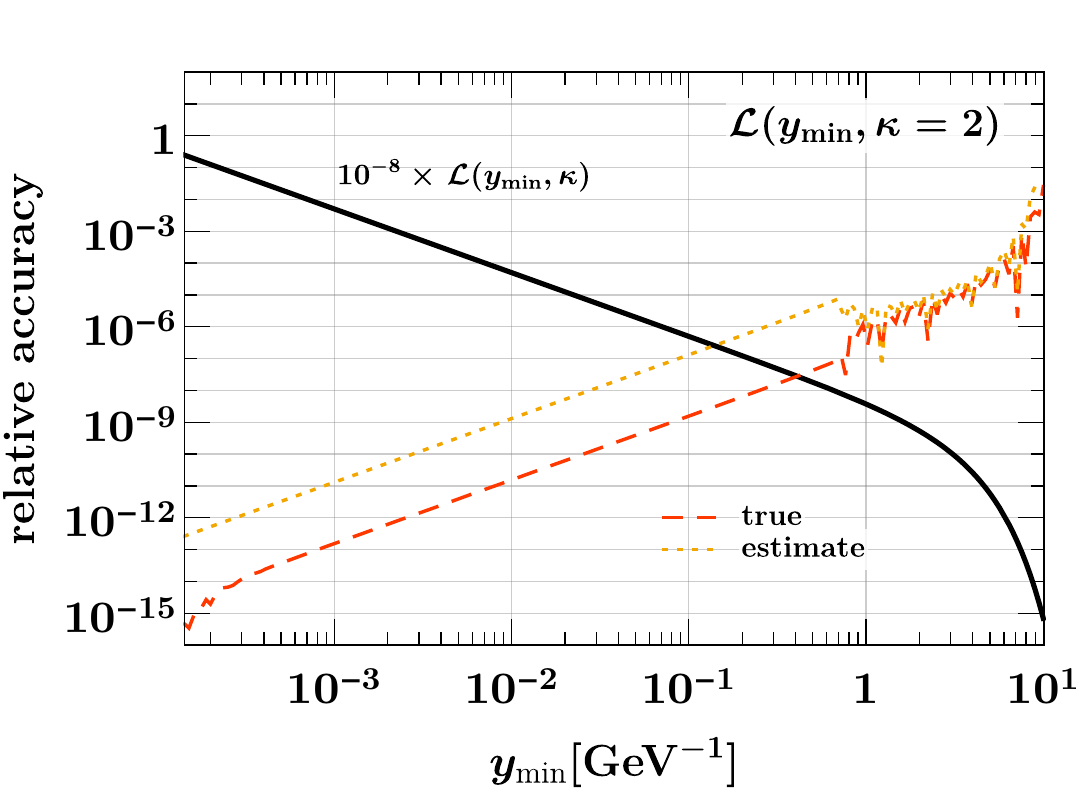}%
   }%
   \caption{\label{fig:integration-y}Left: Relative accuracy of computing
the integrals \eqref{eq:DPD-lumis} with the interpolation grids specified by
\eqss{small-y-grid}{large-y-grid}{y-grids}.  Right: True and
estimated integration accuracy for the
fine grid setting of \eq{y-grids}.  The functions in the legends are the product
$F(y; \kappa_1)\, F(y; \kappa_2)$ of DPDs in the luminosity integral.}
\end{figure}%

The integration accuracy can be computed using the exact expression of
\eqref{eq:DPD-lumis} in terms of an incomplete gamma function.  The result is
shown in the left panels of \fig{integration-y} for the interpolation grids
introduced in \subsec{composite-y-grids}.  We see that for $y_{\min}$ below $1
\GeV^{-1}$, good accuracy is achieved even with the coarse grid.  For cutoff
values above $1 \GeV^{-1}$, the accuracy quickly becomes unsatisfactory unless
the fine grid is used. This reflects the degrading interpolation accuracy in
that region, which was shown in the upper panels of \fig{interpol-y}.  In
physics applications, $y_{\min}$ is of the order of an inverse hard scale, so
that large values of the cutoff will only be needed in special circumstances,
for instance if one is interested in the partial contribution of large $y$ to an
integral.  We see that the fine grid setting is useful in such a case.

In the right panels of \fig{integration-y}, we compare the true integration
accuracy with its estimate obtained from comparing the values obtained with
Clenshaw-Curtis quadrature and with Fej{\'e}r's second rule (see
\subsec{chebyshev}). We see that for $y_{\min}$ below $1 \GeV^{-1}$, one obtains
an overestimate of the true error by one or two orders of magnitude.  This is
qualitatively similar to what we found for integrals of PDFs over~$x$ in
section~3.4 of \refcite{Diehl:2021gvs}.

%%%%%%%%%%%%%%%%%%%%%%%%%%%%%%%%%%%%%%%%%%%%%%%%%%%%%%%%%%%%%%%%%%%%%%%%%%%%%%%%
\section{Cross check: comparison with \dove}
\label{sec:dove}
%%%%%%%%%%%%%%%%%%%%%%%%%%%%%%%%%%%%%%%%%%%%%%%%%%%%%%%%%%%%%%%%%%%%%%%%%%%%%%%%

As a cross check for our implementation of DPD evolution in \chili, we performed
a comparison with \dove, a code that was introduced in \refcite{Gaunt:2009re}
and further developed or adapted in subsequent work in \refscite{Diehl:2014vaa,
Diehl:2017kgu, Cabouat:2019gtm, Cabouat:2020ssr, Diehl:2020xyg}.\footnote{%
Results obtained with an early version of \dove\ are given on the website
\url{https://gsdpdf.hepforge.org}.}
The comparison is done with the version that was used in \refcite{Diehl:2020xyg}
to compute the Gaunt-Stirling sum rules for DPDs.  In the following, we show the
momentum sum rule for a $u$ quark and the number sum rule for the parton
combination $g d_v$, which read
\begin{align}
   \label{eq:dpd-momsum}
   \sum_{a_2}
   \int\limits_0^{1 - x_1} \dd x_2 \; x_2 \,
      F_{u \ms a_2}^{\msbars}(x_1, x_2; \mu)
   &=
   (1 - x_1) \, f_u(x_1; \mu)
   \,,
   \\[-0.3em]
   \label{eq:dpd-numsum}
   \int\limits_0^{1 - x_1} \dd x_2 \;
      F_{g \ms d_v}^{\msbars}(x_1, x_2; \mu)
   &=
   f_{g}(x_1; \mu)
   \,,
\end{align}%
where $F_{g \ms \smash{d_v}}^{} = F_{g \smash{d}}^{} - F_{g \smash{\dbar}}^{}$.
The expressions for general parton combinations are given in equations~(3.1) and
(3.2) of \refcite{Gaunt:2009re}.  As shown in \refcite{Diehl:2018kgr}, the sum
rules hold for the distributions
\begin{align}
   \label{eq:msbar-dpd-def}
   F_{a_1 a_2}^{\msbars}(x_1, x_2; \mu)
   &=
   2 \pi \!\! \int\limits_{b_0 / \nu}^{\infty} \! \dd y \; y \,
      F_{a_1 a_2}(x_1, x_2, y; \mu, \mu)
   + F_{a_1 a_2}^{\text{match}}(x_1, x_2; \nu, \mu)
   \,,
\end{align}
which are integrated over $y$ with an \msbar renormalization prescription for the
$y^{-2}$ singularity due to perturbative splitting.  The matching term $F_{a_1
a_2}^{\text{match}}(x_1, x_2; \nu, \mu)$ relates this prescription with the
cutoff in $y$ used on the r.h.s.\ of \eq{msbar-dpd-def}.  It can be computed in
terms of a PDF and a matching kernel in a similar fashion as the splitting part
in \eq{F-spl}, and its LO expression is given in equation (18) of
\refcite{Diehl:2020xyg}.  The dependence on the matching scale $\nu$ cancels
between the two terms in \eqref{eq:msbar-dpd-def} up to power corrections in
$1/\nu^2$.  We recall that $b_0 = 2 e^{-\gamma_E} \approx 1.12$.

The different roles played by $x_1$ and $x_2$ in the sum rules
\eqref{eq:dpd-momsum} and \eqref{eq:dpd-numsum} motivate us to take different
grids for the two momentum fractions,
\begin{align}
  x_1: & \qquad
  \bigl[ 5 \times 10^{-5}, 5 \times 10^{-3}, 0.5, 0.9, 1.0
  \bigr]_{(12, 12, 16, 8)}
  \,,
  \nn \\
  x_2: & \qquad
  \bigl[ 5 \times 10^{-5}, 5 \times 10^{-3}, 0.5, 1.0
  \bigr]_{(12, 16, 16)}
  \,.
\end{align}
The first grid has an additional small subgrid that leads to a better
interpolation accuracy at large $x_1$, which is not needed for integration over
$x_2$.  The sum rule integrals over $x_2$ are computed using Clenshaw-Curtis
quadrature over the full $x_2$ grid. This implies that the integrals are computed
with a lower cutoff $x_{\min} = 5 \times 10^{-5}$.  The same cutoff is used in
the computation with \dove.

We find that the integration over $y$ in the sum rules can be computed with the
rather small grids
\begin{align}
  &
  \bigl[ b_0 / (170 \GeV), b_0 / (1 \GeV) \bigr]_{(12)}
  && \text{inverse power law transformation with $\alpha = 0.2$}
  \,,
  \\
  &
  \bigl[ b_0 / (1 \GeV), \infty \bigr]_{(8)}
  && \text{Gaussian transformation with $m = M$}
  \,,
\end{align}
where $M = 1 \big/ \sqrt{4 h_{g g}} \approx 0.23 \GeV$ is the Gaussian mass
parameter for gluons in the initial conditions of the model (see \eq{F-int}).

We take the DPD model specified in section~3 of \refcite{Diehl:2020xyg} and
evolve to $\mu_1 = \mu_2 = 144.6 \GeV$ with $\nf{} = 3$ active flavors for each
parton.  At that scale, the momentum and number sum rules are evaluated with $\nu
= 144.6 \GeV$.  These $\mu$ and $\nu$ values correspond to grid points in our
setup of \dove\ and thus avoid additional interpolation.  The detailed grid
settings and integration procedure in \dove\ are as specified in section~4 of
\refcite{Diehl:2020xyg}.

\begin{figure}
   \centering
   \subfigure[
      \label{fig:momsum-u-hi}
      $u$ momentum sum rule
   ]{%
      \includegraphics[height=\HeightTwoSubfigs]{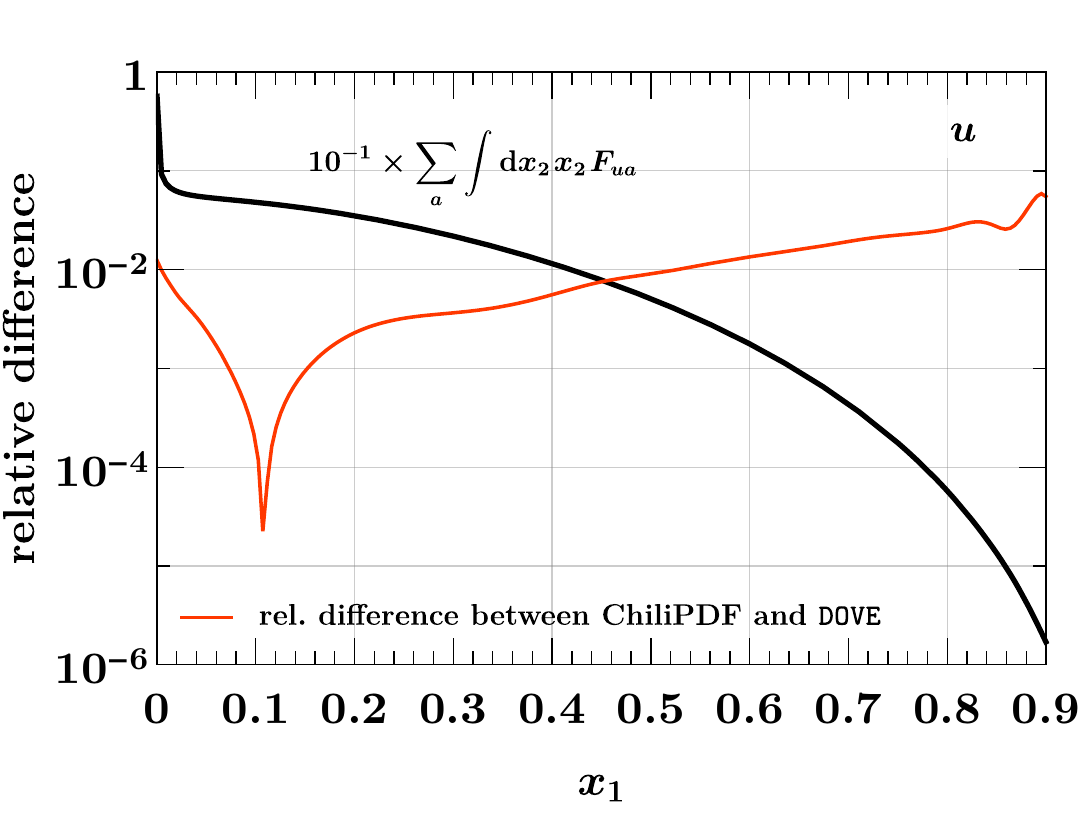}%
   }%
   \hfill
   \subfigure[
      \label{fig:numsum-dminus-g-hi}
      $g d_v$ number sum rule
   ]{%
      \includegraphics[height=\HeightTwoSubfigs]{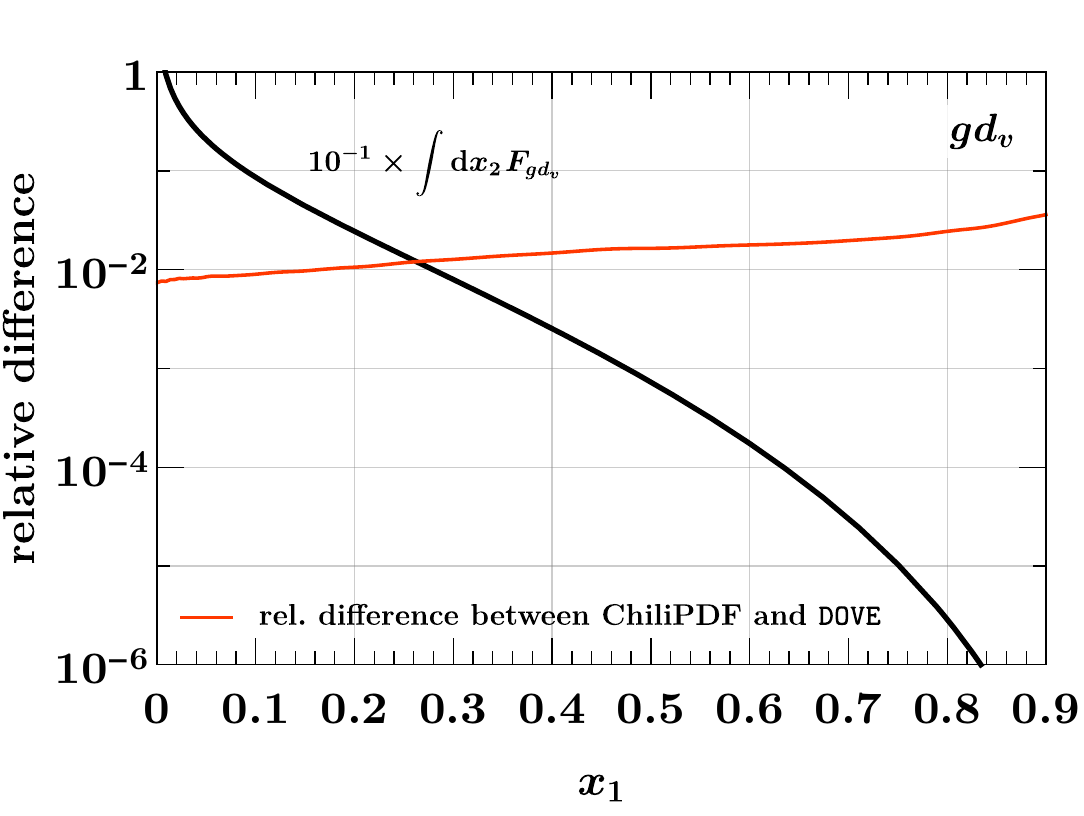}%
   }%
   \caption{\label{fig:dove-comparison}Relative differences between the
left-hand sides of the DPD sum rules \eqref{eq:dpd-momsum} and
\eqref{eq:dpd-numsum}, evaluated using \dove\ and \chili.  Details are given in
the text.}
\end{figure}%

In \fig{dove-comparison}, we show the relative deviation between the left-hand
sides of the sum rules \eqref{eq:dpd-momsum} and \eqref{eq:dpd-numsum} evaluated
in \chili\ and in \dove.  The numerical differences increase with $x_1$ and reach
several percent at $x_1 = 0.9$.  For sum rules with other parton combinations,
we find similar or smaller discrepancies.  We also compared the sum rules
evaluated at smaller values of $\mu_1 = \mu_2$ and $\nu$ and found somewhat
smaller discrepancies than at high scales.

An error estimate for the accuracy of evolution with \dove\ in its original
version is given in figure~9 of \cite{Gaunt:2009re}.  Given that this estimate
was carried out with finer grids than the ones used here, and given that
relatively simple integration rules are used in the evaluation of the sum rules
with \dove, we find the level of agreement between the two codes satisfactory for
the purpose of cross-validation.

%%%%%%%%%%%%%%%%%%%%%%%%%%%%%%%%%%%%%%%%%%%%%%%%%%%%%%%%%%%%%%%%%%%%%%%%%%%%%%%%
\section{Conclusions}
\label{sec:conclusions}
%%%%%%%%%%%%%%%%%%%%%%%%%%%%%%%%%%%%%%%%%%%%%%%%%%%%%%%%%%%%%%%%%%%%%%%%%%%%%%%%

We have shown that discretization on Chebyshev grids allows for a precise and
economical representation of DPDs as functions of the momentum fractions $x_1$,
$x_2$ and the distance $y$ between the partons.  This generalizes the methods we
developed for PDFs in \refcite{Diehl:2021gvs}.  We use grids for $x_1$ that are
independent of $x_2$ and vice versa.  This gives the highest degree of simplicity
and decoupling between operations that involve only one momentum fraction, which
is the case for the Mellin convolutions that appear in DGLAP evolution and in
flavor matching.  This method incurs interpolation errors close to the kinematic
boundary $x_1 + x_2 = 1$ of the DPDs, which we find manageable. We investigated
the interpolation accuracy for typical shapes of DPDs, both for the input
distributions and after evolution.  We find that grids with $p_{x} = 54$ points
for $x \ge 10^{-5}$ yield a very accurate representation of DPDs, with relative
errors that are much smaller than $10^{-3}$ in large parts of the phase space and
remain below $1\%$ for $x_1 + x_2 < 0.8$ (see
\figs{evol-match-accuracy-hi}{evol-match-accuracy-lo}).

For the dependence on $y$, Chebyshev interpolation is used after variable
transformations that are adapted to the functional form and the typical mass
scale for the decrease of the DPDs at large distance.  We find that with $p_{y}
\sim 48$ grid points, very good accuracy can be achieved for interpolation and
integration.

To solve the DGLAP equations for DPDs, we use a high-order Runge-Kutta algorithm
for computing the evolution matrices in
\eqs{evolution-matrix}{dglap-greens-discrete}, which are the discretized version
of the Green functions for the evolution equations.  This provides very high
numerical accuracy for both forward and backward evolution, with errors well
below the ones due to discretization in $x_1$ and $x_2$.  With our algorithm, it
is natural to evolve DPDs separately in the scales $\mu_1$ and $\mu_2$ associated
with the two partons. We note that the computational effort scales linearly with
$p_{y}$ but with the third power of~$p_{x}$.

\rev{In our setup, a DPD with given flavor numbers $(\nf{1}, \nf{2})$ is
specified numerically by its values at all grid points in $(x_1, x_2, y)$ and at
one pair $(\mu_1, \mu_2)$ of scales.  Its values at other scales are computed
``on the fly'' by solving the evolution equations.  DPDs at other values of
$x_1$, $x_2$, and $y$ are computed using the barycentric interpolation formula,
which is fast and numerically stable.  As discussed in the introduction, we avoid
grids in the renormalization scales, which would require a huge amount of
computer memory.  This is a notable difference between our approach and the
widely used method to compute PDFs from grids in $x$ and $\mu$, for instance via
the LHAPDF interface \cite{Buckley:2014ana}, and it reflects the notable
difference in complexity between the distribution functions for one or for two
partons inside a proton.}

With our present implementation of the above methods in \chili, the evolution of
a DPD with $\nf{} = 5$ active flavors (and thus 121 parton flavor combinations)
from $\mu_1 = \mu_2 = 15 \GeV$ to $150 \GeV$ takes 1 to 2 seconds\footnote{%
Timings were obtained on different computers with Intel\textregistered\
Core\texttrademark\ i5, Intel\textregistered\ Core\texttrademark\ i7, or AMD
EPYC\texttrademark\ 7513 32-Core processors.}
for $p_{x} = 54$ grid points in $x_1$ and $x_2$ and $p_{y} = 48$ grid points in
$y$. With grids of this size, the largest part of computing time is spent on
multiplying the evolution matrices with the initial conditions, so that there is
only a weak dependence of the timing on the perturbative order of evolution and
on the initial and final scales. Since the current code is based on a simple
implementation of linear algebra and matrix multiplication, we expect that
significant gains can still be made with a dedicated performance tuning.

We used our code to investigate the impact of higher orders in the DGLAP kernels
on the evolution of DPDs.  For some parton combinations, the change from LO to
NLO evolution produces $\mathcal{O}(1)$ effects at small momentum fractions,
whereas we find only moderate differences when changing from NLO to NNLO kernels.

The numerical delivery of realistic DPDs has been a bottleneck in practice so
far. The presented methods and their implementation provide a practical tool for
computing evolved DPDs from user-given starting conditions, on a par with what
has long been a standard for PDFs.  This increases our ability to use the
predictive power of perturbation theory and thus presents an important ingredient
to making theory predictions for double parton scattering more realistic.

The evolution equations considered in this work apply to DPDs that are summed
separately over the color of each parton.  DPDs with color correlations between
the two partons follow a different evolution pattern, with kernels that have
recently been computed at NLO \cite{Diehl:2022rxb}.  The methods described in
the present paper can be adapted to this case, as will be reported in a future
publication.

%%%%%%%%%%%%%%%%%%%%%%%%%%%%%%%%%%%%%%%%%%%%%%%%%%%%%%%%%%%%%%%%%%%%%%%%%%%%%%%%
\acknowledgments
We thank Florian Fabry, Oskar Grocholski, and Mees van Kampen for their
contributions to \chili.  It is a pleasure to thank Jonathan Gaunt
for valuable discussions and input.

This work is in part supported by the Deutsche Forschungsgemeinschaft (DFG,
German Research Foundation) -- grant number 409651613 (Research Unit FOR 2926)
and grant number 491245950.
The work of RN is supported by the ERC Starting Grant REINVENT-714788.

%%%%%%%%%%%%%%%%%%%%%%%%%%%%%%%%%%%%%%%%%%%%%%%%%%%%%%%%%%%%%%%%%%%%%%%%%%%%%%%%

%%%%%%%%%%%%%%%%%%%%%%%%%%%%%%%%%%%%%%%%%%%%%%%%%%%%%%%%%%%%%%%%%%%%%%%%%%%%%%%%
\appendix
%%%%%%%%%%%%%%%%%%%%%%%%%%%%%%%%%%%%%%%%%%%%%%%%%%%%%%%%%%%%%%%%%%%%%%%%%%%%%%%%

%%%%%%%%%%%%%%%%%%%%%%%%%%%%%%%%%%%%%%%%%%%%%%%%%%%%%%%%%%%%%%%%%%%%%%%%%%%%%%%%
\section{Path independence of evolution and flavor matching}
\label{app:path-indep}
%%%%%%%%%%%%%%%%%%%%%%%%%%%%%%%%%%%%%%%%%%%%%%%%%%%%%%%%%%%%%%%%%%%%%%%%%%%%%%%%
%
In this appendix we show that the set \eqref{eq:double-dglap} of DGLAP equations
has a unique solution for a given initial condition at a point in the $(\mu_1,
\mu_2)$ plane. In other words, the solution is independent of the path in the
$(\mu_1, \mu_2)$ plane that is taken to evolve from the initial condition to the
final scales.  To establish this, it is enough to show that the second
derivatives in the renormalisation scales commute, i.e.\ that
\begin{align}
   \label{eq:DPD-derivative-12}
   \frac{\partial}{\partial L_1} \, \frac{\partial}{\partial L_2} \,
      F^{\nf{1}, \nf{2}}_{a_1 a_2}(L_1, L_2)
   &= \frac{\partial}{\partial L_1} \biggl(
      \sum_{b_2} P^{\nf{2}}_{a_2 b_2}(L_2)
         \conv{2} F^{\nf{1}, \nf{2}}_{a_1 b_2}(L_1, L_2) \biggr)
   \nonumber \\
   &= \sum_{b_2, b_1} P^{\nf{2}}_{a_2 b_2}(L_2)
      \conv{2} P^{\nf{1}}_{a_1 b_1}(L_1)
      \conv{1} F^{\nf{1}, \nf{2}}_{b_1 b_2}(L_1, L_2)
\intertext{is equal to}
   \label{eq:DPD-derivative-21}
   \frac{\partial}{\partial L_2} \, \frac{\partial}{\partial L_1} \,
      F^{\nf{1}, \nf{2}}_{a_1 a_2}(L_1, L_2)
   &= \frac{\partial}{\partial L_2} \biggl(
      \sum_{b_1} P^{\nf{1}}_{a_1 b_1}(L_1)
         \conv{1} F^{\nf{1}, \nf{2}}_{b_1 a_2}(L_1, L_2) \biggr)
   \nonumber \\
   &= \sum_{b_1, b_2} P^{\nf{1}}_{a_1 b_1}(L_1)
      \conv{1} P^{\nf{2}}_{a_2 b_2}(L_2)
      \conv{2} F^{\nf{1}, \nf{2}}_{b_1 b_2}(L_1, L_2)
   \,.
\end{align}%
Here and in the following, we change evolution variables from $\mu_i$ to
\begin{align}
   L_i &= 2 \ln \mu_i
\end{align}
and drop the arguments $x_1, x_2$, and $y$ of the DPDs for brevity.  For
evolution at fixed $(\nf{1}, \nf{2})$, i.e.\ without flavor matching, the
desired equality follows from the fact that the convolution integrals w.r.t.\
the first and second momentum fraction of the DPD commute with each other.  This
can readily be shown from their definition \eqref{eq:convol-def}.  Note that it
is essential for the argument that the evolution kernel for one parton does not
depend on the scale of the other parton.

We now extend this argument, showing that one obtains the same result when
flavor matching along different paths in the $(\nf{1}, \nf{2})$ plane.
Let us see how \eqs{DPD-derivative-12}{DPD-derivative-21} have to be
modified to include the effects from
flavor matching.  We first consider the case of PDFs, where the notation is less
involved. For a set of given matching scales $\tilde{\mu}_{n}$ for the
transition from $n - 1$ to $n$ active flavors, we define the ``canonical''
number of flavors at given $L = 2 \ln \mu$ as
\begin{align}
   \label{eq:canonical-nf}
   n(L) &=
   \begin{cases}
      3 & \text{ for $L < \widetilde{L}_{4}$} \\
      4 & \text{ for $\widetilde{L}_{4} < L < \widetilde{L}_{5}$} \\
      5 & \text{ for $\widetilde{L}_{5} < L < \widetilde{L}_{6}$} \\
      6 & \text{ for $L > \widetilde{L}_{6}$}
   \end{cases}
\end{align}%
where $\widetilde{L}_{n} = 2 \ln \tilde{\mu}_{n}$.  We further define PDFs
\begin{align}
   \label{eq:vfn-pdf-def}
      f^{}_{a}(L)
   &= f_{a}^{n(L)}(L)
      \,,
\end{align}
with the canonical number of flavors at a given renormalisation scale.  From
these distributions, one can obtain $f^{n}_{a}(L)$ for flavor numbers different
from the canonical values by evolution.\footnote{%
This implies that our argument is \emph{not} tied to the use of a specific
``variable flavor number scheme''.}

At the matching scales, the distributions in \eq{vfn-pdf-def} can have
discontinuities, which complicates their description via differential equations.
We therefore introduce a smoothen\-ed version of $f^{}_{a}(L)$.  Taking a single
matching step from $n-1$ to $n$ flavors at $\widetilde{L} = 2 \ln \tilde{\mu}$
for simplicity, one can define a smooth approximation $\theta_\epsilon(L)$ of the
step function that takes the values
\begin{align}
   \theta_\epsilon(t) &=
   \begin{cases}
      0  & \text{ for $t < - \epsilon$} \\
      1  & \text{ for $t >  \epsilon$}
   \end{cases}
\end{align}
and is continuous and monotonic in the intermediate region $- \epsilon < t <
\epsilon$.  Its derivative $\delta_\epsilon(t) = \dd \theta_\epsilon(t) / \dd t$
is then a smooth approximation of the Dirac distribution.  Define now a PDF
$f^{\epsilon}_{a}(L)$ that evolves according to
\begin{align}
   \label{eq:modified-dglap}
      \frac{\dd f^{\epsilon}_{a}(L)}{\dd L}
   &= \sum_{b}  P_{a b}^{n_\epsilon(L)}(L)
         \conv{} f^{\epsilon}_{b}(L)
   \nonumber \\
   & \quad
      + \delta_\epsilon(L - \widetilde{L})
         \biggl[
            \sum_{b} A_{a b}^{n}(m; \widetilde{L})
            \conv{} f^{\epsilon}_{b}(\widetilde{L} - \epsilon)
            - f^{\epsilon}_{a}(\widetilde{L} - \epsilon)
         \biggr]
      \,,
\end{align}%
with
\begin{align}
   n_\epsilon(L) = n - 1 + \theta_\epsilon(L - \widetilde{L})
\end{align}%
interpolating smoothly between $n-1$ and $n$.  The DGLAP kernels for the
evolution of $f^{\epsilon}_{a}(L)$ are thus evaluated with flavor number $n-1$
for $L \le \widetilde{L} - \epsilon$, with flavor number $n$  for $L \ge
\widetilde{L} + \epsilon$, and with fractional flavor number in the intermediate
$L$ region.\footnote{%
PDFs and DGLAP kernels for inactive flavors are set to
zero, so that the range of flavor sums does not explicitly depend on $\nf{}$.}
One furthermore has
\begin{align}
      f^{\epsilon}_{a}(\widetilde{L} + \epsilon)
   &= f^{\epsilon}_{a}(\widetilde{L} - \epsilon)
      + \int_{\widetilde{L} - \epsilon}^{\widetilde{L} + \epsilon}
         \dd L \, \frac{\dd f^{\epsilon}_{a}(L)}{\dd L}
   \nonumber \\
   &= f^{\epsilon}_{a}(\widetilde{L} - \epsilon)
      + \mathcal{O}(\epsilon)
      +  \int_{\widetilde{L} - \epsilon}^{\widetilde{L} + \epsilon}
         \dd L \; \delta_\epsilon(L - \widetilde{L})
   \nonumber \\
   & \qquad \times
      \biggl[
         \sum_{b} A_{a b}^{n}(m; \widetilde{L})
         \conv{} f^{\epsilon}_{b}(\widetilde{L} - \epsilon)
         - f^{\epsilon}_{a}(\widetilde{L} - \epsilon)
      \biggr]
   \nonumber \\[0.1em]
   &= \sum_{b} A_{a b}^{n}(m; \widetilde{L})
         \conv{} f^{\epsilon}_{b}(\widetilde{L} - \epsilon)
      + \mathcal{O}(\epsilon)
   \,,
\end{align}%
where the $\mathcal{O}(\epsilon)$ contribution comes from the integral over the
term with $P^{n_\epsilon(L)}$ in \eq{modified-dglap}.  In the limit $\epsilon
\to 0$, this tends to the flavor matching prescription given in
\eq{pdf-flavor-matching}.

It is straightforward to extend this construction from one to several matching
steps and from PDFs to DPDs.  In analogy to \eq{vfn-pdf-def}, DPDs with
canonical flavor numbers are defined as
\begin{align}
   \label{eq:vfn-dpd-def}
      F^{}_{a_1 a_2}(L_1, L_2)
   &= F_{a_1 a_2}^{n(L_1),\, n(L_2)}(L_1, L_2)
      \,,
\end{align}
and their smoothened version $F^{\epsilon}_{a_1 a_2}(L_1, L_2)$ evolves
according to
\begin{align}
   \label{eq:dpd-modified-dglap}
      \frac{\partial}{\partial L_1} \, F^{\epsilon}_{a_1 a_2}(L_1, L_2)
   &= \sum_{b_1} P^{n_\epsilon(L_1)}_{a_1 b_1}(L_1)
         \conv{1} F^{\epsilon}_{b_1 a_2}(L_1, L_2)
   \nonumber \\
   & \quad
      + \sum_{n} \delta_{\epsilon}(L_1 - \widetilde{L}_{n})
        \sum_{b_1} \Delta A^{n}_{a_1 b_1} \conv{1}
            F^{\epsilon}_{b_1 a_2}(\widetilde{L}_{n} - \epsilon, L_2)
   \,,
   \nonumber \\
   \frac{\partial}{\partial L_2} \, F^{\epsilon}_{a_1 a_2}(L_1, L_2)
   &= \sum_{b_2} P^{n_\epsilon(L_2)}_{a_2 b_2}(L_2)
         \conv{2} F^{\epsilon}_{a_1 b_2}(L_1, L_2)
   \nonumber \\
   &  \quad
      + \sum_{n} \delta_{\epsilon}(L_2 - \widetilde{L}_{n})
        \sum_{b_2} \Delta A^{n}_{a_2 b_2} \conv{2}
            F^{\epsilon}_{a_1 b_2}(L_1, \widetilde{L}_{n} - \epsilon)
   \,,
\end{align}%
where we abbreviated
\begin{align}
      \Delta A^{n}_{a b}(z)
   &= A^{n}_{a b}(z, m_{n}; \widetilde{L}_{n})
         - \delta_{a b} \, \delta(1-z)
      \,.
\end{align}
As at the beginning of this section, one can then show that
\begin{align}
      \frac{\partial}{\partial L_1} \, \frac{\partial}{\partial L_2} \,
      F^{\epsilon}_{a_1 a_2}(L_1, L_2)
   &= \frac{\partial}{\partial L_2} \, \frac{\partial}{\partial L_1} \,
      F^{\epsilon}_{a_1 a_2}(L_1, L_2)
      \,,
\end{align}
so that with an initial condition given in the $(L_1, L_2)$ plane,
\eq{dpd-modified-dglap} has a unique solution independent of the evolution path.
 Taking the limit $\epsilon \to 0$, the desired result is obtained.
%
%%%%%%%%%%%%%%%%%%%%%%%%%%%%%%%%%%%%%%%%%%%%%%%%%%%%%%%%%%%%%%%%%%%%%%%%%%%%%%%%
\addcontentsline{toc}{section}{References}
\bibliographystyle{jhep}
\bibliography{chili-dpd}

\providecommand{\href}[2]{#2}\begingroup\raggedright\begin{thebibliography}{10}

\bibitem{Abe:1997xk}
{\scshape CDF} collaboration, F.~Abe et~al., \emph{{Double parton scattering in
  $\bar{p}p$ collisions at $\sqrt{s}$ = 1.8~TeV}},
  \href{https://doi.org/10.1103/PhysRevD.56.3811}{\emph{Phys. Rev. D}
  {\bfseries 56} (1997) 3811}.

\bibitem{Abazov:2015nnn}
{\scshape D0} collaboration, V.~M. Abazov et~al., \emph{{Study of double parton
  interactions in diphoton + dijet events in $p\bar{p}$ collisions at
  $\sqrt{s}$ = 1.96 TeV}},
  \href{https://doi.org/10.1103/PhysRevD.93.052008}{\emph{Phys. Rev. D}
  {\bfseries 93} (2016) 052008}
  [\href{https://arxiv.org/abs/1512.05291}{{\ttfamily 1512.05291}}].

\bibitem{Aaij:2016bqq}
{\scshape LHCb} collaboration, R.~Aaij et~al., \emph{{Measurement of the
  J/$\psi$ pair production cross-section in pp collisions at $\sqrt{s}$ = 13
  TeV}}, \href{https://doi.org/10.1007/JHEP06(2017)047}{\emph{JHEP} {\bfseries
  06} (2017) 047} [\href{https://arxiv.org/abs/1612.07451}{{\ttfamily
  1612.07451}}].

\bibitem{ATLAS:2019jzd}
{\scshape ATLAS} collaboration, \emph{{Measurement of J/\ensuremath{\psi}
  production in association with a W$^{\pm}$ boson with pp data at 8 TeV}},
  \href{https://doi.org/10.1007/JHEP01(2020)095}{\emph{JHEP} {\bfseries 01}
  (2020) 095} [\href{https://arxiv.org/abs/1909.13626}{{\ttfamily
  1909.13626}}].

\bibitem{CMS:2019jcb}
{\scshape CMS} collaboration, \emph{{Evidence for $\text {W}\text {W}$
  production from double-parton interactions in proton\textendash{}proton
  collisions at $\sqrt{s} = 13 \,\text {TeV} $}},
  \href{https://doi.org/10.1140/epjc/s10052-019-7541-6}{\emph{Eur. Phys. J. C}
  {\bfseries 80} (2020) 41} [\href{https://arxiv.org/abs/1909.06265}{{\ttfamily
  1909.06265}}].

\bibitem{CMS:2021lxi}
{\scshape CMS} collaboration, \emph{{Measurement of double-parton scattering in
  inclusive production of four jets with low transverse momentum in
  proton-proton collisions at $ \sqrt{s} $ = 13 TeV}},
  \href{https://doi.org/10.1007/JHEP01(2022)177}{\emph{JHEP} {\bfseries 01}
  (2022) 177} [\href{https://arxiv.org/abs/2109.13822}{{\ttfamily
  2109.13822}}].

\bibitem{CMS:2022pio}
{\scshape CMS} collaboration, \emph{{Observation of same-sign WW production
  from double parton scattering in proton-proton collisions at $\sqrt{s}$ = 13
  TeV}},  \href{https://arxiv.org/abs/2206.02681}{{\ttfamily 2206.02681}}.

\bibitem{Kulesza:1999zh}
A.~Kulesza and W.~Stirling, \emph{{Like sign $W$ boson production at the LHC as
  a probe of double parton scattering}},
  \href{https://doi.org/10.1016/S0370-2693(99)01512-9}{\emph{Phys. Lett. B}
  {\bfseries 475} (2000) 168}
  [\href{https://arxiv.org/abs/hep-ph/9912232}{{\ttfamily hep-ph/9912232}}].

\bibitem{Gaunt:2010pi}
J.~R. Gaunt, C.-H. Kom, A.~Kulesza and W.~Stirling, \emph{{Same-sign W pair
  production as a probe of double parton scattering at the LHC}},
  \href{https://doi.org/10.1140/epjc/s10052-010-1362-y}{\emph{Eur. Phys. J. C}
  {\bfseries 69} (2010) 53} [\href{https://arxiv.org/abs/1003.3953}{{\ttfamily
  1003.3953}}].

\bibitem{Golec-Biernat:2014nsa}
K.~Golec-Biernat and E.~Lewandowska, \emph{{Electroweak boson production in
  double parton scattering}},
  \href{https://doi.org/10.1103/PhysRevD.90.094032}{\emph{Phys. Rev. D}
  {\bfseries 90} (2014) 094032}
  [\href{https://arxiv.org/abs/1407.4038}{{\ttfamily 1407.4038}}].

\bibitem{Ceccopieri:2017oqe}
F.~A. Ceccopieri, M.~Rinaldi and S.~Scopetta, \emph{{Parton correlations in
  same-sign $W$ pair production via double parton scattering at the LHC}},
  \href{https://doi.org/10.1103/PhysRevD.95.114030}{\emph{Phys. Rev. D}
  {\bfseries 95} (2017) 114030}
  [\href{https://arxiv.org/abs/1702.05363}{{\ttfamily 1702.05363}}].

\bibitem{Cotogno:2018mfv}
S.~Cotogno, T.~Kasemets and M.~Myska, \emph{{Spin on same-sign $W$-boson pair
  production}}, \href{https://doi.org/10.1103/PhysRevD.100.011503}{\emph{Phys.\
  Rev.\ D} {\bfseries 100} (2019) 011503}
  [\href{https://arxiv.org/abs/1809.09024}{{\ttfamily 1809.09024}}].

\bibitem{Cotogno:2020iio}
S.~Cotogno, T.~Kasemets and M.~Myska, \emph{{Confronting same-sign W-boson
  production with parton correlations}},
  \href{https://doi.org/10.1007/JHEP10(2020)214}{\emph{JHEP} {\bfseries 10}
  (2020) 214} [\href{https://arxiv.org/abs/2003.03347}{{\ttfamily
  2003.03347}}].

\bibitem{Khachatryan:2016kod}
{\scshape CMS} collaboration, \emph{{Search for new physics in same-sign
  dilepton events in proton\textendash{}proton collisions at $\sqrt{s} =
  13\,\text {TeV} $}},
  \href{https://doi.org/10.1140/epjc/s10052-016-4261-z}{\emph{Eur. Phys. J. C}
  {\bfseries 76} (2016) 439}
  [\href{https://arxiv.org/abs/1605.03171}{{\ttfamily 1605.03171}}].

\bibitem{Sirunyan:2018yun}
{\scshape CMS} collaboration, \emph{{Search for top quark partners with charge
  5/3 in the same-sign dilepton and single-lepton final states in proton-proton
  collisions at $ \sqrt{s}=13 $ TeV}},
  \href{https://doi.org/10.1007/JHEP03(2019)082}{\emph{JHEP} {\bfseries 03}
  (2019) 082} [\href{https://arxiv.org/abs/1810.03188}{{\ttfamily
  1810.03188}}].

\bibitem{Blok:2010ge}
B.~Blok, Y.~Dokshitzer, L.~Frankfurt and M.~Strikman, \emph{{The Four jet
  production at LHC and Tevatron in QCD}},
  \href{https://doi.org/10.1103/PhysRevD.83.071501}{\emph{Phys. Rev. D}
  {\bfseries 83} (2011) 071501}
  [\href{https://arxiv.org/abs/1009.2714}{{\ttfamily 1009.2714}}].

\bibitem{Gaunt:2011xd}
J.~R. Gaunt and W.~Stirling, \emph{{Double Parton Scattering Singularity in
  One-Loop Integrals}},
  \href{https://doi.org/10.1007/JHEP06(2011)048}{\emph{JHEP} {\bfseries 06}
  (2011) 048} [\href{https://arxiv.org/abs/1103.1888}{{\ttfamily 1103.1888}}].

\bibitem{Ryskin:2011kk}
M.~G. Ryskin and A.~M. Snigirev, \emph{{A Fresh look at double parton
  scattering}}, \href{https://doi.org/10.1103/PhysRevD.83.114047}{\emph{Phys.
  Rev.} {\bfseries D83} (2011) 114047}
  [\href{https://arxiv.org/abs/1103.3495}{{\ttfamily 1103.3495}}].

\bibitem{Diehl:2011yj}
M.~Diehl, D.~Ostermeier and A.~Sch{\"a}fer, \emph{{Elements of a theory for
  multiparton interactions in QCD}},
  \href{https://doi.org/10.1007/JHEP03(2012)089}{\emph{JHEP} {\bfseries 03}
  (2012) 089} [\href{https://arxiv.org/abs/1111.0910}{{\ttfamily 1111.0910}}].

\bibitem{Manohar:2012jr}
A.~V. Manohar and W.~J. Waalewijn, \emph{{A QCD Analysis of Double Parton
  Scattering: Color Correlations, Interference Effects and Evolution}},
  \href{https://doi.org/10.1103/PhysRevD.85.114009}{\emph{Phys. Rev. D}
  {\bfseries 85} (2012) 114009}
  [\href{https://arxiv.org/abs/1202.3794}{{\ttfamily 1202.3794}}].

\bibitem{Ryskin:2012qx}
M.~G. Ryskin and A.~M. Snigirev, \emph{{Double parton scattering in double
  logarithm approximation of perturbative QCD}},
  \href{https://doi.org/10.1103/PhysRevD.86.014018}{\emph{Phys. Rev.}
  {\bfseries D86} (2012) 014018}
  [\href{https://arxiv.org/abs/1203.2330}{{\ttfamily 1203.2330}}].

\bibitem{Gaunt:2012dd}
J.~R. Gaunt, \emph{{Single Perturbative Splitting Diagrams in Double Parton
  Scattering}}, \href{https://doi.org/10.1007/JHEP01(2013)042}{\emph{JHEP}
  {\bfseries 01} (2013) 042} [\href{https://arxiv.org/abs/1207.0480}{{\ttfamily
  1207.0480}}].

\bibitem{Blok:2013bpa}
B.~Blok, Y.~Dokshitzer, L.~Frankfurt and M.~Strikman, \emph{{Perturbative QCD
  correlations in multi-parton collisions}},
  \href{https://doi.org/10.1140/epjc/s10052-014-2926-z}{\emph{Eur. Phys. J. C}
  {\bfseries 74} (2014) 2926}
  [\href{https://arxiv.org/abs/1306.3763}{{\ttfamily 1306.3763}}].

\bibitem{Diehl:2017kgu}
M.~Diehl, J.~R. Gaunt and K.~Sch\"onwald, \emph{{Double hard scattering without
  double counting}}, \href{https://doi.org/10.1007/JHEP06(2017)083}{\emph{JHEP}
  {\bfseries 06} (2017) 083}
  [\href{https://arxiv.org/abs/1702.06486}{{\ttfamily 1702.06486}}].

\bibitem{Fedkevych:2020cmd}
O.~Fedkevych and A.~Kulesza, \emph{{Double parton scattering in four-jet
  production in proton-proton collisions at the LHC}},
  \href{https://doi.org/10.1103/PhysRevD.104.054021}{\emph{Phys. Rev. D}
  {\bfseries 104} (2021) 054021}
  [\href{https://arxiv.org/abs/2008.08347}{{\ttfamily 2008.08347}}].

\bibitem{Bartalini:2017jkk}
P.~Bartalini and J.~R. Gaunt, eds., \emph{{Multiple Parton Interactions at the
  LHC}}, vol.~29. World Scientific Publishing, 2019,
  \href{https://doi.org/10.1142/10646}{10.1142/10646}.

\bibitem{Kasemets:2017vyh}
T.~Kasemets and S.~Scopetta, \emph{{Parton correlations in double parton
  scattering}}, \href{https://doi.org/10.1142/9789813227767_0004}{\emph{Adv.
  Ser. Direct. High Energy Phys.} {\bfseries 29} (2018) 49}
  [\href{https://arxiv.org/abs/1712.02884}{{\ttfamily 1712.02884}}].

\bibitem{Gaunt:2009re}
J.~R. Gaunt and W.~J. Stirling, \emph{{Double Parton Distributions
  Incorporating Perturbative QCD Evolution and Momentum and Quark Number Sum
  Rules}}, \href{https://doi.org/10.1007/JHEP03(2010)005}{\emph{JHEP}
  {\bfseries 03} (2010) 005} [\href{https://arxiv.org/abs/0910.4347}{{\ttfamily
  0910.4347}}].

\bibitem{Golec-Biernat:2015aza}
K.~Golec-Biernat, E.~Lewandowska, M.~Serino, Z.~Snyder and A.~M. Stasto,
  \emph{{Constraining the double gluon distribution by the single gluon
  distribution}},
  \href{https://doi.org/10.1016/j.physletb.2015.09.067}{\emph{Phys. Lett. B}
  {\bfseries 750} (2015) 559}
  [\href{https://arxiv.org/abs/1507.08583}{{\ttfamily 1507.08583}}].

\bibitem{Diehl:2020xyg}
M.~Diehl, J.~R. Gaunt, D.~M. Lang, P.~Pl\"o\ss{}l and A.~Sch\"afer, \emph{{Sum
  rule improved double parton distributions in position space}},
  \href{https://doi.org/10.1140/epjc/s10052-020-8038-z}{\emph{Eur. Phys. J. C}
  {\bfseries 80} (2020) 468}
  [\href{https://arxiv.org/abs/2001.10428}{{\ttfamily 2001.10428}}].

\bibitem{Golec-Biernat:2022wkx}
K.~Golec-Biernat and A.~M. Sta\'sto, \emph{{Momentum sum rule and factorization
  of double parton distributions}},
  \href{https://doi.org/10.1103/PhysRevD.107.054020}{\emph{Phys. Rev. D}
  {\bfseries 107} (2023) 054020}
  [\href{https://arxiv.org/abs/2212.02289}{{\ttfamily 2212.02289}}].

\bibitem{Bali:2021gel}
G.~S. Bali, M.~Diehl, B.~Gl\"a\ss{}le, A.~Sch\"afer and C.~Zimmermann,
  \emph{{Double parton distributions in the nucleon from lattice QCD}},
  \href{https://doi.org/10.1007/JHEP09(2021)106}{\emph{JHEP} {\bfseries 09}
  (2021) 106} [\href{https://arxiv.org/abs/2106.03451}{{\ttfamily
  2106.03451}}].

\bibitem{Chang:2012nw}
H.-M. Chang, A.~V. Manohar and W.~J. Waalewijn, \emph{{Double Parton
  Correlations in the Bag Model}},
  \href{https://doi.org/10.1103/PhysRevD.87.034009}{\emph{Phys. Rev. D}
  {\bfseries 87} (2013) 034009}
  [\href{https://arxiv.org/abs/1211.3132}{{\ttfamily 1211.3132}}].

\bibitem{Rinaldi:2013vpa}
M.~Rinaldi, S.~Scopetta and V.~Vento, \emph{{Double parton correlations in
  constituent quark models}},
  \href{https://doi.org/10.1103/PhysRevD.87.114021}{\emph{Phys. Rev. D}
  {\bfseries 87} (2013) 114021}
  [\href{https://arxiv.org/abs/1302.6462}{{\ttfamily 1302.6462}}].

\bibitem{Broniowski:2013xba}
W.~Broniowski and E.~Ruiz~Arriola, \emph{{Valence double parton distributions
  of the nucleon in a simple model}},
  \href{https://doi.org/10.1007/s00601-014-0840-4}{\emph{Few Body Syst.}
  {\bfseries 55} (2014) 381} [\href{https://arxiv.org/abs/1310.8419}{{\ttfamily
  1310.8419}}].

\bibitem{Rinaldi:2014ddl}
M.~Rinaldi, S.~Scopetta, M.~Traini and V.~Vento, \emph{{Double parton
  correlations and constituent quark models: a Light Front approach to the
  valence sector}}, \href{https://doi.org/10.1007/JHEP12(2014)028}{\emph{JHEP}
  {\bfseries 12} (2014) 028} [\href{https://arxiv.org/abs/1409.1500}{{\ttfamily
  1409.1500}}].

\bibitem{Broniowski:2016trx}
W.~Broniowski, E.~Ruiz~Arriola and K.~Golec-Biernat, \emph{{Generalized Valon
  Model for Double Parton Distributions}},
  \href{https://doi.org/10.1007/s00601-016-1087-z}{\emph{Few Body Syst.}
  {\bfseries 57} (2016) 405}
  [\href{https://arxiv.org/abs/1602.00254}{{\ttfamily 1602.00254}}].

\bibitem{Kasemets:2016nio}
T.~Kasemets and A.~Mukherjee, \emph{{Quark-gluon double parton distributions in
  the light-front dressed quark model}},
  \href{https://doi.org/10.1103/PhysRevD.94.074029}{\emph{Phys. Rev. D}
  {\bfseries 94} (2016) 074029}
  [\href{https://arxiv.org/abs/1606.05686}{{\ttfamily 1606.05686}}].

\bibitem{Rinaldi:2016jvu}
M.~Rinaldi, S.~Scopetta, M.~C. Traini and V.~Vento, \emph{{Correlations in
  Double Parton Distributions: Perturbative and Non-Perturbative effects}},
  \href{https://doi.org/10.1007/JHEP10(2016)063}{\emph{JHEP} {\bfseries 10}
  (2016) 063} [\href{https://arxiv.org/abs/1608.02521}{{\ttfamily
  1608.02521}}].

\bibitem{Rinaldi:2016mlk}
M.~Rinaldi and F.~A. Ceccopieri, \emph{{Relativistic effects in model
  calculations of double parton distribution function}},
  \href{https://doi.org/10.1103/PhysRevD.95.034040}{\emph{Phys. Rev. D}
  {\bfseries 95} (2017) 034040}
  [\href{https://arxiv.org/abs/1611.04793}{{\ttfamily 1611.04793}}].

\bibitem{Corke:2011yy}
R.~Corke and T.~Sjostrand, \emph{{Multiparton Interactions with an x-dependent
  Proton Size}}, \href{https://doi.org/10.1007/JHEP05(2011)009}{\emph{JHEP}
  {\bfseries 05} (2011) 009} [\href{https://arxiv.org/abs/1101.5953}{{\ttfamily
  1101.5953}}].

\bibitem{Blok:2015afa}
B.~Blok and P.~Gunnellini, \emph{{Dynamical approach to MPI in W+dijet and
  Z+dijet production within the PYTHIA event generator}},
  \href{https://doi.org/10.1140/epjc/s10052-016-4035-7}{\emph{Eur. Phys. J. C}
  {\bfseries 76} (2016) 202}
  [\href{https://arxiv.org/abs/1510.07436}{{\ttfamily 1510.07436}}].

\bibitem{Blok:2015rka}
B.~Blok and P.~Gunnellini, \emph{{Dynamical approach to MPI four-jet production
  in Pythia}}, \href{https://doi.org/10.1140/epjc/s10052-015-3520-8}{\emph{Eur.
  Phys. J. C} {\bfseries 75} (2015) 282}
  [\href{https://arxiv.org/abs/1503.08246}{{\ttfamily 1503.08246}}].

\bibitem{Cabouat:2019gtm}
B.~Cabouat, J.~R. Gaunt and K.~Ostrolenk, \emph{{A Monte-Carlo Simulation of
  Double Parton Scattering}},
  \href{https://doi.org/10.1007/JHEP11(2019)061}{\emph{JHEP} {\bfseries 11}
  (2019) 061} [\href{https://arxiv.org/abs/1906.04669}{{\ttfamily
  1906.04669}}].

\bibitem{Cabouat:2020ssr}
B.~Cabouat and J.~R. Gaunt, \emph{{Combining single and double parton
  scatterings in a parton shower}},
  \href{https://doi.org/10.1007/JHEP10(2020)012}{\emph{JHEP} {\bfseries 10}
  (2020) 012} [\href{https://arxiv.org/abs/2008.01442}{{\ttfamily
  2008.01442}}].

\bibitem{Cafarella:2008du}
A.~Cafarella, C.~Coriano and M.~Guzzi, \emph{{Precision Studies of the NNLO
  DGLAP Evolution at the LHC with CANDIA}},
  \href{https://doi.org/10.1016/j.cpc.2008.06.004}{\emph{Comput. Phys. Commun.}
  {\bfseries 179} (2008) 665}
  [\href{https://arxiv.org/abs/0803.0462}{{\ttfamily 0803.0462}}].

\bibitem{Salam:2008qg}
G.~P. Salam and J.~Rojo, \emph{{A Higher Order Perturbative Parton Evolution
  Toolkit (HOPPET)}},
  \href{https://doi.org/10.1016/j.cpc.2008.08.010}{\emph{Comput. Phys. Commun.}
  {\bfseries 180} (2009) 120}
  [\href{https://arxiv.org/abs/0804.3755}{{\ttfamily 0804.3755}}].

\bibitem{Botje:2010ay}
M.~Botje, \emph{{QCDNUM: Fast QCD Evolution and Convolution}},
  \href{https://doi.org/10.1016/j.cpc.2010.10.020}{\emph{Comput. Phys. Commun.}
  {\bfseries 182} (2011) 490}
  [\href{https://arxiv.org/abs/1005.1481}{{\ttfamily 1005.1481}}].

\bibitem{Bertone:2013vaa}
V.~Bertone, S.~Carrazza and J.~Rojo, \emph{{APFEL: A PDF Evolution Library with
  QED corrections}},
  \href{https://doi.org/10.1016/j.cpc.2014.03.007}{\emph{Comput. Phys. Commun.}
  {\bfseries 185} (2014) 1647}
  [\href{https://arxiv.org/abs/1310.1394}{{\ttfamily 1310.1394}}].

\bibitem{Bertone:2017gds}
V.~Bertone, \emph{{APFEL++: A new PDF evolution library in C++}},
  \href{https://doi.org/10.22323/1.297.0201}{\emph{PoS} {\bfseries DIS2017}
  (2018) 201} [\href{https://arxiv.org/abs/1708.00911}{{\ttfamily
  1708.00911}}].

\bibitem{Weinzierl:2002mv}
S.~Weinzierl, \emph{{Fast evolution of parton distributions}},
  \href{https://doi.org/10.1016/S0010-4655(02)00584-2}{\emph{Comput. Phys.
  Commun.} {\bfseries 148} (2002) 314}
  [\href{https://arxiv.org/abs/hep-ph/0203112}{{\ttfamily hep-ph/0203112}}].

\bibitem{Vogt:2004ns}
A.~Vogt, \emph{{Efficient evolution of unpolarized and polarized parton
  distributions with QCD-PEGASUS}},
  \href{https://doi.org/10.1016/j.cpc.2005.03.103}{\emph{Comput. Phys. Commun.}
  {\bfseries 170} (2005) 65}
  [\href{https://arxiv.org/abs/hep-ph/0408244}{{\ttfamily hep-ph/0408244}}].

\bibitem{Candido:2022tld}
A.~Candido, F.~Hekhorn and G.~Magni, \emph{{EKO: evolution kernel operators}},
  \href{https://doi.org/10.1140/epjc/s10052-022-10878-w}{\emph{Eur. Phys. J. C}
  {\bfseries 82} (2022) 976}
  [\href{https://arxiv.org/abs/2202.02338}{{\ttfamily 2202.02338}}].

\bibitem{Buckley:2014ana}
A.~Buckley, J.~Ferrando, S.~Lloyd, K.~Nordstr\"om, B.~Page, M.~R\"ufenacht
  et~al., \emph{{LHAPDF6: parton density access in the LHC precision era}},
  \href{https://doi.org/10.1140/epjc/s10052-015-3318-8}{\emph{Eur. Phys. J. C}
  {\bfseries 75} (2015) 132} [\href{https://arxiv.org/abs/1412.7420}{{\ttfamily
  1412.7420}}].

\bibitem{Diehl:2021gvs}
M.~Diehl, R.~Nagar and F.~J. Tackmann, \emph{{ChiliPDF: Chebyshev interpolation
  for parton distributions}},
  \href{https://doi.org/10.1140/epjc/s10052-022-10223-1}{\emph{Eur. Phys. J. C}
  {\bfseries 82} (2022) 257}
  [\href{https://arxiv.org/abs/2112.09703}{{\ttfamily 2112.09703}}].

\bibitem{Diehl:2013mla}
M.~Diehl and T.~Kasemets, \emph{{Positivity bounds on double parton
  distributions}}, \href{https://doi.org/10.1007/JHEP05(2013)150}{\emph{JHEP}
  {\bfseries 05} (2013) 150} [\href{https://arxiv.org/abs/1303.0842}{{\ttfamily
  1303.0842}}].

\bibitem{Kirschner:1979im}
R.~Kirschner, \emph{{Generalized {Lipatov-Altarelli-Parisi} Equations and Jet
  Calculus Rules}},
  \href{https://doi.org/10.1016/0370-2693(79)90300-9}{\emph{Phys. Lett. B}
  {\bfseries 84} (1979) 266}.

\bibitem{Shelest:1982dg}
V.~P. Shelest, A.~M. Snigirev and G.~M. Zinovev, \emph{{The Multiparton
  Distribution Equations in {QCD}}},
  \href{https://doi.org/10.1016/0370-2693(82)90049-1}{\emph{Phys. Lett. B}
  {\bfseries 113} (1982) 325}.

\bibitem{Snigirev:2003cq}
A.~M. Snigirev, \emph{{Double parton distributions in the leading logarithm
  approximation of perturbative QCD}},
  \href{https://doi.org/10.1103/PhysRevD.68.114012}{\emph{Phys. Rev. D}
  {\bfseries 68} (2003) 114012}
  [\href{https://arxiv.org/abs/hep-ph/0304172}{{\ttfamily hep-ph/0304172}}].

\bibitem{Ceccopieri:2010kg}
F.~A. Ceccopieri, \emph{{An update on the evolution of double parton
  distributions}},
  \href{https://doi.org/10.1016/j.physletb.2011.02.047}{\emph{Phys. Lett. B}
  {\bfseries 697} (2011) 482}
  [\href{https://arxiv.org/abs/1011.6586}{{\ttfamily 1011.6586}}].

\bibitem{Buza:1996wv}
M.~Buza, Y.~Matiounine, J.~Smith and W.~L. van Neerven, \emph{{Charm
  electroproduction viewed in the variable flavor number scheme versus fixed
  order perturbation theory}},
  \href{https://doi.org/10.1007/BF01245820}{\emph{Eur. Phys. J. C} {\bfseries
  1} (1998) 301} [\href{https://arxiv.org/abs/hep-ph/9612398}{{\ttfamily
  hep-ph/9612398}}].

\bibitem{Moch:2004pa}
S.~Moch, J.~A.~M. Vermaseren and A.~Vogt, \emph{{The Three loop splitting
  functions in QCD: The Nonsinglet case}},
  \href{https://doi.org/10.1016/j.nuclphysb.2004.03.030}{\emph{Nucl. Phys.}
  {\bfseries B688} (2004) 101}
  [\href{https://arxiv.org/abs/hep-ph/0403192}{{\ttfamily hep-ph/0403192}}].

\bibitem{Vogt:2004mw}
A.~Vogt, S.~Moch and J.~A.~M. Vermaseren, \emph{{The Three-loop splitting
  functions in QCD: The Singlet case}},
  \href{https://doi.org/10.1016/j.nuclphysb.2004.04.024}{\emph{Nucl. Phys.}
  {\bfseries B691} (2004) 129}
  [\href{https://arxiv.org/abs/hep-ph/0404111}{{\ttfamily hep-ph/0404111}}].

\bibitem{Moch:2014sna}
S.~Moch, J.~A.~M. Vermaseren and A.~Vogt, \emph{{The Three-Loop Splitting
  Functions in QCD: The Helicity-Dependent Case}},
  \href{https://doi.org/10.1016/j.nuclphysb.2014.10.016}{\emph{Nucl. Phys. B}
  {\bfseries 889} (2014) 351}
  [\href{https://arxiv.org/abs/1409.5131}{{\ttfamily 1409.5131}}].

\bibitem{Moch:2015usa}
S.~Moch, J.~A.~M. Vermaseren and A.~Vogt, \emph{{On ${\gamma}_5$ in
  higher-order QCD calculations and the NNLO evolution of the polarized valence
  distribution}},
  \href{https://doi.org/10.1016/j.physletb.2015.07.027}{\emph{Phys. Lett. B}
  {\bfseries 748} (2015) 432}
  [\href{https://arxiv.org/abs/1506.04517}{{\ttfamily 1506.04517}}].

\bibitem{Blumlein:2021enk}
J.~Bl\"umlein, P.~Marquard, C.~Schneider and K.~Sch\"onwald, \emph{{The
  three-loop unpolarized and polarized non-singlet anomalous dimensions from
  off shell operator matrix elements}},
  \href{https://doi.org/10.1016/j.nuclphysb.2021.115542}{\emph{Nucl. Phys. B}
  {\bfseries 971} (2021) 115542}
  [\href{https://arxiv.org/abs/2107.06267}{{\ttfamily 2107.06267}}].

\bibitem{Blumlein:2021ryt}
J.~Bl\"umlein, P.~Marquard, C.~Schneider and K.~Sch\"onwald, \emph{{The
  three-loop polarized singlet anomalous dimensions from off-shell operator
  matrix elements}}, \href{https://doi.org/10.1007/JHEP01(2022)193}{\emph{JHEP}
  {\bfseries 01} (2022) 193}
  [\href{https://arxiv.org/abs/2111.12401}{{\ttfamily 2111.12401}}].

\bibitem{Vogelsang:1997ak}
W.~Vogelsang, \emph{{Next-to-leading order evolution of transversity
  distributions and Soffer's inequality}},
  \href{https://doi.org/10.1103/PhysRevD.57.1886}{\emph{Phys. Rev. D}
  {\bfseries 57} (1998) 1886}
  [\href{https://arxiv.org/abs/hep-ph/9706511}{{\ttfamily hep-ph/9706511}}].

\bibitem{Vogelsang:1998yd}
W.~Vogelsang, \emph{{$Q^2$ evolution of spin dependent parton densities}},
  {\emph{Acta Phys. Polon. B} {\bfseries 29} (1998) 1189}
  [\href{https://arxiv.org/abs/hep-ph/9805295}{{\ttfamily hep-ph/9805295}}].

\bibitem{Ablinger:2014lka}
J.~Ablinger, J.~Bl\"umlein, A.~De~Freitas, A.~Hasselhuhn, A.~von Manteuffel,
  M.~Round et~al., \emph{{The Transition Matrix Element $A_{gq}(N)$ of the
  Variable Flavor Number Scheme at $O(\alpha_s^3)$}},
  \href{https://doi.org/10.1016/j.nuclphysb.2014.02.007}{\emph{Nucl. Phys. B}
  {\bfseries 882} (2014) 263}
  [\href{https://arxiv.org/abs/1402.0359}{{\ttfamily 1402.0359}}].

\bibitem{Ablinger:2014vwa}
J.~Ablinger, A.~Behring, J.~Bl\"umlein, A.~De~Freitas, A.~Hasselhuhn, A.~von
  Manteuffel et~al., \emph{{The 3-Loop Non-Singlet Heavy Flavor Contributions
  and Anomalous Dimensions for the Structure Function $F_2(x,Q^2)$ and
  Transversity}},
  \href{https://doi.org/10.1016/j.nuclphysb.2014.07.010}{\emph{Nucl. Phys. B}
  {\bfseries 886} (2014) 733}
  [\href{https://arxiv.org/abs/1406.4654}{{\ttfamily 1406.4654}}].

\bibitem{Behring:2014eya}
A.~Behring, I.~Bierenbaum, J.~Bl\"umlein, A.~De~Freitas, S.~Klein and
  F.~Wi\ss{}brock, \emph{{The logarithmic contributions to the $O(\alpha^3_s)$
  asymptotic massive Wilson coefficients and operator matrix elements in deeply
  inelastic scattering}},
  \href{https://doi.org/10.1140/epjc/s10052-014-3033-x}{\emph{Eur. Phys. J. C}
  {\bfseries 74} (2014) 3033}
  [\href{https://arxiv.org/abs/1403.6356}{{\ttfamily 1403.6356}}].

\bibitem{Diehl:2019rdh}
M.~Diehl, J.~R. Gaunt, P.~Pl\"o\ss{}l and A.~Sch\"afer, \emph{{Two-loop
  splitting in double parton distributions}},
  \href{https://doi.org/10.21468/SciPostPhys.7.2.017}{\emph{SciPost Phys.}
  {\bfseries 7} (2019) 017} [\href{https://arxiv.org/abs/1902.08019}{{\ttfamily
  1902.08019}}].

\bibitem{Diehl:2021wpp}
M.~Diehl, J.~R. Gaunt and P.~Pl{\"o}{\ss}l, \emph{{Two-loop splitting in double
  parton distributions: the colour non-singlet case}},
  \href{https://doi.org/10.1007/JHEP08(2021)040}{\emph{JHEP} {\bfseries 08}
  (2021) 040} [\href{https://arxiv.org/abs/2105.08425}{{\ttfamily
  2105.08425}}].

\bibitem{Bailey:2020ooq}
S.~Bailey, T.~Cridge, L.~A. Harland-Lang, A.~D. Martin and R.~S. Thorne,
  \emph{{Parton distributions from LHC, HERA, Tevatron and fixed target data:
  MSHT20 PDFs}},
  \href{https://doi.org/10.1140/epjc/s10052-021-09057-0}{\emph{Eur. Phys. J. C}
  {\bfseries 81} (2021) 341}
  [\href{https://arxiv.org/abs/2012.04684}{{\ttfamily 2012.04684}}].

\bibitem{ATLAS:2014yjd}
{\scshape ATLAS} collaboration, \emph{{Measurement of the production cross
  section of prompt $J/\psi$ mesons in association with a $W^\pm$ boson in $pp$
  collisions at $\sqrt{s} =$ 7 TeV with the ATLAS detector}},
  \href{https://doi.org/10.1007/JHEP04(2014)172}{\emph{JHEP} {\bfseries 04}
  (2014) 172} [\href{https://arxiv.org/abs/1401.2831}{{\ttfamily 1401.2831}}].

\bibitem{Trefethen}
L.~N. Trefethen, \emph{{Approximation Theory and Approximation Practice}}.
  Society for Industrial and Applied Mathematics, 2012.

\bibitem{PRINCE198167}
P.~Prince and J.~Dormand, \emph{{High order embedded Runge-Kutta formulae}},
  \href{https://doi.org/https://doi.org/10.1016/0771-050X(81)90010-3}{\emph{Journal
  of Computational and Applied Mathematics} {\bfseries 7} (1981) 67 }.

\bibitem{Giele:2002hx}
W.~Giele et~al., \emph{{The QCD / SM working group: Summary report}},  in
  \emph{{Physics at TeV colliders. Proceedings, Euro Summer School, Les
  Houches, France, May 21-June 1, 2001}},
  \href{https://arxiv.org/abs/hep-ph/0204316}{{\ttfamily hep-ph/0204316}}.

\bibitem{Dittmar:2005ed}
M.~Dittmar et~al., \emph{{Working Group I: Parton distributions: Summary report
  for the HERA LHC Workshop Proceedings}},
  \href{https://arxiv.org/abs/hep-ph/0511119}{{\ttfamily hep-ph/0511119}}.

\bibitem{Diehl:2014vaa}
M.~Diehl, T.~Kasemets and S.~Keane, \emph{{Correlations in double parton
  distributions: effects of evolution}},
  \href{https://doi.org/10.1007/JHEP05(2014)118}{\emph{JHEP} {\bfseries 05}
  (2014) 118} [\href{https://arxiv.org/abs/1401.1233}{{\ttfamily 1401.1233}}].

\bibitem{Diehl:2022dia}
M.~Diehl, R.~Nagar and P.~Pl\"o\ss{}l, \emph{{Quark mass effects in double
  parton distributions}},  \href{https://arxiv.org/abs/2212.07736}{{\ttfamily
  2212.07736}}.

\bibitem{Diehl:2018kgr}
M.~Diehl, P.~Pl\"o\ss{}l and A.~Sch\"afer, \emph{{Proof of sum rules for double
  parton distributions in QCD}},
  \href{https://doi.org/10.1140/epjc/s10052-019-6777-5}{\emph{Eur. Phys. J. C}
  {\bfseries 79} (2019) 253}
  [\href{https://arxiv.org/abs/1811.00289}{{\ttfamily 1811.00289}}].

\bibitem{Diehl:2022rxb}
M.~Diehl, F.~Fabry and A.~Vladimirov, \emph{{Two-loop evolution kernels for
  colour dependent double parton distributions}},
  \href{https://doi.org/10.1007/JHEP05(2023)067}{\emph{JHEP} {\bfseries 05}
  (2023) 067} [\href{https://arxiv.org/abs/2212.11843}{{\ttfamily
  2212.11843}}].

\end{thebibliography}\endgroup

\end{document}